\documentclass[preprint,onecolumn,nofootinbib]{revtex4}
\pdfoutput=1
\usepackage[colorlinks=true,linkcolor=blue,urlcolor=blue,filecolor=black,citecolor=red,pdfstartview=FitV,pdftitle={},pdfsubject={},pdfkeywords={},pdfpagemode=None,bookmarksopen=true]{hyperref}
\usepackage{graphicx}
\usepackage{amsmath}
\usepackage{amsfonts}
\usepackage{amssymb}
\usepackage{color}%
\usepackage{dcolumn}
\newcommand{\f}{\begin{equation}}
\newcommand{\ff}{\end{equation}}
\newcommand{\fa}{\begin{eqnarray}}
\newcommand{\ffa}{\end{eqnarray}}

\begin{document}
\title{Transport phenomena and Weyl correction in effective holographic theory of momentum dissipation}
\author{Jian-Pin Wu}
\email{jianpinwu@mail.bnu.edu.cn}
\affiliation{
Center for Gravitation and Cosmology, College of Physical Science and Technology, Yangzhou University, Yangzhou 225009, China}
\begin{abstract}

We construct a higher derivative theory involving an axionic field and the Weyl tensor in four dimensional spacetime.
Up to the first order of the coupling parameters, the charged black brane solution with momentum dissipation in a perturbative manner is constructed.
Metal-insulator transitions are implemented when varying the system parameters at zero temperature.
Also, we study the transports including DC conductivity and optical conductivity at zero charge density.
We observe the exact particle-vortex duality for some specific momentum dissipation strength.

\end{abstract}
\maketitle
\section{Introduction}

The quantum critical (QC) system has long been a central and challenging subject in condensed matter physics \cite{Sachdev:2000qpt}.
It is believed to account for the most interesting phenomena, such as the strange metal and pseudo-gap phase, in strongly correlated quantum materials.
The QC system is associated with a QC phase transition and a QC phase.
Since the QC system is strongly correlated, the conventional perturbative tools in traditional field theory, unfortunately, lose their power.
We need to develop novel non-perturbative techniques and methods.

The AdS/CFT correspondence \cite{Maldacena:1997re,Gubser:1998bc,Witten:1998qj,Aharony:1999ti},
mapping a strongly coupled quantum field
theory to a weakly coupled gravitational theory in the large N limit,
provides a powerful tool to the study of QC physics and has led to great progress.
Especially, the metal-insulator transition (MIT), a special example of the QC phase transition,
has been widely studied in the holographic framework; for instance see \cite{Donos:2012js,Donos:2013eha,Donos:2014uba,Ling:2015ghh,Ling:2015dma,Ling:2015epa,Ling:2015exa,Ling:2016wyr,Ling:2016dck,Baggioli:2014roa,Baggioli:2016oqk,Baggioli:2016oju,Donos:2014oha,Kiritsis:2015oxa}
and the references therein.
To implement an MIT in a holographic framework, the key point is to deform the infrared (IR) geometry
to a new fixed point by the introduction of momentum dissipation \cite{Donos:2012js,Donos:2013eha}.

Holographic QC phase at zero density has also been intensely explored in \cite{Myers:2010pk,Ritz:2008kh,WitczakKrempa:2012gn,WitczakKrempa:2013ht,Witczak-Krempa:2013nua,Witczak-Krempa:2013aea,Katz:2014rla,Sachdev:2011wg,Hartnoll:2016apf,Bai:2013tfa,Myers:2016wsu,Lucas:2017dqa}.
By studying transport phenomena, in particular the optical conductivity, from a probe Maxwell field coupled to the Weyl tensor $C_{\mu\nu\rho\sigma}$
on top of the Schwarzschild-AdS (SS-AdS) black brane background \cite{Myers:2010pk,Sachdev:2011wg,Hartnoll:2016apf,Ritz:2008kh,WitczakKrempa:2012gn,
WitczakKrempa:2013ht,Witczak-Krempa:2013nua,Witczak-Krempa:2013aea,Katz:2014rla},
one observed a non-trivial frequency dependent conductivity attributed to the introduction of the Weyl tensor.
It exhibits a peak, which resembles the particle response and we refer to this as the Damle-Sachdev (DS) peak \cite{Damle:1997rxu},
or a dip, which is similar to the behavior of the vortex response,
and is analogous to the one in the superfluid-insulator quantum critical point (QCP)
\footnote{Those kinds of peak and dip features have also been observed in probe branes and DBI action \cite{Chen:2017dsy}
and just higher terms in $F^2$ with $F$ being the Maxwell field strength \cite{Baggioli:2016oju}.} \cite{Myers:2010pk,Sachdev:2011wg,Hartnoll:2016apf}.

But the peak is not the standard Drude peak and the DC conductivity has a bound, which cannot approach zero.
When higher derivative (HD) terms are introduced, an arbitrarily sharp Drude-like peak can be observed at low frequency in the optical conductivity
and the bound of conductivity is violated such that a zero DC conductivity can be obtained at a specific parameter
\footnote{We would like to point out that this bound in the conductivity is formalized in ``almost" general theories in \cite{Grozdanov:2015qia,Ikeda:2016rqh}. But in more generic theories \cite{Baggioli:2016oqk,Gouteraux:2016wxj,Baggioli:2016pia},
this bound is also violated.} \cite{Witczak-Krempa:2013aea}.
Another step forward is the construction of a neutral scalar hair black brane by coupling the Weyl tensor with a neutral scalar field,
which provides a framework to describe the QC phase and a transition away from QCP \cite{Myers:2016wsu,Lucas:2017dqa}.

In this paper, we shall construct a higher derivative theory including the four derivative terms,
a simple summation of the Weyl tensor as well as a term from the trace of axions coupling with the gauge field,
and a six derivative term, a mixed term of the product of the Weyl tensor and the axionic field coupling with the gauge field,
and we obtain a charged black brane solution in a perturbative manner.
By using a perturbative method, some charged black brane solutions from higher derivative gravity theory
have been constructed; for instance see \cite{Ling:2016dck,Myers:2009ij,Liu:2008kt,Cai:2011uh,Dey:2015poa,Dey:2015ytd}
and the references therein.
Especially, in \cite{Ling:2016dck}, it is the first time that an MIT is realized in the framework of higher derivative gravity.
Along the line of \cite{Ling:2016dck}, we shall study the MIT physics of our present model.
Also, we explore the QC phase of this model at zero charge density.

We organize this paper as follows.
In Sect. \ref{sec-HF}, we construct the higher derivative model coupling axionic field and Weyl tensor with the gauge field.
Then the perturbative black brane solution is obtained in Sect. \ref{sec-BBS}.
In Sect. \ref{sec-DC-finite}, we calculate the DC conductivity at finite charge density and study the MIT at zero temperature.
The conductivity at zero charge density is explored in Sect. \ref{sec-cond}.
A brief discussion is presented in Sect. \ref{sec-discussion}.
The constraint on the coupling parameters is obtained in Appendix \ref{sec-Bounds}.

\section{Holographic model}\label{sec-HF}

We construct a higher derivative holographic effective theory including metric, axions and gauge field as follows:
\begin{subequations}
\label{action}
\begin{align}
\label{ac-ax}
&S_0=\int \mathrm{d}^4x\sqrt{-g}\left(R+\frac{6}{L^2}-\bar{\Phi}\right)
\,,
\
\\
\label{ac-SA}
&S_A=\int \mathrm{d}^4x\sqrt{-g}\left(-\frac{L^2}{8g_F^2}F_{\mu\nu}X^{\mu\nu\rho\sigma}F_{\rho\sigma}\right)
\,,
\end{align}
\end{subequations}
where
\begin{subequations}
\label{X-Phib}
\begin{align}
\label{X}
&X_{\mu\nu}^{\ \ \rho\sigma}=I_{\mu\nu}^{\ \ \rho\sigma}
-4\gamma_{1,0}L^2\bar{\Phi}I_{\mu\nu}^{\ \ \rho\sigma}
-8\gamma_{0,1}L^2C_{\mu\nu}^{\ \ \rho\sigma}
-8\gamma_{1,1}L^4\bar{\Phi}C_{\mu\nu}^{\ \ \rho\sigma}\,,
\
\\
\label{Phib}
&
\bar{\Phi}\equiv Tr[\Phi]\equiv \Phi^{\mu}_{\ \mu}\,,~~~~~~~~\Phi^{\mu}_{\ \nu}=\frac{1}{2}\sum_{I=x,y}\partial^{\mu}\phi_I\partial_{\nu}\phi_I\,.
\end{align}
\end{subequations}
A pair of spatial linear dependent axionic fields, $\phi_I=\alpha x_I$ with $I=x,y$ and $\alpha$ being a constant,
are introduced in the above action, which are responsible for dissipating the momentum of the dual boundary field.
$L$ is the radius of the AdS spacetimes. $g_F$ and $\gamma_{m,n}$ with $m,n=0,1$ are the dimensionless coupling parameters.
In what follows, we shall set $g_F=1$.
$
\Phi^{\mu}_{\ \nu}
$ is the second order derivative term with respect to axions.
The first term in the tensor $X$ gives the standard Maxwell term.
$I_{\mu\nu}^{\ \ \rho\sigma}$ is an identity matrix defined as
$I_{\mu\nu}^{\ \ \rho\sigma}=\delta_{\mu}^{\ \rho}\delta_{\nu}^{\ \sigma}-\delta_{\mu}^{\ \sigma}\delta_{\nu}^{\ \rho}$.
The second term can be classified as a four derivative term,
which is the term with $n=0,m=1$ in
\cite{Gouteraux:2016wxj,Baggioli:2016pia} (Eq. (2.13) in \cite{Gouteraux:2016wxj}).
The third term is also a four derivative one, constructed by the Weyl tensor, which has been well studied in \cite{Ling:2016dck,Wu:2016jjd}.
For consistency with the current literature \cite{Myers:2010pk,Wu:2016jjd,Fu:2017oqa}, we denote $\gamma_{0,1}=\gamma$ in what follows.
The last term is a $6$ derivative term constructed by axions and the Weyl tensor.
More higher derivative terms can be constructed in terms of axions, the Weyl tensor and the gauge field,
which we leave for future study.
It is easy to see that the new tensor $X$ possesses the same symmetry as
$
X_{\mu\nu\rho\sigma}=X_{[\mu\nu][\rho\sigma]}=X_{\rho\sigma\mu\nu}
$
, like in \cite{Myers:2010pk,Wu:2016jjd,Fu:2017oqa}.

The equations of motion (EOMs) can be straightforwardly derived from the above action \eqref{action},
\begin{subequations}
\label{eom}
\begin{align}
&
\nabla_{\mu}\Big[\nabla^{\mu}\phi_I\Big(1-\gamma_{1,0}L^4F^2-\gamma_{1,1}L^6C^{\mu\nu\rho\sigma}F_{\mu\nu}F_{\rho\sigma}\Big)\Big]=0
\,,
\label{eom-phi}
\\
&
\nabla_{\nu}(X^{\mu\nu\rho\sigma}F_{\rho\sigma})=0
\,,
\label{eom-max}
\\
&
R_{\mu\nu}-\frac{1}{2}R g_{\mu\nu}-\frac{3}{L^2}g_{\mu\nu}
-\frac{L^2}{2}(1-4\gamma_{1,0}L^2\bar{\Phi})\Big(F_{\mu\rho}F_{\nu}^{\ \rho}-\frac{1}{4}g_{\mu\nu}F_{\rho\sigma}F^{\rho\sigma}\Big)
\nonumber
\\
&
-\frac{L^2}{2}(1-\gamma_{1,0}L^4F^2-\gamma_{1,1}L^6C^{\mu\nu\rho\sigma}F_{\mu\nu}F_{\rho\sigma})\Big(\sum_{I=x,y}\partial_\mu\phi_I\partial_\nu\phi_I\Big)
+\frac{1}{2}g_{\mu\nu}\bar{\Phi}
\nonumber
\\
&
-L^4(\gamma+\gamma_{1,1}L^2\bar{\Phi})(G_{1\mu\nu}+G_{2\mu\nu}+G_{3\mu\nu})
=0\,,
\label{eom-ein}
\end{align}
\end{subequations}
where
\begin{subequations}
\label{Weyl-tensor}
\begin{align}
\label{G1}
G_{1\mu\nu}&=\frac{1}{2}g_{\mu\nu}R_{\alpha\beta\rho\sigma}F^{\alpha\beta}F^{\rho\sigma}
-3R_{(\mu|\alpha\beta\lambda|}F_{\nu)}^{\ \alpha}F^{\beta\lambda}
-2\nabla_{\alpha}\nabla_{\beta}(F^{\alpha}_{\ (\nu}F^{\beta}_{\ \mu)})
\,,
\
\\
\nonumber
G_{2\mu\nu}&=-g_{\mu\nu}R_{\alpha\beta}F^{\alpha\lambda}F^{\beta}_{\ \lambda}
+g_{\mu\nu}\nabla_{\alpha}\nabla_{\beta}(F^{\alpha}_{\ \lambda}F^{\beta\lambda})
+\Box(F_{\mu}^{\ \lambda}F_{\nu\lambda})
-2\nabla_\alpha\nabla_{(\mu}(F_{\nu)\beta}F^{\alpha\beta})
\
\\
\label{G2}
&
+2R_{\nu\alpha}F_{\mu}^{\ \beta}F^{\alpha}_{\ \beta}
+2R_{\alpha\beta}F^{\alpha}_{\ \mu}F^{\beta}_{\ \nu}
+2R_{\alpha\mu}F^{\alpha\beta}F_{\nu\beta}
\,,
\
\\
\label{G3}
G_{3\mu\nu}&=
\frac{1}{6}g_{\mu\nu}RF^2-\frac{1}{3}R_{\mu\nu}F^2-\frac{2}{3}RF^{\alpha}_{\ \mu}F_{\alpha\nu}
+\frac{1}{3}\nabla_{(\nu}\nabla_{\mu)} F^2-\frac{1}{3}g_{\mu\nu}\Box F^2\,.
\end{align}
\end{subequations}

Following Ref. \cite{Myers:2010pk}, we can construct the electromagnetic (EM) dual theory of \eqref{action} with \eqref{X-Phib}, which is
\fa
\label{ac-SB}
S_B=\int \mathrm{d}^4x\sqrt{-g}\Big(-\frac{L^2}{8\hat{g}_F}G_{\mu\nu}\widehat{X}^{\mu\nu\rho\sigma}G_{\rho\sigma}\Big)\,,
\ffa
where $\hat{g}_F^2\equiv 1/g_F^2$ and $G_{\mu\nu}\equiv\partial_{\mu}B_{\nu}-\partial_{\nu}B_{\mu}$.
The tensor $\widehat{X}$ is defined by
\fa
&&
\widehat{X}_{\mu\nu}^{\ \ \rho\sigma}=-\frac{1}{4}\varepsilon_{\mu\nu}^{\ \ \alpha\beta}(X^{-1})_{\alpha\beta}^{\ \ \gamma\lambda}\varepsilon_{\gamma\lambda}^{\ \ \rho\sigma}\,,
\label{X-hat}
\
\\
&&
\frac{1}{2}(X^{-1})_{\mu\nu}^{\ \ \rho\sigma}X_{\rho\sigma}^{\ \ \alpha\beta}\equiv I_{\mu\nu}^{\ \ \alpha\beta}\,,
\label{X-ne-def}
\ffa
where $\varepsilon_{\mu\nu\rho\sigma}$ is a volume element.
The tensor $\widehat{X}$ possesses the same symmetry as $X$, i.e.,
$
\widehat{X}_{\mu\nu\rho\sigma}=\widehat{X}_{[\mu\nu][\rho\sigma]}=\widehat{X}_{\rho\sigma\mu\nu}
$.

When $X_{\mu\nu}^{\ \ \rho\sigma}=I_{\mu\nu}^{\ \ \rho\sigma}$, the modified Maxwell theory \eqref{ac-SA} is reduced the standard Maxwell one.
In this case, one can easily deduce that $X^{-1}=X$ and so $\widehat{X}_{\mu\nu}^{\ \ \rho\sigma}=I_{\mu\nu}^{\ \ \rho\sigma}$ from Eqs. \ref{X-ne-def} and \ref{X-hat}.
Hence, the actions \eqref{ac-SA} and \eqref{ac-SB} are identical, which demonstrates that the standard Maxwell theory is self-dual.

It has been shown in \cite{Myers:2010pk} that, when the higher derivative term $\gamma$ is introduced, the EM self-duality is violated.
Here, we demonstrate that even if only the $\gamma_{1,0}$ coupling term is introduced, the EM self-duality is also violated.
We first evaluate the inverse of $X$ in terms of \eqref{X-ne-def}, which is
\fa
\label{X-ne-gamma10}
(X^{-1})_{\mu\nu}^{\ \ \rho\sigma}
=\frac{1}{1-4\gamma_{1,0}L^2\bar{\Phi}}I_{\mu\nu}^{\ \ \rho\sigma}\,.
\ffa
Immediately, from Eq. \eqref{X-hat}, we find
\fa
\label{X-hat-gamma10}
\widehat{X}_{\mu\nu}^{\ \ \rho\sigma}=(X^{-1})_{\mu\nu}^{\ \ \rho\sigma}=\frac{1}{1-4\gamma_{1,0}L^2\bar{\Phi}}I_{\mu\nu}^{\ \ \rho\sigma}\,.
\ffa
Since $\widehat{X}\neq X$, the EM self-duality is violated.

\section{Black brane solution}\label{sec-BBS}

Since the EOM \eqref{eom} are a set of third order differential equations with high nonlinearity,
it has been hard to solve it analytically or even numerically so far.
So following the strategy in \cite{Ling:2016dck} (also see \cite{Myers:2009ij,Liu:2008kt,Cai:2011uh,Dey:2015poa,Dey:2015ytd}),
we shall construct analytical solutions up to the first order of those coupling parameters. \footnote{When the Weyl terms are turned off,
i.e., $\gamma=0$ and $\gamma_{1,1}=0$, the black brane can be worked out analytically \cite{Gouteraux:2016wxj,Baggioli:2016pia}.
We shall make a qualitative comparison on the DC conductivity between \cite{Gouteraux:2016wxj,Baggioli:2016pia} and our present results in Sec. \ref{DC-without-Weyl}.}
To this end, we take the following ansatz:
\begin{subequations}
\label{bg-ansatz-r}
\begin{align}
\label{bg-g-r}
&\mathrm{d}s^2=-\frac{r^2}{L^2}f(r)\mathrm{d}t^2+\frac{L^2}{r^2f(r)}\mathrm{d}r^2+\frac{r^2}{L^2}g(r)(\mathrm{d}x^2+\mathrm{d}y^2)\,,
\
\\
&
\label{bg-At-r}
A=A_t(r)\mathrm{d}t\,,
\end{align}
\end{subequations}
where the UV boundary is at $r\rightarrow\infty$.
Note that, when we take the following ansatz of $\phi_I$: $\phi_I=\alpha x_I$, Eq. \eqref{eom-phi} satisfied automatically.
So we only need to expand the functions $f(r)$, $g(r)$ and $A_t(r)$ in powers of $\gamma_{0,1}$, $\gamma$ and $\gamma_{1,1}$ up to the first order as
\begin{subequations}
\label{per-r}
\begin{align}
&
\label{f-per-r}
f(r)=f_0(r)+\gamma_{1,0}Y_{1,0}(r)+\gamma Y(r)+\gamma_{1,1} Y_{1,1}(r)\,,
\
\\
&
\label{g-per-r}
g(r)=1+\gamma G(r)+\gamma_{1,1} G_{1,1}(r)
\
\\
&
\label{At-per-r}
A_t(r)=A_{t0}(r)+\gamma_{1,0}H_{1,0}(r)+\gamma H(r)+\gamma_{1,1} H_{1,1}(r)\,,
\end{align}
\end{subequations}
where $f_0(r)$ and $A_{t0}(r)$ are the zeroth order solutions, which have been worked out in \cite{Andrade:2013gsa},
while $Y_{i,j}(r)$, $G_{i,j}(r)$ and $H_{i,j}(r)$ are the first order solutions of $\gamma_{i,j}$.
Note that we do not include the correction from $\gamma_{0,1}$ into the function of $g(r)$,
so that we can make a direct comparison with the analytical solution in \cite{Gouteraux:2016wxj,Baggioli:2016pia}.

By directly solving Eq. \eqref{eom} to the zeroth and first order of the coupling parameters, we can determine these functions:
\begin{subequations}
\label{sol0}
\begin{align}
&
\label{solb-r}
f_0(r)=
1-\frac{M}{r^3}+\frac{q^2}{r^4}-\frac{\alpha^2L^4}{2r^2}\,,
~~~~~~
A_{t0}(r)=\mu-\frac{2 q}{rL^2}\,,
\
\\
&
\label{solp10-r}
Y_{1,0}(r)=
-\frac{4\alpha ^2 q^2L^4}{3r^6}\,,
~~~~~~
H_{1,0}(r)=\frac{8\alpha ^2 qL^2}{3r^3}\,,
\
\\
&
Y(r)=\frac{c_0q^2}{r^5}-\frac{c_0M}{2r^4}
+\frac{c_1\alpha^2L^4}{2r^2}-\frac{c_0}{r}
+\frac{20M q^2}{9r^7}-\frac{104q^4}{45r^8}
+\frac{10\alpha^2q^2L^4}{9r^6}-\frac{32 q^2 }{9r^4}\,,
\nonumber
\\
&
G(r)=-\frac{c_0}{r}+c_1+\frac{4 q^2}{9r^4}\,,
\nonumber
\\
&
\label{solp01-r}
H(r)=-\frac{c_0 q}{r^2L^2}-\frac{4 M q}{r^4L^2}
+\frac{296 q^3}{45r^5L^2}-\frac{8\alpha ^2 qL^2}{9r^3}\,,
\
\\
&
Y_{1,1}(r)=\frac{d_0 M }{2r^4}-\frac{d_0 q^2}{r^5}
+\frac{\alpha^2 d_1 L^4}{2r^2}+\frac{d_0}{r}+\frac{64\alpha ^2 M q^2L^4}{45r^9}
   -\frac{496\alpha ^2 q^4L^4}{315r^{10}}
   +\frac{28\alpha ^4 q^2L^8}{45r^8}
   -\frac{32\alpha ^2 q^2L^4}{45r^6}\,,
\nonumber
\\
&
G_{1,1}(r)=\frac{d_0}{r}+d_1+\frac{8\alpha ^2 q^2L^4}{45r^6}\,,
\nonumber
\\
&
\label{solp11}
H_{1,1}(r)=\frac{d_0 q}{r^2L^2}-\frac{8\alpha ^2 M qL^2}{3r^6}
   +\frac{208\alpha ^2 q^3L^2}{45r^7}-\frac{8\alpha ^4 qL^6}{15r^5}\,.
\end{align}
\end{subequations}
$(\mu, q, M, c_0, c_1, d_0, d_1)$ are seven integration constants, which are not independent from one another.
Below, we shall derive the relations among them.

First, we can make the coordinate transformations
\begin{subequations}
\label{coortran-vr}
\begin{align}
&
\label{coortran-r}
r\to r+\frac{1}{2}\gamma c_0-\frac{1}{2} d_0 \gamma _{1,1} \,,
\
\\
&
\label{coortran-xy}
(x,y)\to (x,y) \left(-\frac{1}{2} d_1 \gamma
   _{1,1}-\frac{\gamma  c_1}{2}+1\right)\,,
\end{align}
\end{subequations}
and a redefinition of the axionic charge $\alpha$
\fa
\label{redef-alpha}
\alpha \to \alpha  \left(\frac{1}{2} d_1 \gamma
   _{1,1}+\frac{\gamma  c_1}{2}+1\right)\,,
\ffa
such that the integration constants $(c_0, c_1, d_0, d_1)$ can be eliminated.
Using the conditions that $f$ and $A_{t}$ vanish at the horizon $r=r_h$, we obtain the relations for $(\mu,q,M)$:
\begin{subequations}
\label{rel-qM}
\begin{align}
q&=\frac{\mu r_h L^2}{2}-\gamma _{1,0}\frac{2 \alpha ^2 \mu  L^6}{3 r_h}
+\gamma \left(\frac{5 \alpha ^2 \mu  L^6}{18 r_h}+\frac{29 \mu ^3 L^6}{180 r_h}-\mu  L^2 r_h \right)
\nonumber
\\
&
+\gamma_{1,1} \left(\frac{\alpha ^4 \mu  L^{10}}{5 r_h^3}+\frac{11 \alpha ^2 \mu ^3 L^{10}}{90 r_h^3}-\frac{2 \alpha ^2 \mu
    L^6}{3 r_h}\right)
\,,\label{rel-q}
\
\\
M&=r_h^3-\frac{1}{2} \alpha ^2 L^4 r_h+\frac{1}{4} \mu ^2 L^4 r_h
-\gamma _{1,0}\frac{\alpha ^2 \mu ^2 L^8}{3 r_h}
+\gamma  \left(\frac{5 \alpha ^2 \mu ^2 L^8}{18 r_h}+\frac{7 \mu ^4 L^8}{45 r_h}-\frac{4}{3} \mu ^2 L^4 r_h\right)
\nonumber
\\
&
\label{rel-M}
+\gamma_{1,1} \left(\frac{8 \alpha ^4 \mu ^2 L^{12}}{45 r_h^3}+\frac{71 \alpha ^2 \mu ^4 L^{12}}{630 r_h^3}-\frac{22 \alpha
   ^2 \mu ^2 L^8}{45 r_h}\right)
   \,.
\end{align}
\end{subequations}

It is convenient to work with dimensionless quantities.
So we make the following rescaling:
\begin{align}
\label{rescaling}
r\rightarrow r_hr\,,~~(t,\vec{x})\rightarrow \frac{L^2}{r_h}(t,\vec{x})\,,~~
A_t\rightarrow \frac{r_h}{L^2}A_t\,,~~
M\rightarrow Mr_h^2\,,~~
Q\rightarrow Qr_h^2\,,~~
\alpha\rightarrow \frac{r_h}{L^2}\alpha\,.
\end{align}
Under this rescaling, we can set $L=1$ and $r_h=1$.
Then the dimensionless temperature can be given by
\fa
\label{tem}
T=-\frac{2 \alpha ^2+\mu ^2-12}{16 \pi }
-\gamma_{1,0}\frac{\alpha ^2 \mu ^2}{12 \pi }
+\gamma \frac{\mu ^2 \left(\mu ^2-60\right)}{720 \pi }
+\gamma_{1,1}\frac{\alpha ^2 \mu ^2 \left(8 \alpha ^2+3 \mu
   ^2-84\right)}{360 \pi }\,.
\ffa
Note that all the above quantities $q$, $M$ and $T$ have been expanded to the first order of the coupling parameters $(\gamma_{1,0}, \gamma, \gamma_{1,1})$.
This black brane is characterized by two parameters, i.e., the temperature $T/\mu$ and the strength of the momentum dissipation $\alpha/\mu$.
$\mu$ is interpreted as the chemical potential of the dual field and can be treated as the unit for the grand canonical system.
For later convenience, we denote $\bar{T}\equiv T/\mu$ and $\bar{\alpha}\equiv \alpha/T$.

In addition, for the convenience of calculation,
we shall work with the coordinate $u=1/r$.
Then, in terms of $\mu$, we reexpress $f(u)$, $g(u)$ and $A_t(u)$ as follows:
\begin{subequations}
\label{fgAt-mu}
\begin{align}
&
f(u)=(1-u)p(u)\,,
\label{fu-p}
\\
&
p(u)=-\frac{1}{4} \mu ^2 u^3-\frac{\alpha ^2
   u^2}{2}+u^2+u+1
  -\frac{1}{3}\gamma_{1,0} \alpha ^2 \mu ^2 u^3 \left(u^2+u-1\right)
\nonumber
\\
&
+\gamma\frac{1}{180} \mu ^2 u^3 \left(2 \mu ^2 \left(13
   u^4-14\right)+50 \alpha ^2
   \left(u^3-1\right)+\left(\mu ^2-100\right)
   \left(u^3+u^2+u\right)+240\right)
\nonumber
\\
&
+\gamma_{1,1}\frac{1}{630} \alpha ^2 \mu ^2 u^3 \Big(14 \alpha ^2
   \left(8 u^5+u^4+u^3+u^2+u-8\right)-28 \left(8
   u^5+8 u^4+8 u^3+4 u^2+4 u-11\right)
\nonumber
\\
&
\label{pu}
   +\mu ^2
   \left(62 u^6+6 u^5+6 u^4+6 u^3+6 u^2+6
   u-71\right)\Big)
   \,,
\
\\
&
\label{gu-p}
g(u)=\frac{2}{45} \alpha ^2 \mu ^2 u^6 \gamma
   _{1,1}+\frac{1}{9} \gamma  \mu ^2 u^4+1\,,
\
\\
&
A_{t}(u)=\mu  (1-u)
\Big[1+\frac{4}{3} \alpha ^2 u (u+1) \gamma _{1,0}
+\gamma  \Big(2 u \left(u^2+u+1\right)
-\frac{1}{9}
   \alpha ^2 u (u (9 u+5)+5)
\nonumber
\\
&
   -\frac{1}{90} \mu ^2 u (u
   (u (74 u+29)+29)+29)\Big)
+\frac{1}{45} \alpha ^2 u \gamma _{1,1} \Big(60
   \left(u^4+u^3+u^2+u+1\right)
\nonumber
\\
&
\label{Atu-p}
   -6 \alpha ^2 (u (u (u
   (5 u+3)+3)+3)+3)-\mu ^2 (u (u (u (u (26
   u+11)+11)+11)+11)+11)\Big)
\Big]\,.
\end{align}
\end{subequations}

\section{DC conductivity at finite density}\label{sec-DC-finite}

\subsection{The derivation of the DC conductivity}

In this section, we follow the procedure in \cite{Donos:2014uba,Donos:2014cya,Blake:2014yla,Ling:2016dck} to calculate the DC conductivity.
To this end, we turn on the following consistent perturbations
\begin{align}
\label{DCperturbations}
\delta g_{tx}=\frac{1}{u^2}h_{tx}(u)[1+\gamma G(u)+\gamma_{1,1}G_{1,1}(u)]\,,~~\delta A_{x}=-E_x t+a_x(u)\,,~~\delta \phi_{x}= \chi_{x}(u)\,.
\end{align}
Then, one can define a radial conserved current in the bulk as
\fa
\label{Jx-def}
J^x=\frac{1}{2}\sqrt{-g}X^{\mu\nu\rho\sigma}F_{\rho\sigma}\,.
\ffa
Up to the first order of the coupling parameters, this conserved current can be evaluated as
\fa
J^x&=&
-Q h_{\text{tx}}(u)+f(u) a_x'(u)-4 \alpha ^2 u^2 \gamma _{1,0} f(u) a_x'(u)
\nonumber
\\
&&
\label{Jx}
-\frac{2}{3} u^2 f(u) \left(\alpha ^2 u^2 \gamma _{1,1}+\gamma \right) \left(f''(u) a_x'(u)+3 A_t'(u)
   h_{\text{tx}}''(u)\right)\,.
\ffa
We have defined $Q=J^t$ in the above equation. It is the conserved electric charge density.
Once $J^x$ is at hand, the DC conductivity can be evaluated in terms of Ohm's law
\fa
\label{DC-def}
\sigma_{\mathrm{DC}}=\frac{J^x}{E_x}\,.
\ffa

Since $J^x$ is a radial conserved quantity, the DC conductivity can be evaluated at the horizon $u=1$.
Fist, we extract the value of $h_{tx}$ at the horizon from the $t$, $x$ component of the Einstein equation, which reads
\begin{align}
&
\gamma_{1,0}\Big(
\frac{1}{4} h_{\text{tx}} \left(G_{1,1} \left(\left(A'\right)_t^2-2 f''+8 f'+12\right)-2 f'
   G'_{1,1}\right)+\frac{1}{3} \alpha ^2 u^2 A'_t \left(2 f f'' a'_x+h_{\text{tx}} A'_t \left(f''-2
   f'\right)\right)
\Big)
   \nonumber
\\
&
-\frac{1}{6} f a'_x A'_t \left(4 \alpha ^2 u^2 \gamma _{1,1} f''+4 \gamma  f''+3\right)
+h_{tx}\Big(\frac{1}{6} \left(3 \gamma _{1,1} f' G'_{1,1}-2 \gamma  \left(A'\right)_t^2 \left(f''-2 f'\right)+3
   \gamma  f' G'\right)
   \nonumber
\\
&
+\frac{1}{6} \alpha ^2 \left(2 u^2 \gamma _{1,1} \left(A'\right)_t^2 \left(2
   f'-f''\right)+3\right)-\frac{1}{4} \left(A'\right)_t^2+\frac{f''}{2}-2 f'-3
\Big)
   =0
   \,.
   \label{htx-ax}
\end{align}
Notice that the above equation has taken value at $u=1$.
In addition, we also need to add a regular boundary condition of $a_x$ at the horizon, which is
\fa
\label{ax}
a_x'=\frac{E_x}{f}\,.
\ffa
Collecting Eqs. \eqref{Jx}-\eqref{ax},
we can obtain the DC conductivity:
\begin{align}
\sigma_{0}&=1+\frac{1}{\bar{\alpha }^2}
+\gamma_{1,0}\Big(\frac{28 \mu ^2}{3}-\frac{8}{3} \mu ^4 \bar{\alpha }^2-\frac{3 \mu ^4}{5 \bar{\alpha }^2}-4 \mu ^2 \bar{\alpha }^2-\frac{4
   \mu ^2}{5 \bar{\alpha }^2}-\frac{38 \mu ^4}{15}\Big)
   +\gamma\Big(4-\frac{4}{3} \mu ^2 \bar{\alpha }^2+\frac{\frac{8 \mu ^2}{15}-4}{\bar{\alpha }^2}+\frac{\mu ^2}{9}\Big)
   \nonumber
   \\
   &
   +\gamma_{1,1}\Big(-\frac{4}{3} \mu ^4 \bar{\alpha }^4-\frac{1}{5} \mu ^4 \bar{\alpha }^2+\frac{\mu ^4}{3 \bar{\alpha
   }^2}+4 \mu ^2 \bar{\alpha }^2-\frac{4 \mu ^2}{\bar{\alpha }^2}+\frac{10 \mu ^4}{9}-\frac{8 \mu
   ^2}{3}\Big)\,.
\label{dc}
\end{align}

When $\gamma_{1,0}=0$ and $\gamma_{1,1}=0$, the result \eqref{dc} reduces to Eq.(37) in \cite{Ling:2016dck}.
To compare with our present results, involving more coupling terms, we would like to present a brief review \cite{Ling:2016dck}:
\begin{itemize}
  \item There is a relation,
  \fa
  \label{mir-sym}
  \sigma_{0}(\gamma,T)\simeq\text{const.}-\sigma_{0}(-\gamma,T)\,,
  \ffa
  which can be seen to hold when $\bar{\alpha}$ is fixed.
  It can be viewed as a special particle-vortex duality in \cite{Burgess:2000kj,Murugan:2016zal}.
  \item A metal-insulator transition (MIT) happens at zero temperature for a given nonzero $\gamma$ when we change the axionic charge $\bar{\alpha}$.
  \item There is a mirror symmetry at zero temperature \footnote{This mirror symmetry also may be applicable at finite temperature.}
  \fa
  \label{mir-sym-psigma-pt}
  \frac{\partial\sigma_0}{\partial\bar{T}}(\gamma,\bar{\alpha})=-\frac{\partial\sigma_0}{\partial\bar{T}}(-\gamma,\bar{\alpha})\,.
  \ffa
\end{itemize}

Next, we shall analyze the behavior of the DC conductivity and explore the MIT.
Before proceeding, we introduce the definition of metallic phase and insulating phase adopted in many holography references \cite{Baggioli:2016rdj,Donos:2012js,Donos:2013eha,Donos:2014uba,Ling:2015ghh,Ling:2015dma,Ling:2015epa,Ling:2015exa,Ling:2016wyr,Ling:2016dck,Baggioli:2014roa,Baggioli:2016oqk,Baggioli:2016oju,Donos:2014oha,Kiritsis:2015oxa}:
\begin{itemize}
  \item Metallic phase: $\partial_T\sigma_0<0$.
  \item Insulating phase: $\partial_T\sigma_0>0$.
  \item Critical point (line): $\partial_T\sigma_0=0$.
\end{itemize}

\subsection{DC conductivity without Weyl term}\label{DC-without-Weyl}

\begin{figure}
\center{
\includegraphics[scale=0.55]{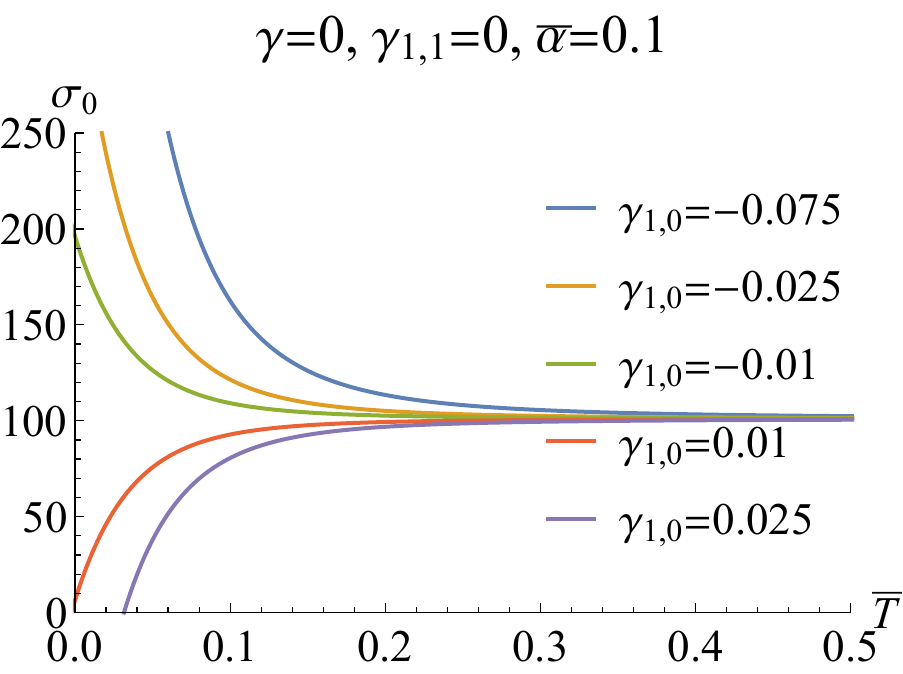}\ \hspace{0.8cm}
\includegraphics[scale=0.55]{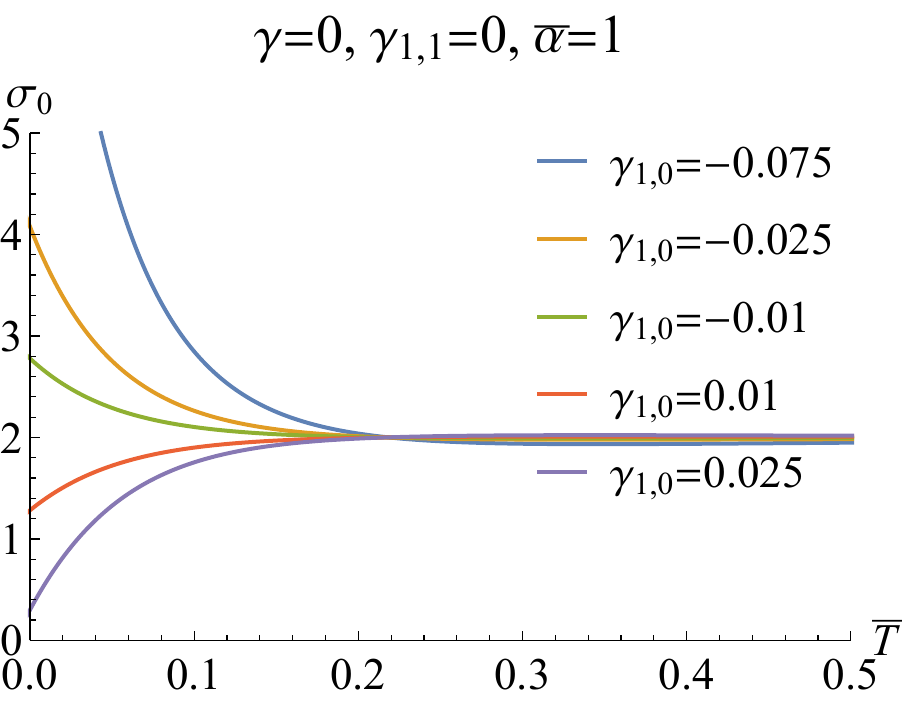}\ \hspace{0.8cm}\ \\
\caption{\label{dcvst_gamma10} DC conductivity $\sigma_0$ as a function of the temperature $\bar{T}$ with different $\gamma_{1,0}$ and $\bar{\alpha}$.}}
\end{figure}

In Appendix \ref{sec-Bounds}, we analyze the causality and instabilities of the vector modes at zero density.
When we only consider the $\gamma_{1,0}$ term, the analysis and the requirement of the positive DC conductivity indicate $-3/40\leq\gamma_{1,0}\leq 1/40$.
But it is hard to analyze the causality and instabilities of the vector modes at finite density even if we have an analytical perturbative black brane solution.
We shall leave this problem for future study.
Here, we only approximately impose a further constraint from the requirement of the positive DC conductivity at finite density.

Figure \ref{dcvst_gamma10} show the DC conductivity $\sigma_0$ as a function of the temperature $\bar{T}$ with different $\gamma_{1,0}$ and $\bar{\alpha}$.
We find that, when $\gamma_{1,0}=0.025$, $\sigma_0$ is negative for small $\bar{\alpha}$ and low temperature $\bar{T}$.
Further detailed exploration indicates that the positive definiteness of the DC conductivity constrains $\gamma_{1,0}$ in the range
\fa
\label{gamma10-range-finite}
-3/40\leq\gamma_{1,0}\leq 1/100\,.
\ffa

\begin{figure}
\center{
\includegraphics[scale=0.55]{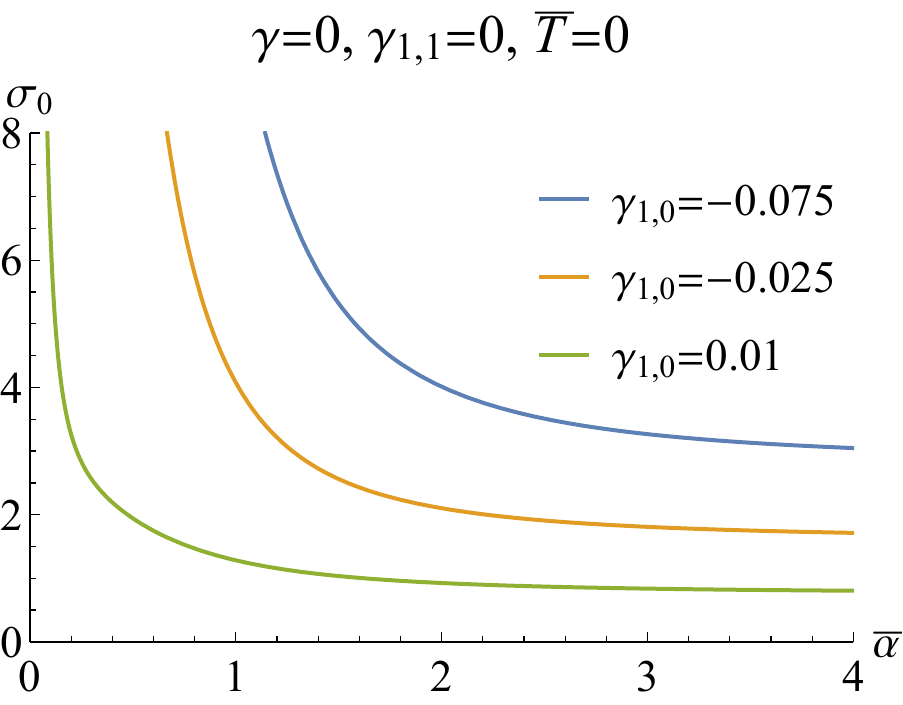}\ \hspace{0.8cm}
\includegraphics[scale=0.55]{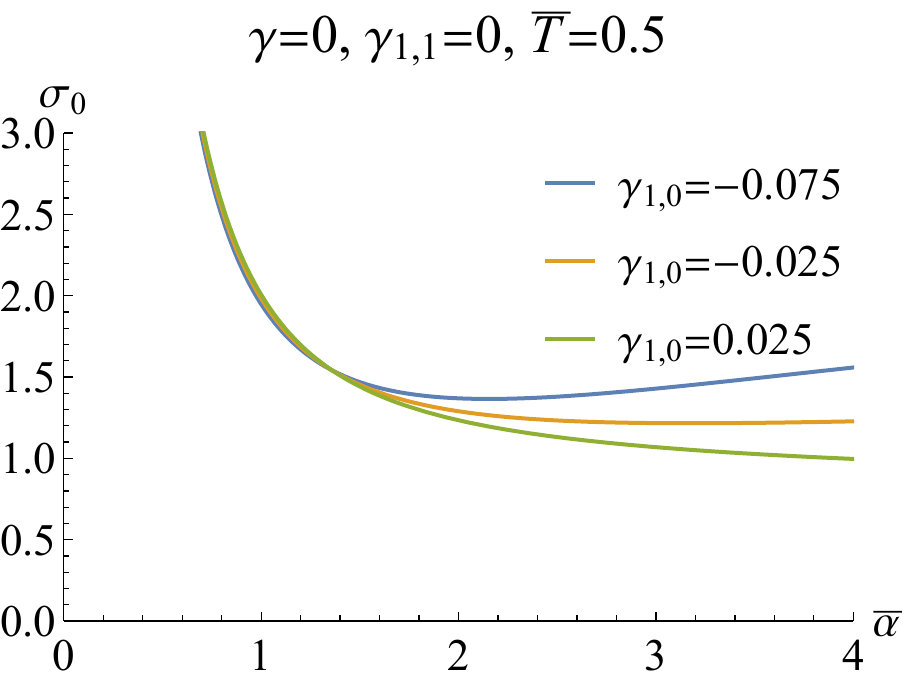}\ \hspace{0.8cm}\ \\
\caption{\label{dcvsalpha_gamma10} DC conductivity $\sigma_0$ as a function of $\bar{\alpha}$ with different $\gamma_{1,0}$ at zero temperature (left plot) and finite temperature (right plot), respectively.}}
\end{figure}
\begin{figure}
\center{
\includegraphics[scale=0.6]{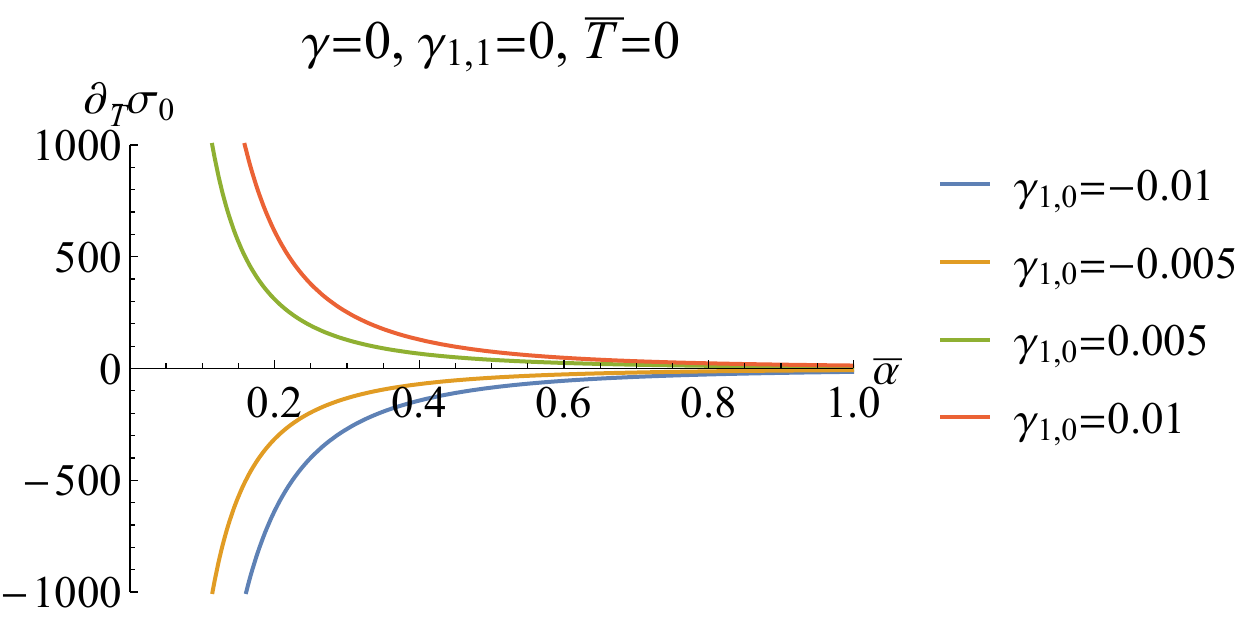}\ \hspace{0.8cm}\ \\
\caption{\label{pdcvsalpha_gamma10} $\partial_{\bar{T}}\sigma_0$ as a function of $\bar{\alpha}$ at zero temperature for different $\gamma_{1,0}$.}}
\end{figure}

Also, we, respectively, show the DC conductivity as a function of $\bar{\alpha}$ for $\gamma_{1,0}$ belonging to the range \eqref{gamma10-range-finite} at zero temperature and finite temperature in Fig. \ref{dcvsalpha_gamma10}.
Figures \ref{dcvst_gamma10} and \ref{dcvsalpha_gamma10} show that our result is qualitatively the same as that found in \cite{Gouteraux:2016wxj}:
\begin{itemize}
  \item At zero temperature, the DC conductivity monotonously decreases in terms of $\bar{\alpha}$.
  \item At finite temperature, the DC conductivity is qualitatively similar to that at zero temperature when $\gamma_{1,0}>0$.
  Meanwhile for $-3/40\leq\gamma_{1,0}<0$, the DC conductivity no longer monotonously decreases but has a minimum at some finite value of $\bar{\alpha}$.
  \item When $\bar{\alpha}$ is fixed, the DC conductivity monotonously decreases in terms of $\bar{T}$ for $\gamma_{1,0}>0$, which demonstrates a metal phase.
  When the sign of $\gamma_{1,0}$ changes, an opposite behavior is found, which is an insulator phase.
\end{itemize}
Therefore, our system up to the first order of the coupling parameters captures the main properties as shown in \cite{Gouteraux:2016wxj}.

Finally, we present some comments on comparing with the $4$ derivative Weyl term studied in \cite{Ling:2016dck}.
\begin{itemize}
  \item Different from that for the four derivative Weyl term, no MIT happens for a given nonzero $\gamma_{1,0}$ when changing $\bar{\alpha}$ (see Fig. \ref{pdcvsalpha_gamma10}).
  But the mirror symmetry on $\frac{\partial\sigma_0}{\partial\bar{T}}(\bar{\alpha})$ \eqref{mir-sym-psigma-pt} at zero temperature holds when the sign of $\gamma_{1,0}$ changes.
  \item Equation \eqref{mir-sym} holds
 when the sign of $\gamma_{1,0}$ changes and $\bar{\alpha}$ is fixed.
\end{itemize}

\subsection{DC conductivity from four derivative theory}

\begin{figure}
\center{
\includegraphics[scale=0.55]{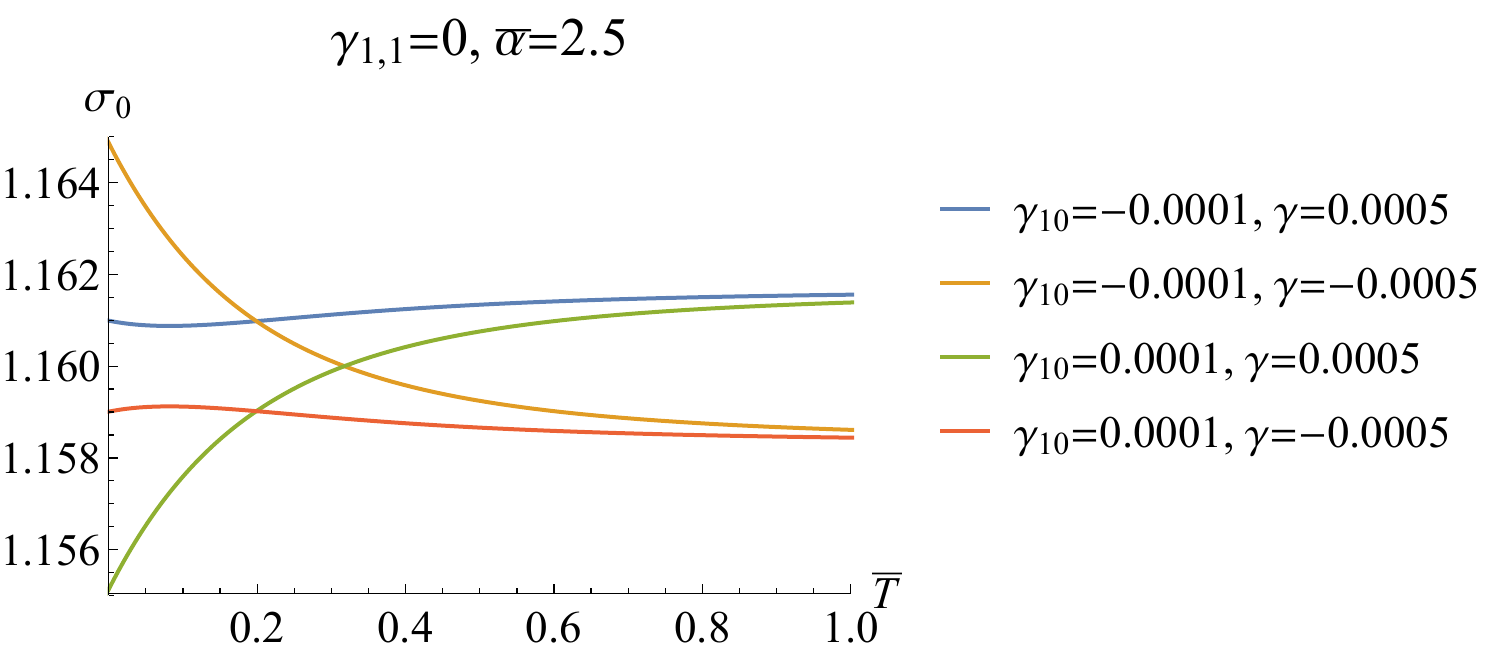}\ \hspace{0.8cm}\ \\
\caption{\label{dcvst_gamma10_alphab_d_gamma}Left plot: $\sigma_0$ as a function of $\bar{T}$ with $\gamma_{1,0}=0.0005$, $\bar{\alpha}=2.5$ and for different $\gamma$.
Right plot: $\sigma_0$ as a function of $\bar{T}$ with $\gamma=10^{-4}$, $\bar{\alpha}=2.5$ and for different $\gamma_{1,0}$.}}
\end{figure}
\begin{figure}
\center{
\includegraphics[scale=0.55]{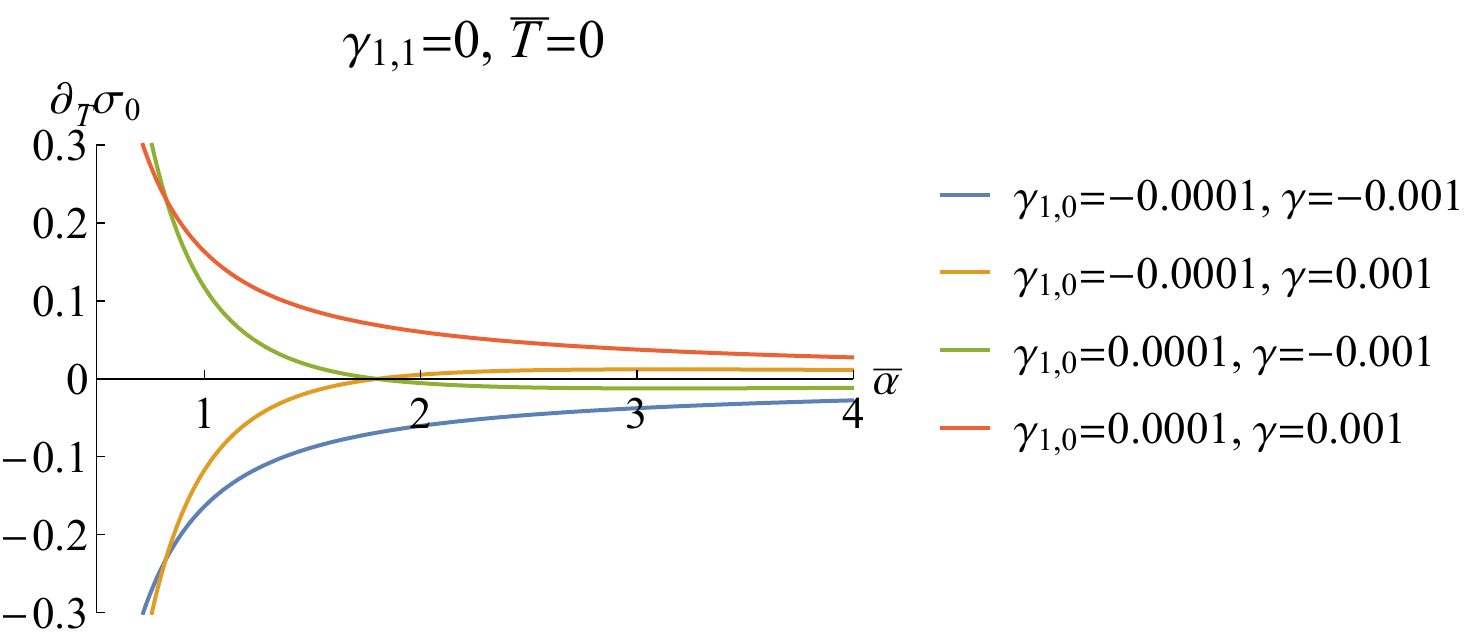}\ \hspace{0.8cm}
\includegraphics[scale=0.55]{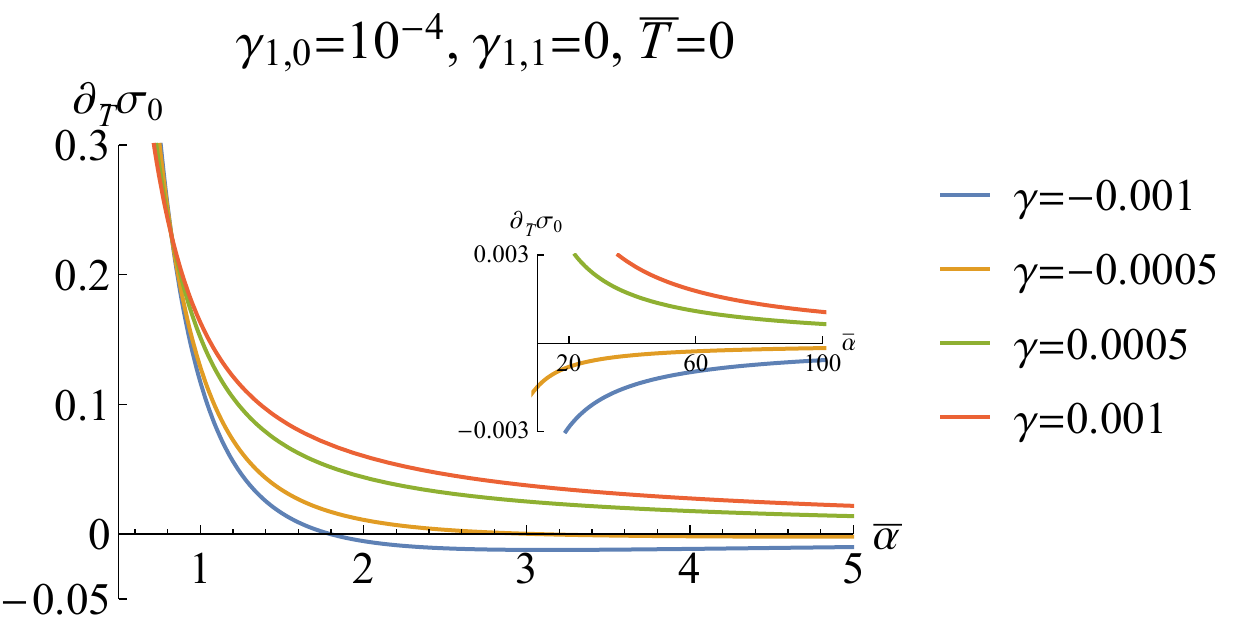}\ \hspace{0.8cm}\ \\
\caption{\label{pdcvsalpha_gamma10_pn_gamma} $\partial_{\bar{T}}\sigma_0$ as a function of $\bar{\alpha}$ at zero temperature.}}
\end{figure}

When only the four derivative Weyl term $\gamma$ is involved, an MIT occurs at zero temperature by varying the axionic charge $\bar{\alpha}$.
In particular, the quantum critical line is independent of the coupling parameter $\gamma$ \cite{Ling:2016dck}.

\begin{figure}
\center{
\includegraphics[scale=0.55]{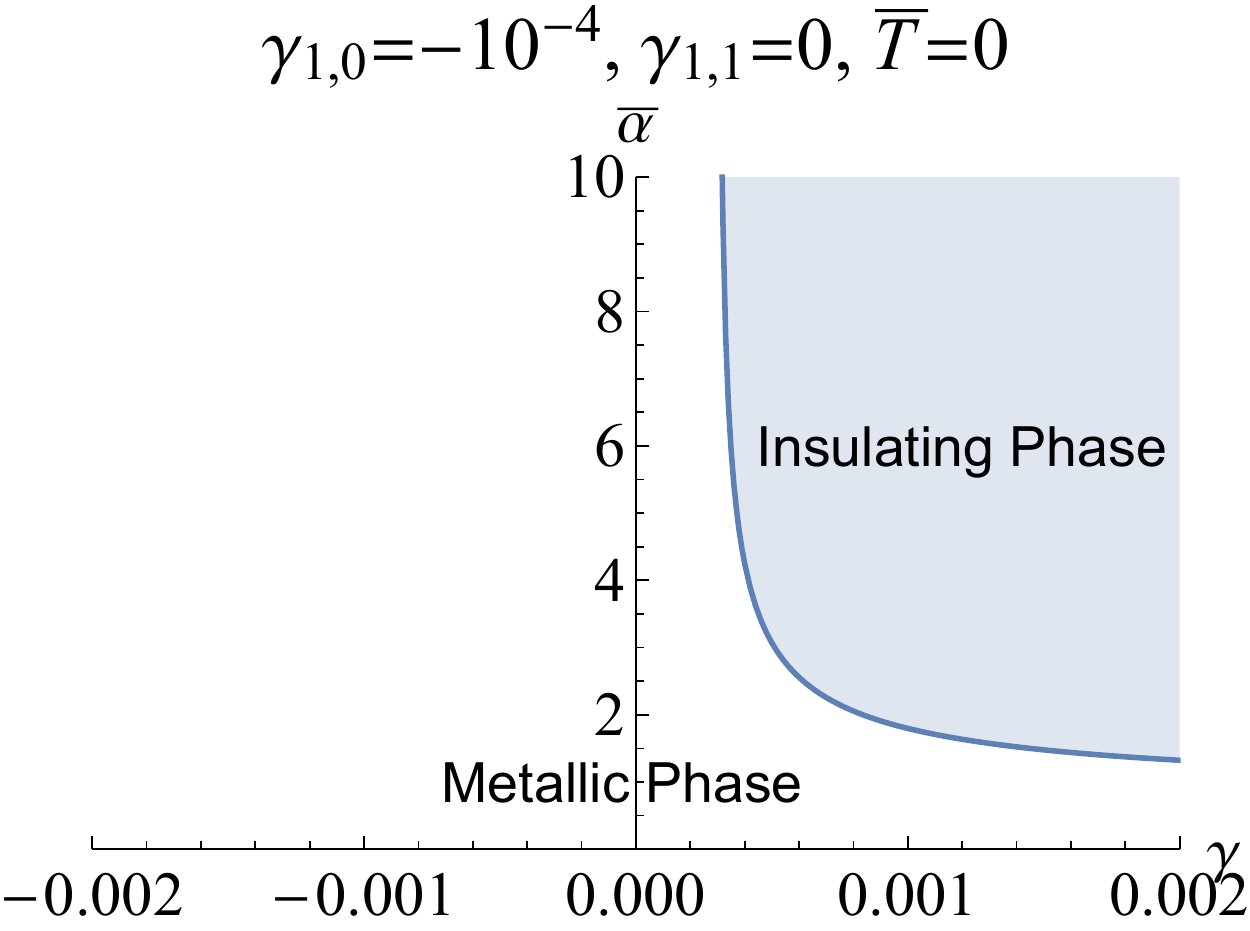}\ \hspace{0.8cm}
\includegraphics[scale=0.55]{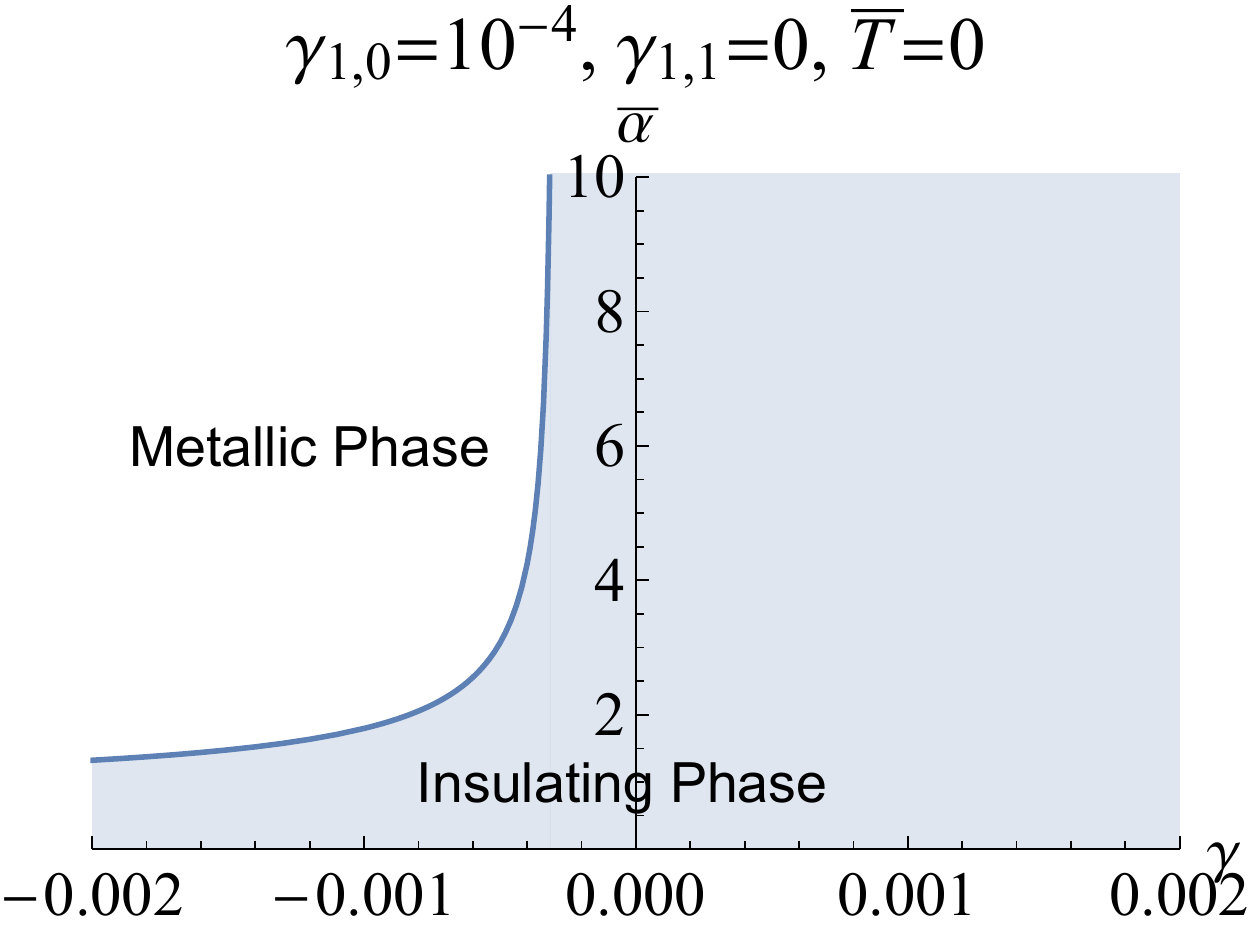}\ \hspace{0.8cm}\ \\
\caption{\label{pd_gamma10} Phase diagram over $(\gamma,\bar{\alpha})$ plane for the MIT from four derivative theory at zero temperature 
(left plot for $\gamma_{1,0}=-10^{-4}$ and right plot for $\gamma_{1,0}=10^{-4}$).}}
\end{figure}

In this section, we consider the mixed effect on DC conductivity in the four derivative theory including both $\gamma_{1,0}$ and $\gamma$ terms.
The main properties are summarized as follows:
\begin{itemize}
  \item Equations \eqref{mir-sym} and \eqref{mir-sym-psigma-pt} hold for fixed $\bar{\alpha}$ and changing the signs of $\gamma$ and $\gamma_{1,0}$ (Fig. \ref{dcvst_gamma10_alphab_d_gamma} and left plot in Fig. \ref{pdcvsalpha_gamma10_pn_gamma}).
  \item For positive (negative) small $\gamma_{1,0}$, a MIT can be observed for negative (positive) $\gamma$ (see right plot in FIG.\ref{pdcvsalpha_gamma10_pn_gamma}).
  But different from the case only involving the four derivative term in \cite{Ling:2016dck}, the quantum critical line is dependent on $\gamma$ (Fig.\ref{pd_gamma10}).
  It provides a new platform of QCP such that we can study the holographic entanglement entropy and the butterfly effect close to QCP as in \cite{Ling:2016wyr,Ling:2016ibq,Ling:2016dck}.
  We shall explore them in our present model in the future.
\end{itemize}

Before proceeding, we present some comments on the phase diagram for the MIT from four derivative theory at zero temperature (Fig. \ref{pd_gamma10}).
For $\gamma_{1,0}<0$ and $\gamma>0$, with the increase of the strength of momentum dissipation, there is a phase transition from metallic phase to insulating one.
This phenomenon is consistent with that of the usual charged particle excitations.
On the other hand, for $\gamma_{1,0}>0$ and $\gamma<0$, we find that with the increase of the strength of momentum dissipation, the phase transition is opposite,
i.e., the stronger momentum dissipates, the more insulating is the material.
A better description of this phenomenon is provided by considering the excitations of vortices.
Just as described \cite{Myers:2010pk}, the EM duality of the bulk theory, which is related by changing the sign of $\gamma$,
corresponds to the particle-vortex duality in the dual holographic CFT.
Figure \ref{pd_gamma10} shows such a duality; when we change the sign of $\gamma$, there is a duality between metallic and insulating phase.
In fact, the phenomena can be easily concluded from Eq. \eqref{mir-sym}.
Finally, we would like to mention two corresponding examples.
One is the transition observed in \cite{Myers:2010pk} from the Drude-like peak at low frequency optical conductivity, which is interpreted as the charged particle excitations,
to the dip, which resembles the excitations of vortices.
Another one is the observation in \cite{Wu:2016jjd} that
the momentum dissipation drives the Drude-like peak into the dip of the low frequency optical conductivity for $\gamma>0$.
Meanwhile for $\gamma<0$, the opposite scenario appears.
When the sign of $\gamma$ changes, an approximate duality in optical conductivity is also observed for fixed strength of momentum dissipation.
This duality is also observed in the next section.

\subsection{DC conductivity from six derivative theory}

\begin{figure}
\center{
\includegraphics[scale=0.55]{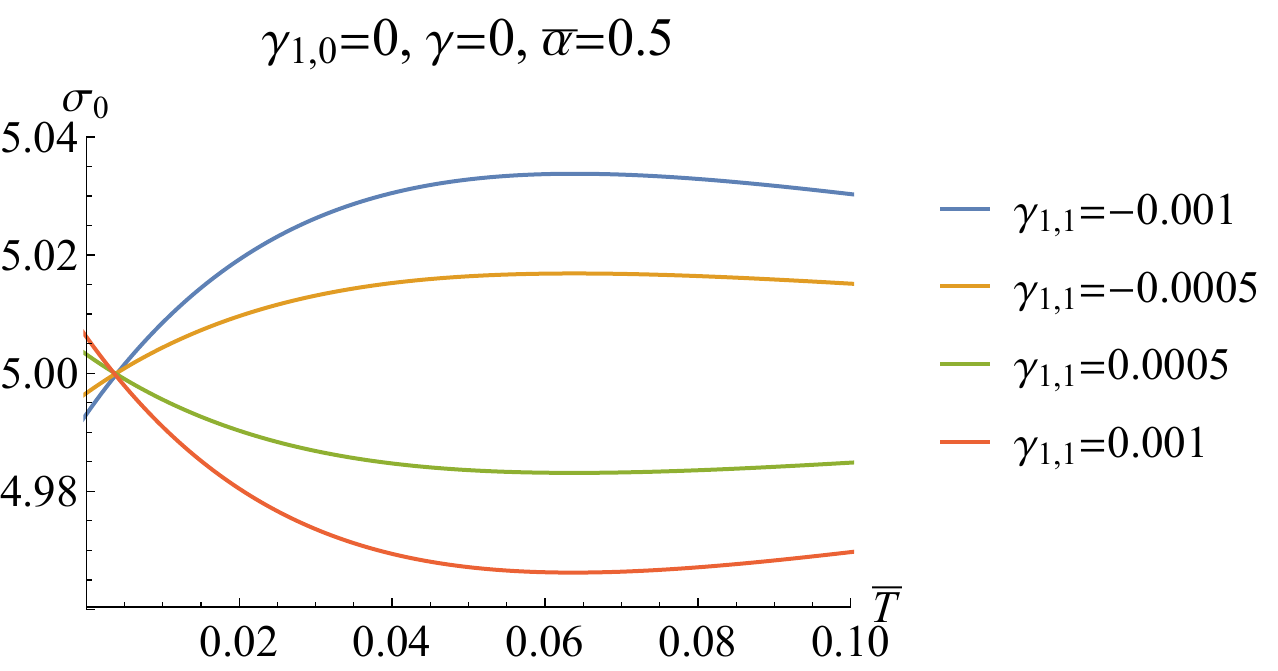}\ \hspace{0.8cm}
\includegraphics[scale=0.55]{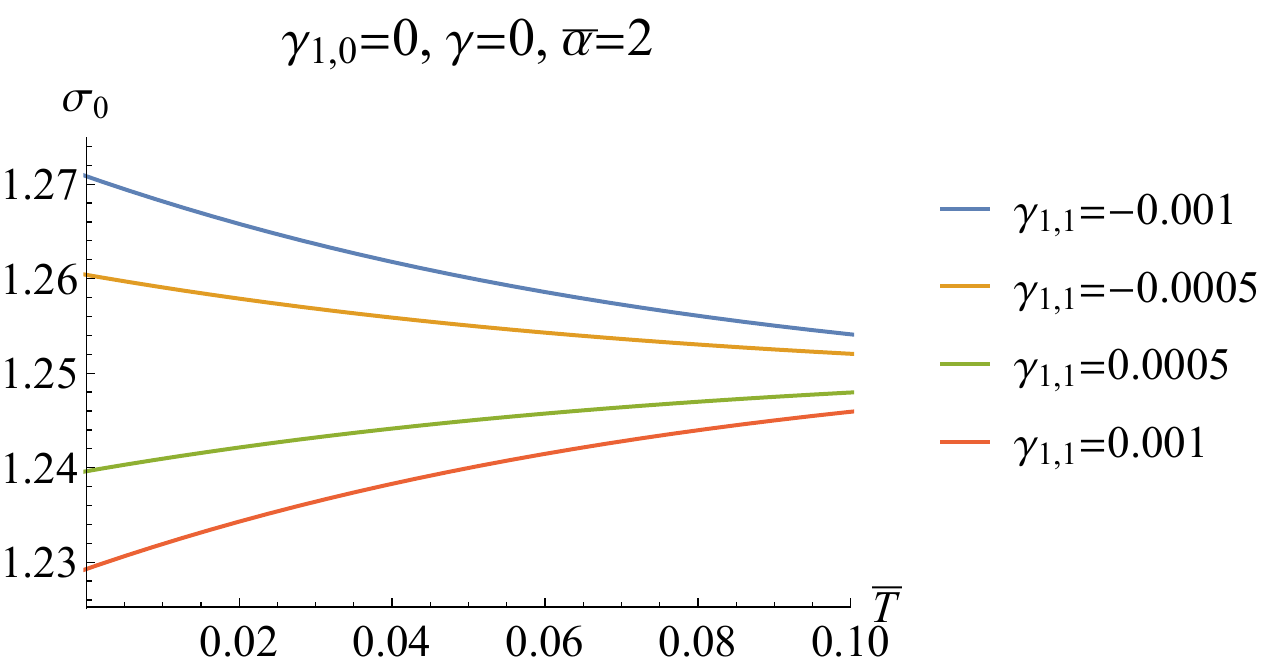}\ \hspace{0.8cm}\ \\
\caption{\label{dcvst_gamma11} DC conductivity $\sigma_0$ as a function of the temperature $\bar{T}$ for some representative $\bar{\alpha}$ and $\gamma_{1,1}$.}}
\end{figure}

\begin{figure}
\center{
\includegraphics[scale=0.6]{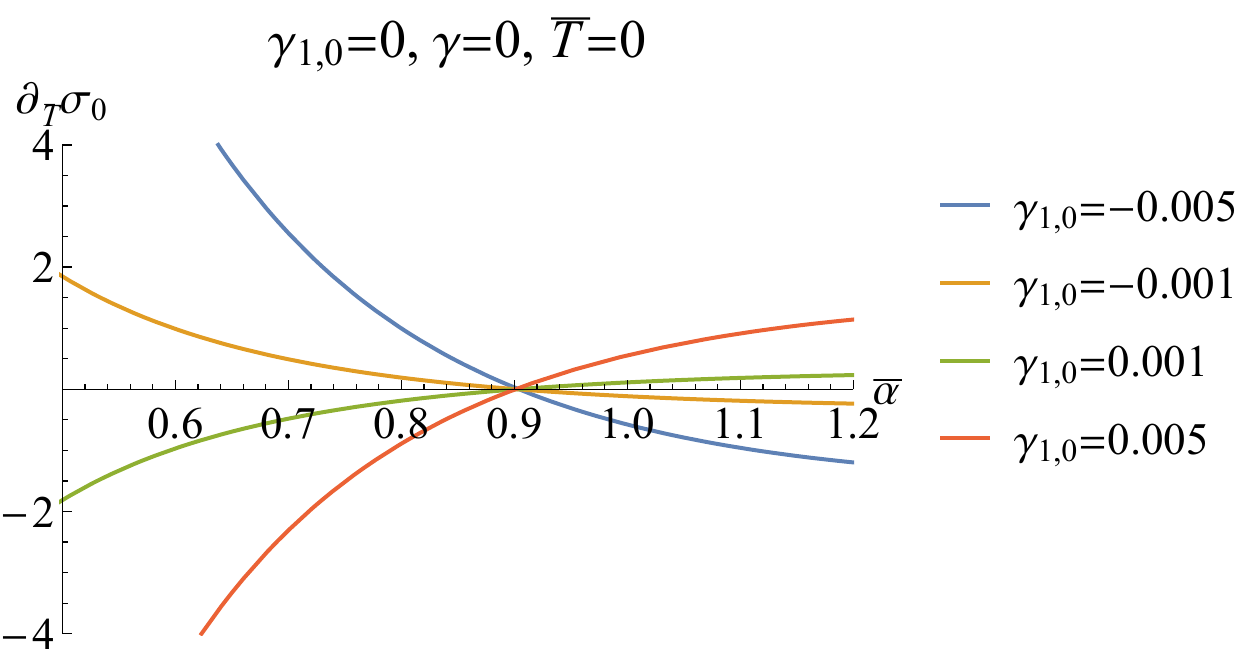}\ \hspace{0.8cm}
\includegraphics[scale=0.5]{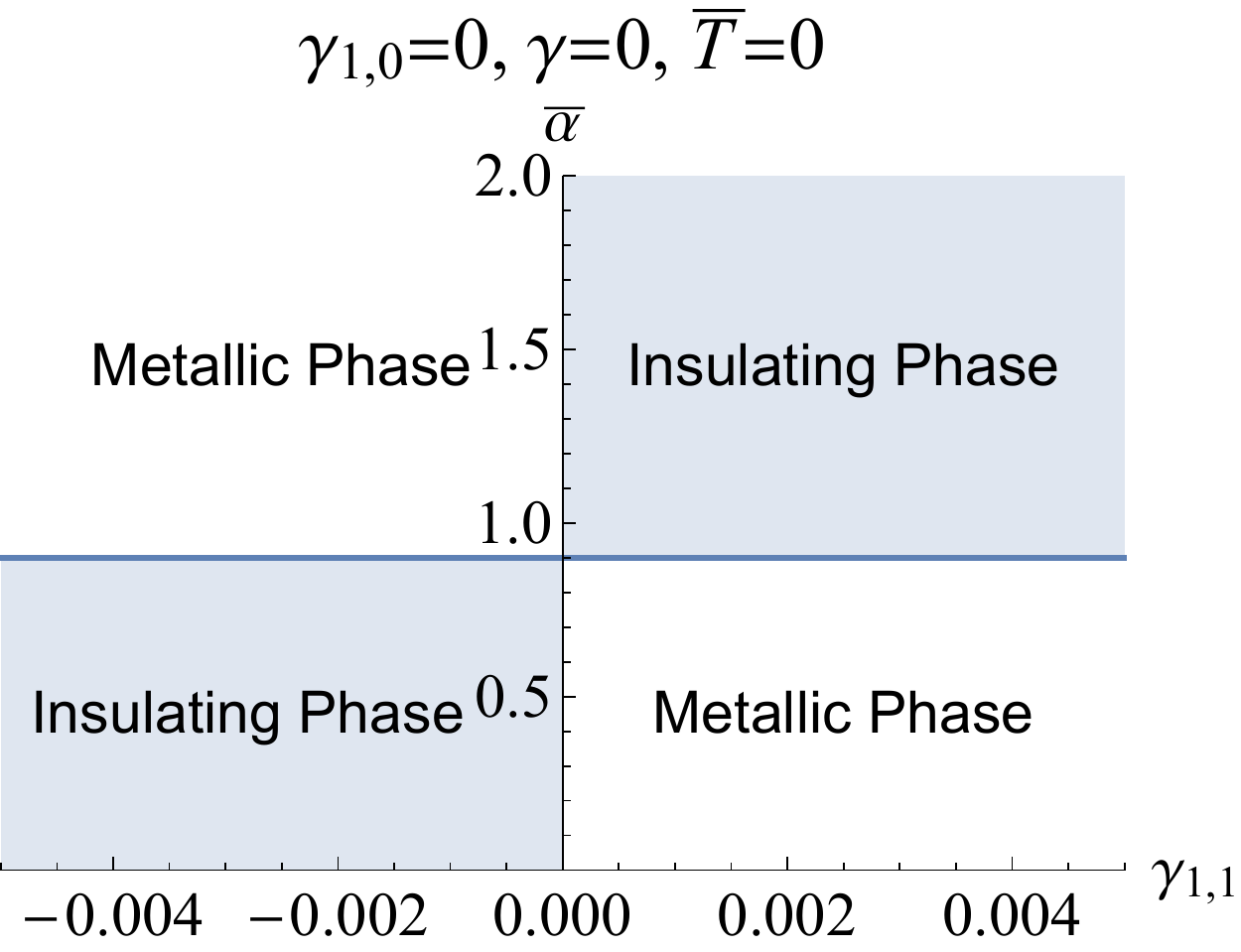}\ \hspace{0.8cm}\ \\
\caption{\label{pd_gamma11} Left plot: $\partial_{\bar{T}}\sigma_0$ as a function of $\bar{\alpha}$ at zero temperature.
Right plot: Phase diagram over $(\gamma_{1,1},\bar{\alpha})$ plane for the MIT from six derivative theory at zero temperature.}}
\end{figure}

Now, we turn to a study of the effect of the six derivative term. For simplicity, we turn off the four derivative terms, i.e., we set $\gamma_{1,0}=0$ and $\gamma=0$.
Figure. \ref{dcvst_gamma11} exhibits the DC conductivity $\sigma_0$ as a function of the temperature $\bar{T}$ for some representative $\bar{\alpha}$ and $\gamma_{1,1}$.
The left plot in Fig. \ref{pd_gamma11} shows $\partial_{\bar{T}}\sigma_0$ as a function of $\bar{\alpha}$ at zero temperature,
while right plot shows the phase diagram in the $(\gamma_{1,1},\bar{\alpha})$ plane for the MIT at zero temperature.
We find that the properties of DC transport from six derivative theory is very similar to that from four derivative theory only involving the Weyl term \cite{Ling:2016dck}; here the mirror symmetries \eqref{mir-sym} and \eqref{mir-sym-psigma-pt} hold for fixed $\bar{\alpha}$ and changing the signs of $\gamma_{1,1}$ in
the phase diagram over $(\gamma_{1,1},\bar{\alpha})$ plane for the MIT at zero temperature.
One difference is that the quantum critical line is shifted to $\bar{\alpha}\simeq 0.9$. \footnote{The quantum critical line from four derivative theory only involving the Weyl term is located at $\bar{\alpha}\simeq 0.82$.}
It is also interesting to explore the DC conductivity at finite density from the six derivative theory only involving Weyl terms
and compare the results with present results, including the mixed effect of both axions and Weyl tensor.
We leave this problem for future study.


\section{Transports at zero density}\label{sec-cond}

In this section, we study the transports at zero density.
In this case, the black brane solution reduces to the neutral one \cite{Andrade:2013gsa}
\fa
&&
\label{bl-br}
\mathrm{d}s^2=\frac{1}{u^2}\Big(-f(u)\mathrm{d}t^2+\frac{1}{f(u)}\mathrm{d}u^2+\mathrm{d}x^2+\mathrm{d}y^2\Big)\,,
\nonumber
\\
&&
f(u)=(1-u)p(u)\,,~~~~~~~
p(u)=\frac{\sqrt{1+6\hat{\alpha}^2}-2\hat{\alpha}^2-1}{\hat{\alpha}^2}u^2+u+1\,.
\ffa
Note that we have parameterized this black brane solution by one scaling-invariant quantity $\hat{\alpha}=\alpha/4\pi T$
with $T=p(1)/4\pi$.
Based on this neutral geometry background, we shall study the transport starting from four derivative and six derivative theory, respectively.

\subsection{Four derivative theory}

\begin{figure}
\center{
\includegraphics[scale=0.5]{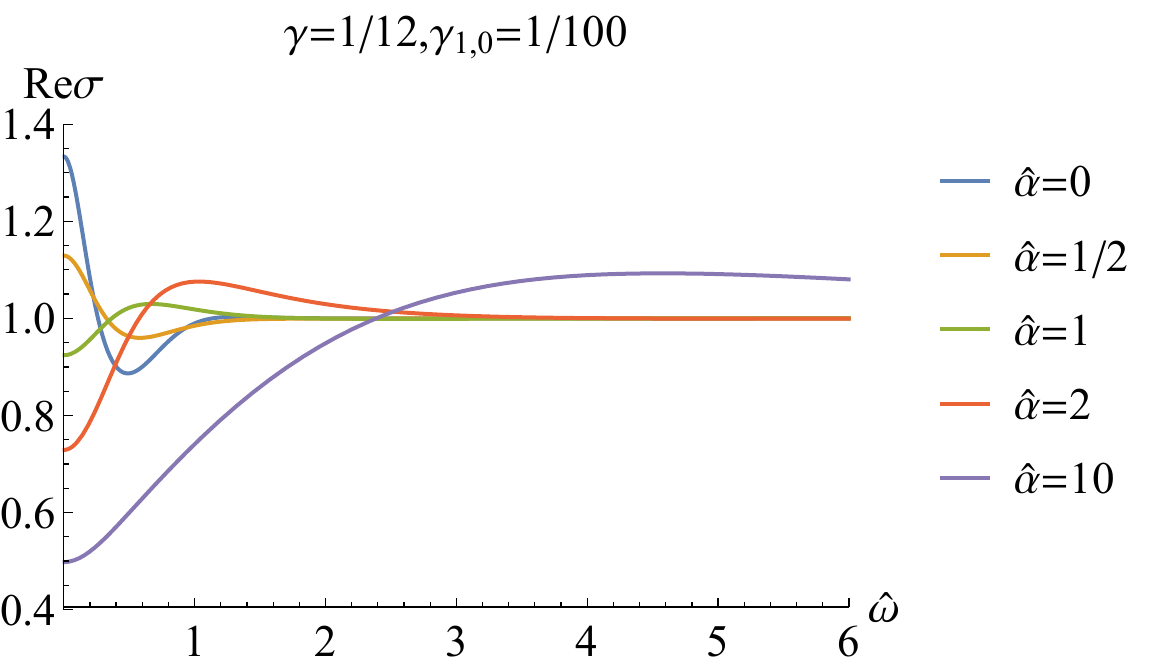}\ \hspace{0.8cm}
\includegraphics[scale=0.5]{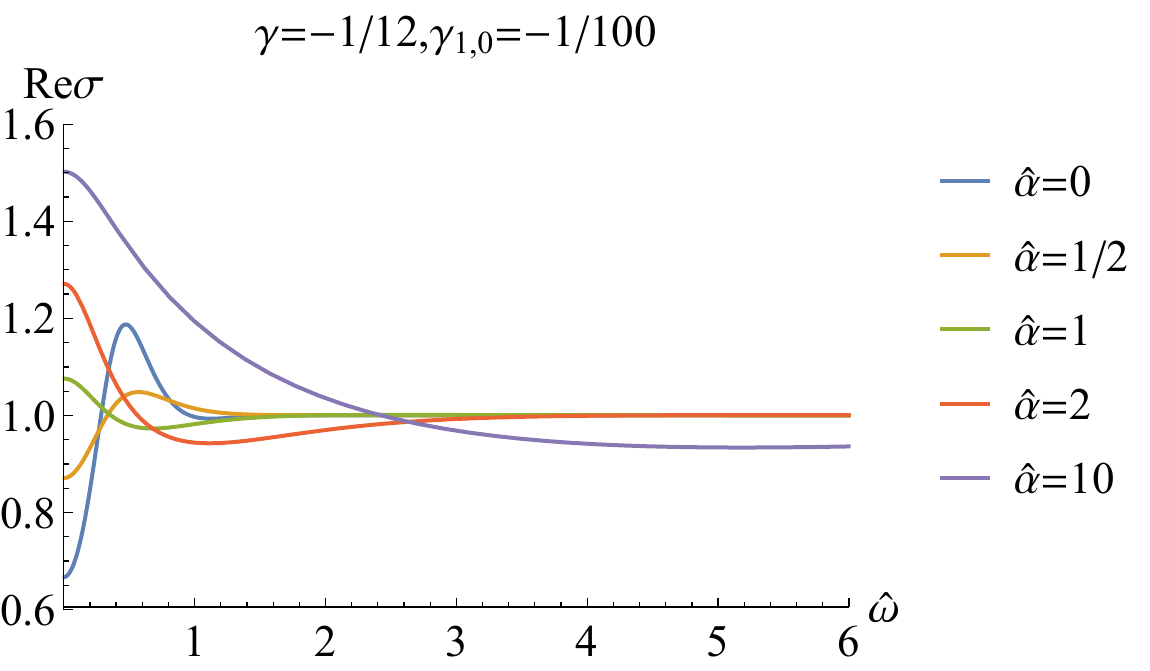}\ \\
\caption{\label{fig-case1} The optical conductivity $\sigma(\hat{\omega})$ as a function of $\hat{\omega}$
with representative $\gamma_{1,0}$, $\gamma$ and $\hat{\alpha}$.
}}
\end{figure}

In this section, we study the properties of the conductivity in four derivative theory
and see how the new higher derivative coupling term $\gamma_{1,0}$ affects them.
Figure. \ref{fig-case1} shows the optical conductivity $\sigma(\hat{\omega})$ as a function of $\hat{\omega}$
with representative $\gamma_{1,0}$, $\gamma$ and $\hat{\alpha}$.
Comparing Fig. \ref{fig-case1} with Fig. 1 in our previous work \cite{Wu:2016jjd},
we observe that, for the system with positive (negative) $\gamma$ and $\gamma_{1,0}$,
the transition from peak (dip) to dip (peak) appears to go easier
with the increase of $\hat{\alpha}$.
\begin{figure}
\center{
\includegraphics[scale=0.5]{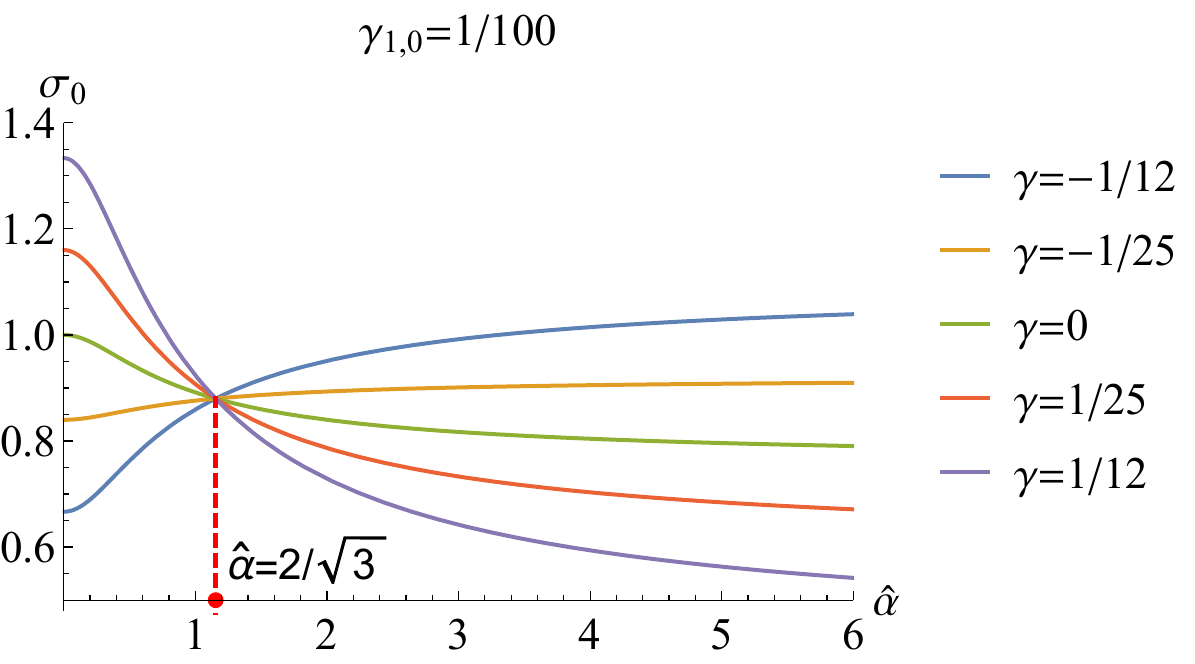}\ \hspace{0.8cm}
\includegraphics[scale=0.5]{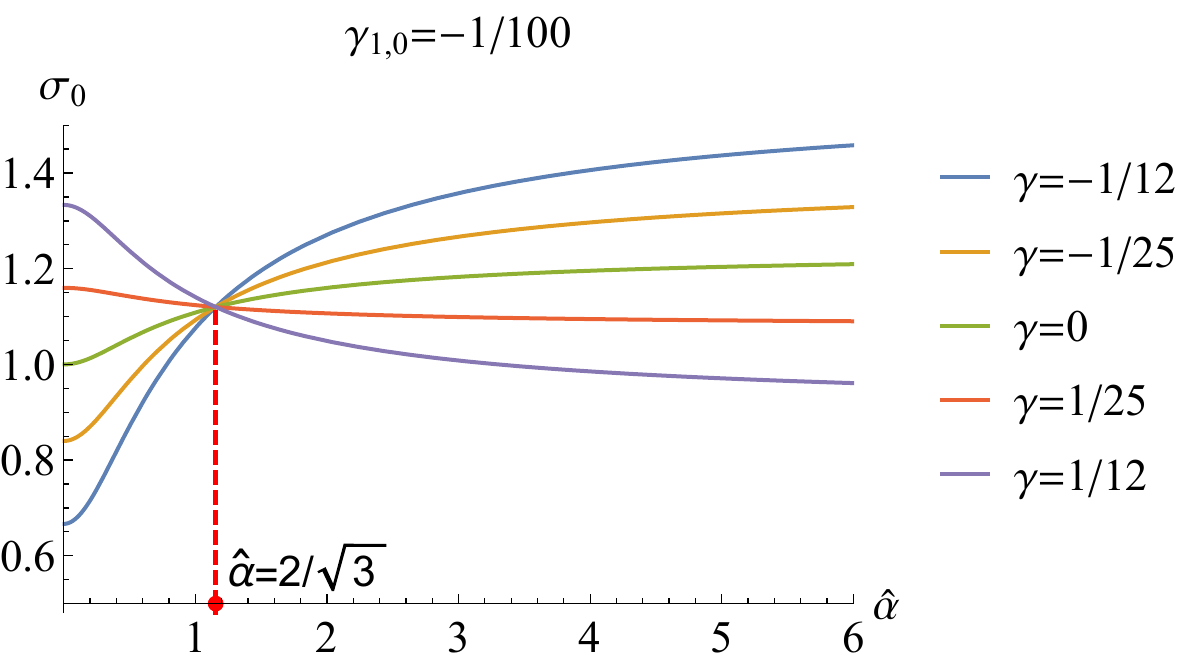}\ \\
\caption{\label{dc_I_gamma10} The DC conductivity $\sigma_0$ versus $\hat{\alpha}$ for the representative $\gamma$ and $\gamma_{1,0}$.
}}
\end{figure}
\begin{figure}
\center{
\includegraphics[scale=0.5]{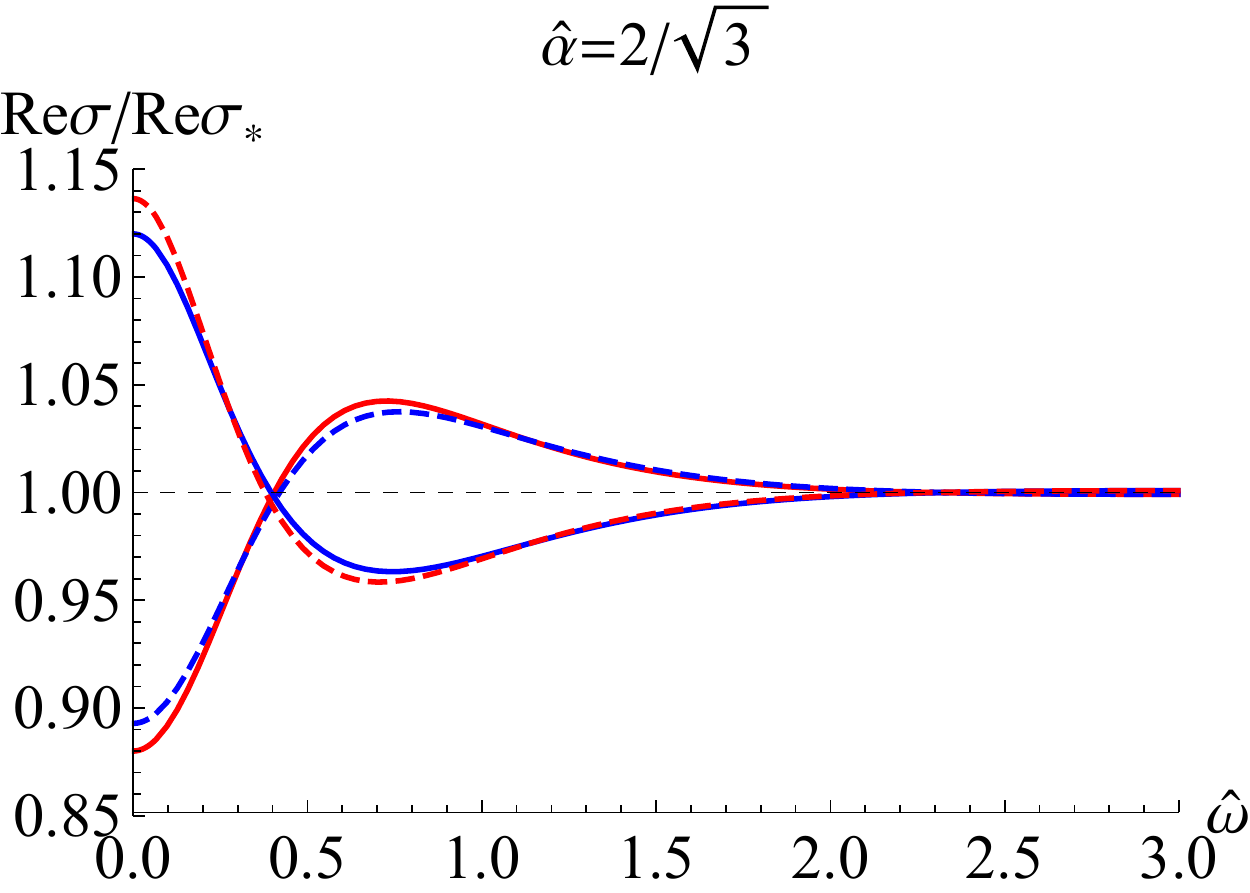}\ \hspace{0.8cm}
\includegraphics[scale=0.5]{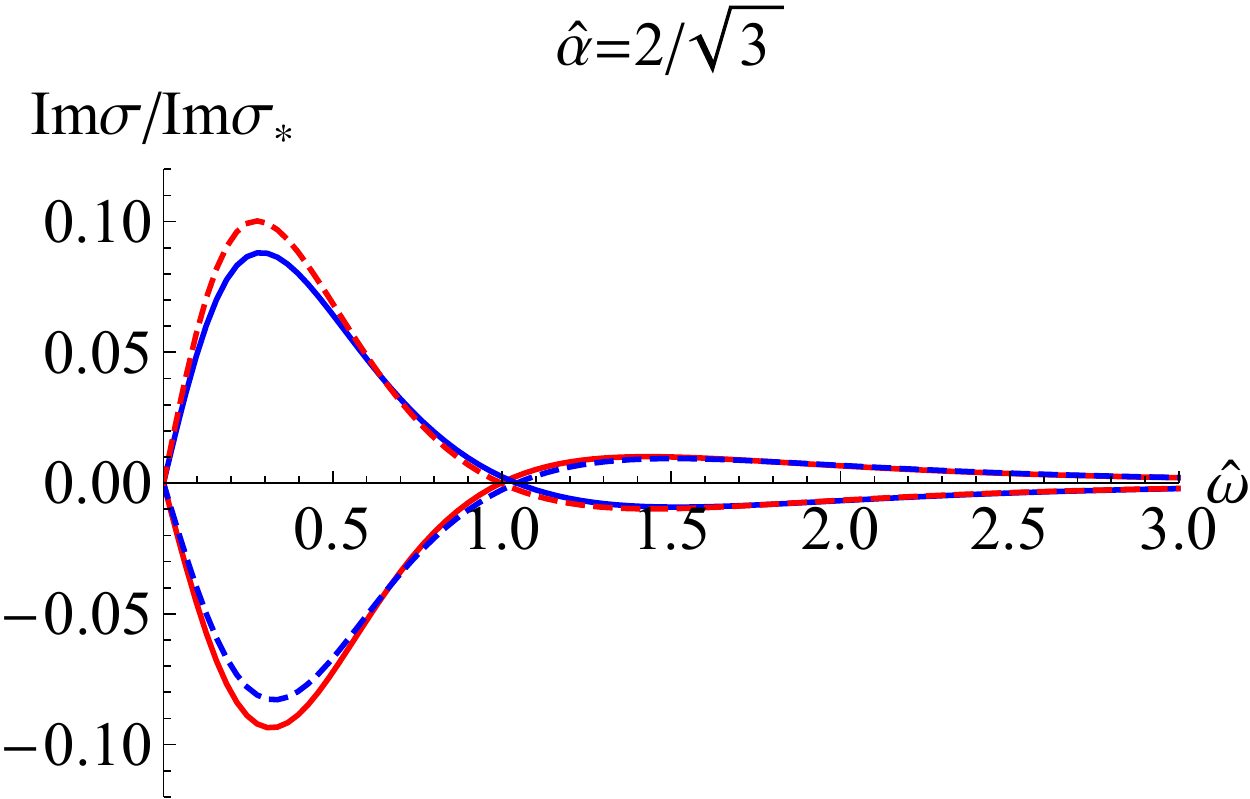}\ \\
\caption{\label{case1_dual_2osqrt3} The optical conductivity as a function of $\hat{\omega}$
for various values of $\gamma$, $\gamma_{1,0}$ and fixed $\hat{\alpha}=2/\sqrt{3}$.
The solid and dashed curves are the conductivity of the original EM theory and its dual theory, respectively
(red for $\gamma=1/12$ and $\gamma_{1,0}=1/100$ and blue for $\gamma=-1/12$ and $\gamma_{1,0}=-1/100$).
}}
\end{figure}

As have been revealed in \cite{Wu:2016jjd},
the particle-vortex duality is recovers with the change of $\gamma\rightarrow -\gamma$
for a specific value of $\hat{\alpha}=2/\sqrt{3}$.
Now we want to explore if this phenomenon is generic when a new higher derivative coupling term $\gamma_{1,0}$ is taken into account.
Figure. \ref{dc_I_gamma10} shows the DC conductivity $\sigma_0$ as a function of $\hat{\alpha}$ for the representative $\gamma$ and $\gamma_{1,0}$.
We find that, for a given $\gamma_{1,0}$, all the lines of $\sigma_0(\hat{\alpha})$ with different $\gamma$ intersect at one point $\hat{\alpha}=2/\sqrt{3}$,
which is similar to that found for only the Weyl term $\gamma$ being involved.
It indicates that $\sigma_0(\hat{\alpha})$ is independent of $\gamma$ for $\hat{\alpha}=2/\sqrt{3}$,
which can also be deduced from the expression for DC conductivity (\ref{dc-gamma10-gamma}).
But we note that the value of $\sigma_0(\hat{\alpha}=2/\sqrt{3},\gamma)$ is not equal to unity.
Also, the relation $\sigma_0(\hat{\alpha}=2/\sqrt{3},\gamma)=\frac{1}{\sigma_0(\hat{\alpha}=2/\sqrt{3},-\gamma)}$ does not hold.
It indicates the exact duality of the DC conductivity only with the Weyl term for $\hat{\alpha}=2/\sqrt{3}$ is violated when the $\gamma_{1,0}$ term is taken into account.
Furthermore, we study the optical conductivities of both the original EM theory and its dual theory for the specific value of $\hat{\alpha}=2/\sqrt{3}$,
shown in Fig. \ref{case1_dual_2osqrt3},
and we find that the exact particle-vortex duality is indeed violated when $\gamma\rightarrow-\gamma$ and $\gamma_{1,0}\rightarrow-\gamma_{1,0}$.
It is easy to check that if we fix $\gamma_{1,0}$, the particle-vortex duality is also violated when $\gamma\rightarrow-\gamma$.

\subsection{Six derivative theory}

\begin{figure}
\center{
\includegraphics[scale=0.5]{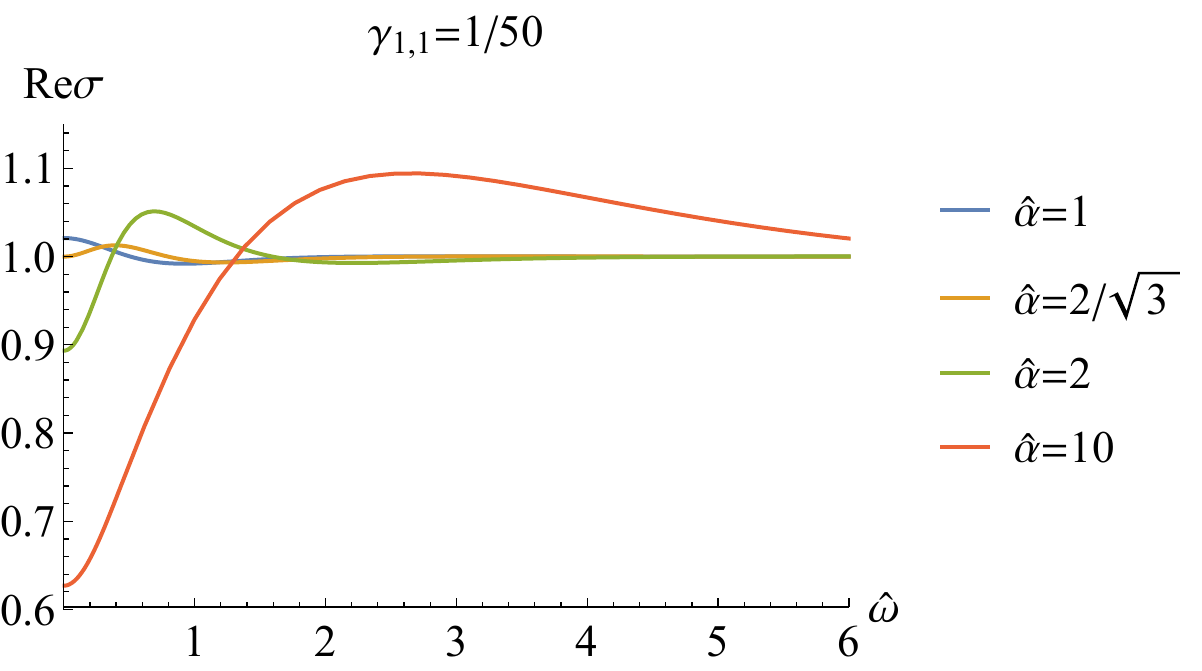}\ \hspace{0.8cm}
\includegraphics[scale=0.5]{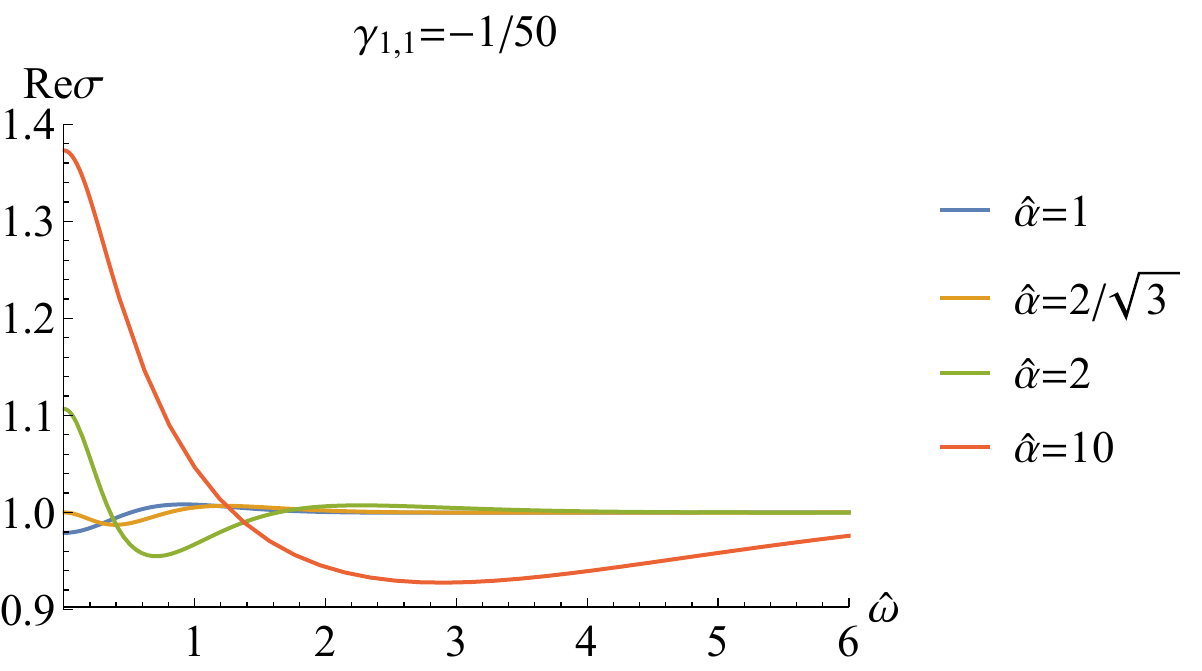}\ \\
\caption{\label{fig-case2} The optical conductivity $\sigma(\hat{\omega})$ as the function of $\hat{\omega}$
with representative $\gamma_{1,1}$ and $\hat{\alpha}$.
}}
\end{figure}

Now, we turn to a study of the case in six derivative theory.
Figure \ref{fig-case2} shows the optical conductivity with $\gamma_{1,1}$ being turned on.
We observe that for positive $\gamma_{1,1}$ and small $\hat{\alpha}$,
a small peak is displayed in the low frequency region.
With the increase of $\hat{\alpha}$, the small peak starts to develop into a dip (left plot in Fig. \ref{fig-case2}).
Meanwhile for negative $\gamma_{1,1}$,
an opposite scenario is found (right plot in Fig. \ref{fig-case2}).
The phenomenon is similar to that with the $\gamma$ term.

\begin{figure}
\center{
\includegraphics[scale=0.5]{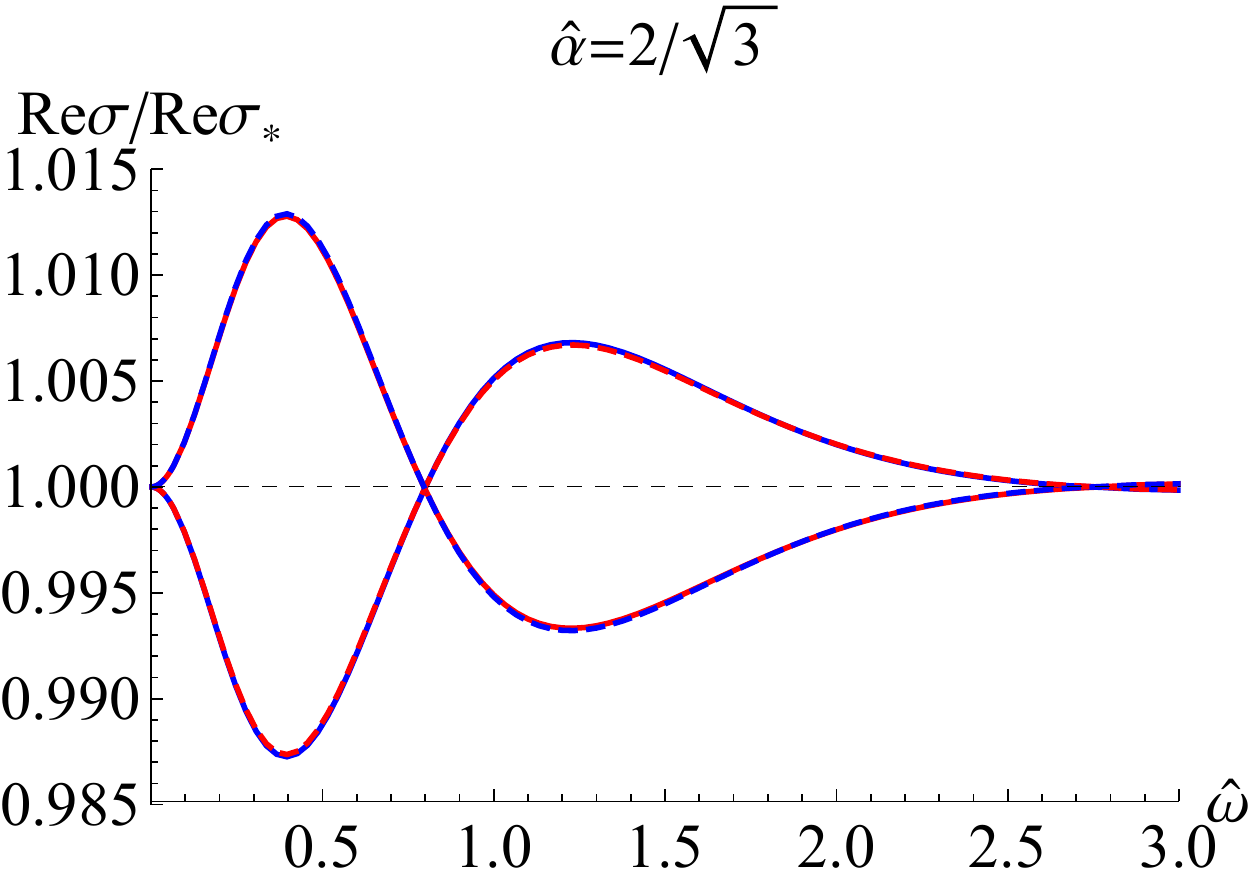}\ \hspace{0.8cm}
\includegraphics[scale=0.5]{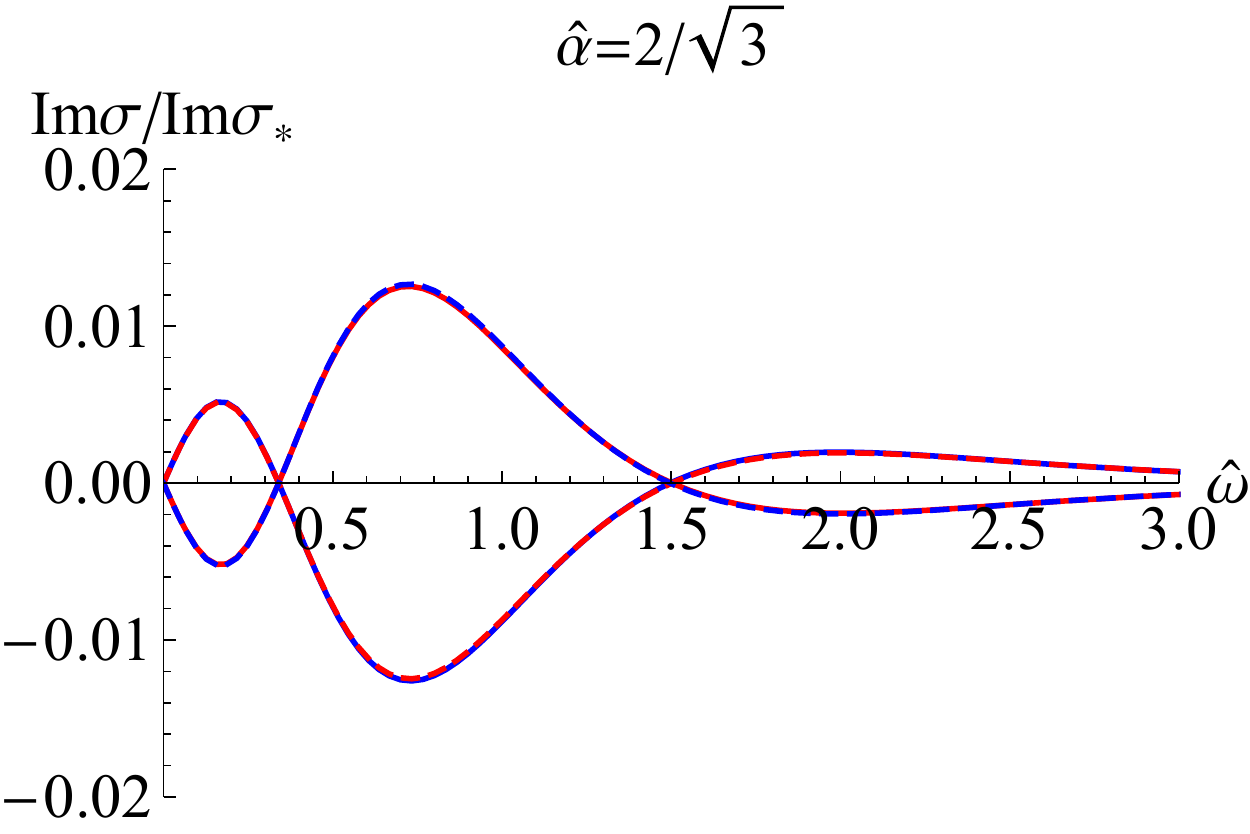}\ \\
\includegraphics[scale=0.5]{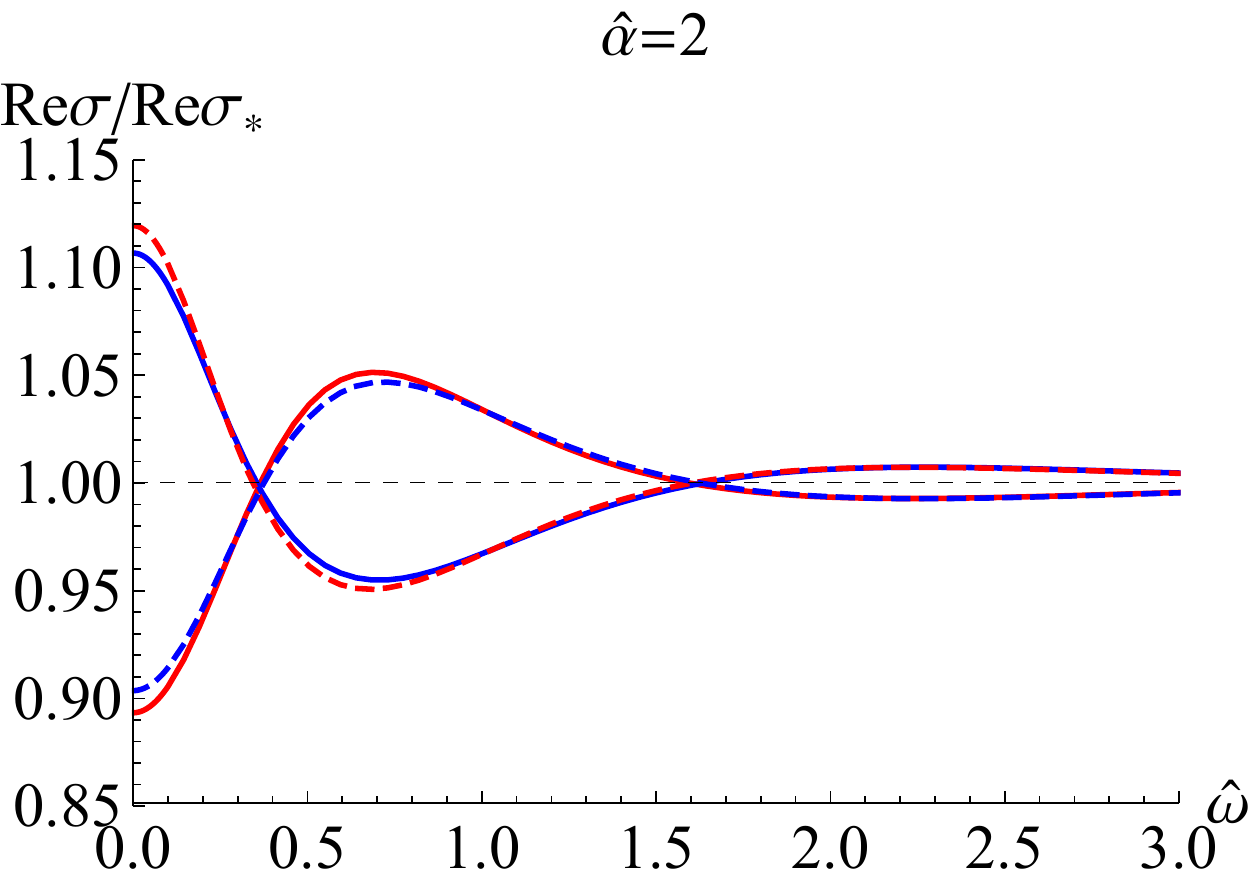}\ \hspace{0.8cm}
\includegraphics[scale=0.5]{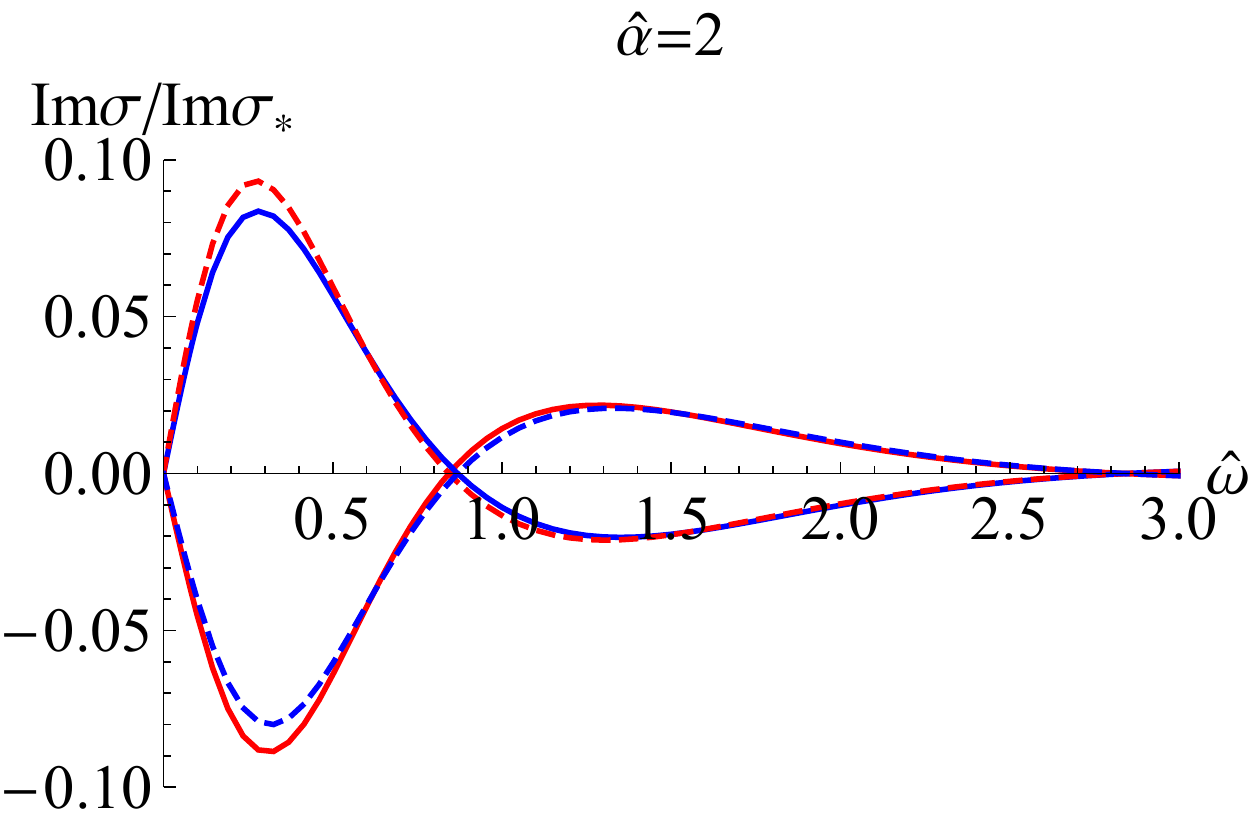}\ \\
\caption{\label{fig-case2-dual} The optical conductivity as the function of $\hat{\omega}$
for various values of $\gamma_{1,1}$ and $\hat{\alpha}$.
The solid and dashed curves are the conductivity of the original EM theory and its dual theory, respectively
(red for $\gamma_{1,1}=1/50$ and blue for $\gamma_{1,1}=-1/50$).
}}
\end{figure}

\begin{figure}
\center{
\includegraphics[scale=0.5]{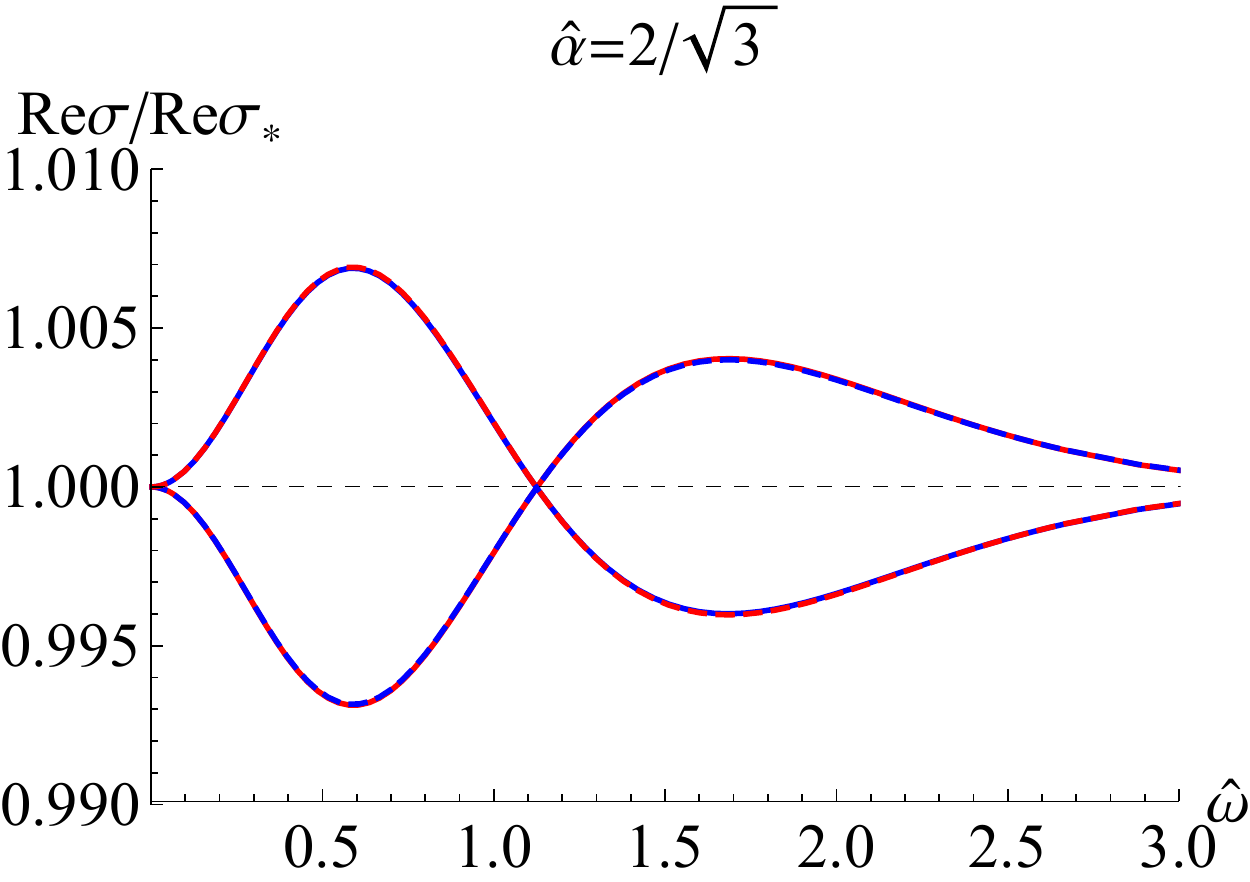}\ \hspace{0.8cm}
\includegraphics[scale=0.5]{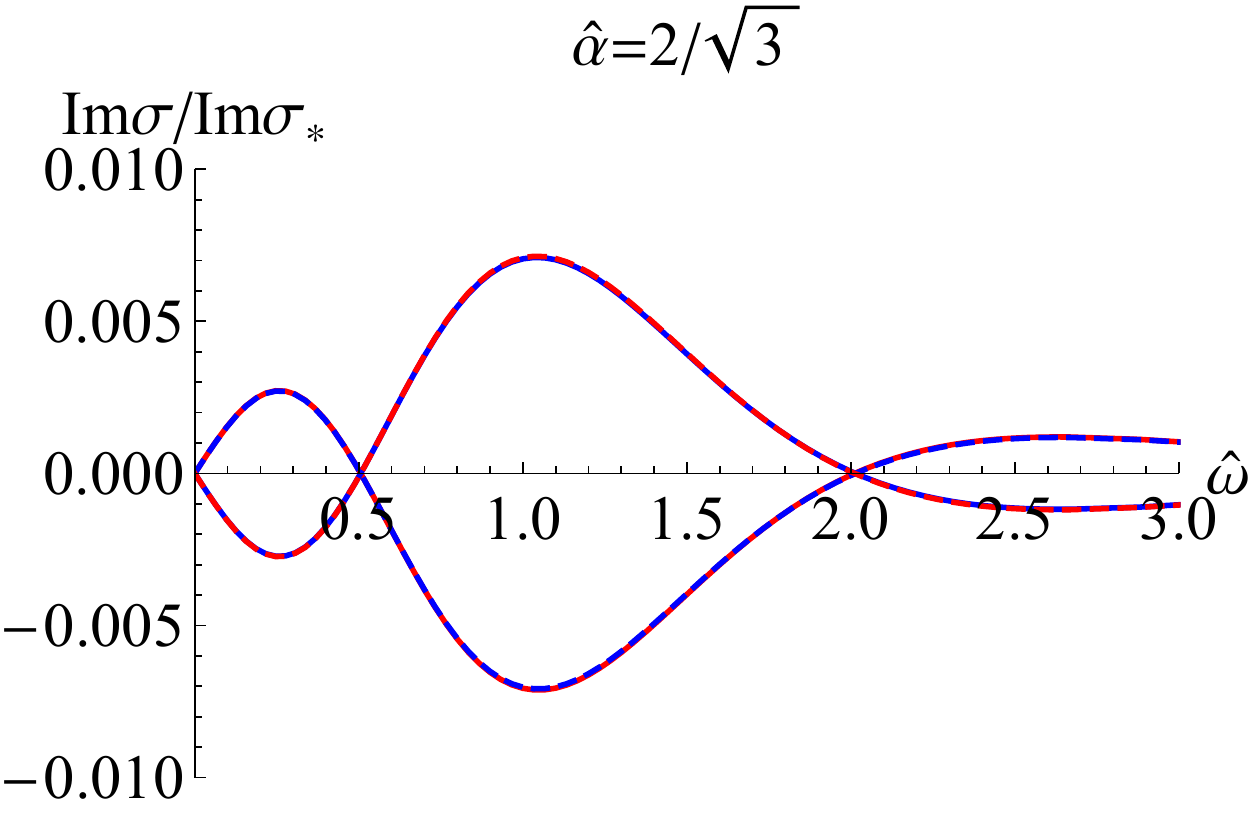}\ \\
\caption{\label{fig-case3-dual} The optical conductivity as the function of $\hat{\omega}$
for $\gamma_{1}=\pm 0.02$ and $\hat{\alpha}=2/\sqrt{3}$.
The solid and dashed curves are the conductivity of the original EM theory and its dual theory, respectively
(red for $\gamma_{1}=0.02$ and blue for $\gamma_{1}=-0.02$).
}}
\end{figure}

Also, we note that, for the specific value of $\hat{\alpha}=2/\sqrt{3}$,
the DC conductivity $\sigma_0=1$ and is independent of $\gamma_{1,1}$ (see Fig. \ref{sigmavsalphabeta11}),
which is similar to that with only the Weyl term \cite{Wu:2016jjd}.
Furthermore, we study the particle-vortex duality of this case,
which is shown in Fig. \ref{fig-case2-dual}.
It is obvious that for small $\gamma_{1,1}$, the particle-vortex duality approximately holds.
Meanwhile, for the specific value of $\hat{\alpha}=2/\sqrt{3}$, the duality exactly holds.
Though here we do not work out the analytical understanding on the particle-vortex duality
for the specific value of $\hat{\alpha}=2/\sqrt{3}$,
it seems to originate from the Weyl term.
The additional $\gamma_{1,0}$ term violates this exact duality.
Further, we examine the duality from another six derivative term with $X_{\mu\nu}^{\ \ \rho\sigma}=-4\gamma_1C^2I_{\mu\nu}^{\ \ \rho\sigma}$,
of which the original theory has been studied in our previous work \cite{Fu:2017oqa}.
Again, the particle-vortex duality exactly holds for  $\hat{\alpha}=2/\sqrt{3}$ when $\gamma_1\rightarrow-\gamma_1$ (see Fig. \ref{fig-case3-dual}).
In future, we will further test the robustness of this phenomena by exploring that with the higher order terms of the Weyl coupling.

\section{Discussions}\label{sec-discussion}

In this work, we extend our previous work \cite{Ling:2016dck,Wu:2016jjd}
to constructing a higher derivative theory including the coupling among the axionic field, the Weyl tensor and the gauge field.
To be more specific, we construct four derivative terms, a simple summation of the Weyl term $C_{\mu\nu\rho\sigma}$ coupling with the gauge field,
as well as a term from the trace of axions coupling with the gauge field,
and a six derivative term, a mixed term by the product of Weyl tensor and the axionic field, coupling with the gauge field.

Following the strategy in \cite{Ling:2016dck}, we construct the charged black brane solution with momentum dissipation in a perturbative manner up to the first order of the coupling parameters.
We study the QCP from four and six derivative theory, respectively.
For four derivative theory, because of the introduction of $\gamma_{1,0}$, the quantum critical line is independent of $\gamma$,
which is different from the case only involving the $4$ derivative term in \cite{Wu:2016jjd}.
It provides a new platform of QCP such that we can study holographic entanglement entropy and butterfly effect close to QCP,
which may inspire new insight.
For six derivative theory, the quantum critical line is independent of the coupling parameter $\gamma_{1,1}$,
which is similar that in \cite{Ling:2016dck}.

Also, we study the transport phenomena including DC conductivity and optical conductivity at zero charge density,
which is away from the QC phase.
For four derivative theory,
the momentum dissipation makes the transition from peak (dip) to dip (peak) easier,
comparing with that in our previous work \cite{Wu:2016jjd}.
In addition, we find that for the specific value of $\hat{\alpha}=2/\sqrt{3}$,
the exact particle-vortex duality, holding for only the $\gamma$ term, survives \cite{Wu:2016jjd} and is violated when the $\gamma_{1,0}$ term is turned on.
For the six derivative theory, the particle-vortex duality exactly holds for $\hat{\alpha}=2/\sqrt{3}$.
Meanwhile the effect of the momentum dissipation on the transition between the gap and the dip
is similar to that in four derivative theory.

It is definitely a novelty and an interesting matter to compute the optical conductivity at finite chemical potential $\mu$.
However, even if we have obtained the perturbative black brane solution to the first order of $\gamma$ in Sect. \ref{sec-BBS},
we still need to solve the linear perturbative differential equations beyond the second order to obtain the optical conductivity.
It is a hard task and so we shall leave it for the future.
In addition, this simple model including the mixed terms between the Weyl tensor and the axions can be straightforwardly generalized to
include the charge complex scalar field such that we can study the superconducting phase.
It is also interesting and valuable to further explore the transport of our present model at full momentum and energy spaces,
which certainly will reveal more information of the systems.
This work deserves further study and we plan to
publish our results in the near future.

\begin{acknowledgments}
We in particular thank Zhenhua Zhou for the very helpful discussion on the calculation of DC conductivity at finite density.
We are grateful to Peng Liu and Wei-Jia Li for helpful discussions.
We also thank the anonymous referee for his/her very valuable suggestions, greatly improving our manuscript.
This work is supported by the Natural Science Foundation of China under Grant nos. 11775036, 11305018 and the Natural Science Foundation of Liaoning Province under Grant no. 201602013.

\end{acknowledgments}

\begin{appendix}

\section{Bounds on the coupling}\label{sec-Bounds}

In this appendix, we explore the constraints on the coupling parameters.
We mainly examine the causality of the dual boundary theory, the instabilities of the vector modes and the positive definiteness of the DC conductivity at zero charge density.
We also discuss the constraint from the requirement that the graviton mass is real, i.e., $m_g^2>0$.

\subsection{Bounds on the coupling at zero charge density}\label{sec-Bounds-zero}

To examine the causality of the dual boundary theory and the instabilities of the vector modes,
we decompose the perturbations of gauge field in the Fourier space as
$
A_{\mu}(t,x,y,u)\sim \mathrm{e}^{i\bf{q}\cdot\bf{x}}A_{\mu}(u,\bf{q}),
$
with $\textbf{q}\cdot\textbf{x}=-\omega t+ q^x x + q^y y$,
and write down the EOMs as follows:
\fa
&&
A'_t+\frac{\hat{q}f}{\hat{\omega}}\frac{X_5}{X_3}A'_x=0\,,
\label{Ma-AtI}
\\
&&
A''_t+\frac{X'_3}{X_3}A'_t
-\frac{\mathfrak{p}^2\hat{q}}{f}\frac{X_1}{X_3}\big(\hat{q}A_t+\hat{\omega}A_x\big)
=0\,,
\label{Ma-AtII}
\\
&&
A''_x+\Big(\frac{f'}{f}
+\frac{X'_5}{X_5}\Big)A'_x
+\frac{\mathfrak{p}^2\hat{\omega}}{f^2}\frac{X_1}{X_5}\big(\hat{q}A_t+\hat{\omega}A_x\big)=0\,,
\label{Ma-Ax}
\\
&&
A''_y
+\Big(\frac{f'}{f}+\frac{X'_6}{X_6}\Big)A'_y
+\frac{\mathfrak{p}^2}{f^2}\Big(\hat{\omega}^2\frac{X_2}{X_6}-\hat{q}^2f\frac{X_4}{X_6}\Big)A_y
=0\,,
\label{Ma-Ay}
\ffa
where the prime denotes the derivative with respect to $u$ and
the dimensionless frequency and momentum
$
\hat{\omega}\equiv\frac{\omega}{4\pi T}=\frac{\omega}{\mathfrak{p}}\,,
\hat{q}\equiv\frac{q}{4\pi T}=\frac{q}{\mathfrak{p}},
$ with
$
\mathfrak{p}\equiv p(1)=4\pi T
$, 
are introduced.
Due to the rotational symmetry in $xy$-plane, we have set $\textbf{q}^{\mu}=(\omega,q,0)$.
Also we choose the gauge as $A_{u}(u,\textbf{q})=0$.
At the same time, a tensor $X_{\mu\nu}^{\ \ \rho\sigma}$ defined as
$
X_{A}^{\ B}=\{X_1(u),X_2(u),X_3(u),X_4(u),X_5(u),X_6(u)\},
$
with
$
A,B\in\{tx,ty,tu,xy,xu,yu\},
$
has been introduced to simplify the expression of the perturbative EOMs.
Since the background is rotationally symmetric in the $xy$-plane, we have $X_1(u)=X_2(u)$ and $X_5(u)=X_6(u)$.
Combining Eqs. (\ref{Ma-AtI}) and (\ref{Ma-AtII}), one has a decoupled EOM for $A_t(u,\hat{\bf{q}})$, which is
\fa
A'''_t+\Big(\frac{f'}{f}-\frac{X'_1}{X_1}+2\frac{X'_3}{X_3}\Big)A''_t
+\Big(-\frac{\mathfrak{p}^2\hat{q}^2X_1}{fX_3}+\frac{\mathfrak{p}^2\hat{\omega}^2X_1}{f^2X_5}+\frac{f'X'_3}{fX_3}-\frac{X_1'X_3'}{X_1X_3}+\frac{X_3''}{X_3}\Big)A'_t=0\,.
\label{Ma-At-dec}
\ffa
By making a transform as $A_{\mu}\rightarrow B_{\mu}$ and $X_i\rightarrow \widehat{X}_i$,
we can obtain the EOMs of the dual EM theory from the above equations.
Note that from Eq. (\ref{X-hat}), it is easy to deduce that $\widehat{X}_A^{\ B}$ is also diagonal with $\widehat{X}_i=1/X_i$.

Since $A_x$ can be expressed by $A_t$ in terms of Eq. (\ref{Ma-AtI}),
there are only two independent vector modes, $A_t$ and $A_y$, which correspond to EOMs (\ref{Ma-At-dec}) and (\ref{Ma-Ay}).
They can be formulated in Schr\"odinger form as
\fa
-\partial_z^2\psi_i(z)+V_i(u)\psi_i(z)=\hat{\omega}^2\psi_i(z)\,.
\label{Sch-form}
\ffa
Notice that we have made a coordinate transformation, $dz/du=\mathfrak{p}/f$,
and a separation of variable, $A_i(u)=G_i(u)\psi_i(u)$, where $A_{\bar{t}}(u):=A'_t(u)$ and $i=\bar{t},y$.
For later convenience, we decompose the effective potential
$V_i(u)$ into both a momentum dependent part and an independent one,
\fa
V_i(u)=\hat{q}^2V_{0i}(u)+V_{1i}(u)\,,
\label{Vi-V0-V1}
\ffa
where \cite{Witczak-Krempa:2013aea}
\fa
&&
V_{0\bar{t}}=f\frac{X_1}{X_3}\,,
\,\,\,\,\,\,\,\,\,\,
V_{0y}=f\frac{X_3}{X_1}\,,
\label{V0t-V0y}
\\
&&
V_{1\bar{t}}=\frac{f}{4\mathfrak{p}^2X_1^2}[3f(X_1')^2-2X_1(fX_1')']\,,
\label{V1t}
\\
&&
V_{1y}=\frac{f}{4\mathfrak{p}^2X_1^2}[-f(X_1')^2+2X_1(fX_1')']\,.
\label{V1t}
\ffa

Before proceeding, we present the main ingredients constraining the coupling parameters as follows.
\begin{itemize}
  \item If $V_i(u)$ satisfies,
  \fa
0\leq V_{i}(u)\leq 1\,.
\label{V0i-constraint}
\ffa
  the modes meet the requirements of both causality and the stability of the dual boundary theory \cite{Buchel:2009tt,Brigante:2008gz,Myers:2007we}.
  \item When $V_i(u)$ violates the lower bound, the modes may be instable.
  We need further analyze the zero energy bound state of the potential.
  \item An additional condition is the requirement of positive definiteness of the real part of the conductivity,
  especially the DC conductivity.
\end{itemize}
Next, we analyze the constraint on the coupling parameters.

\subsubsection{Four derivative theory}

When only the coupling parameter $\gamma_{1,0}$ survives,
some related discussions have been explored in \cite{Gouteraux:2016wxj}.
But here one only discusses the Schr$\ddot{o}$dinger potential of the perturbation $A_x$.
Here, we shall present a more detailed discussion in our present framework.

\begin{figure}
\center{
\includegraphics[scale=0.5]{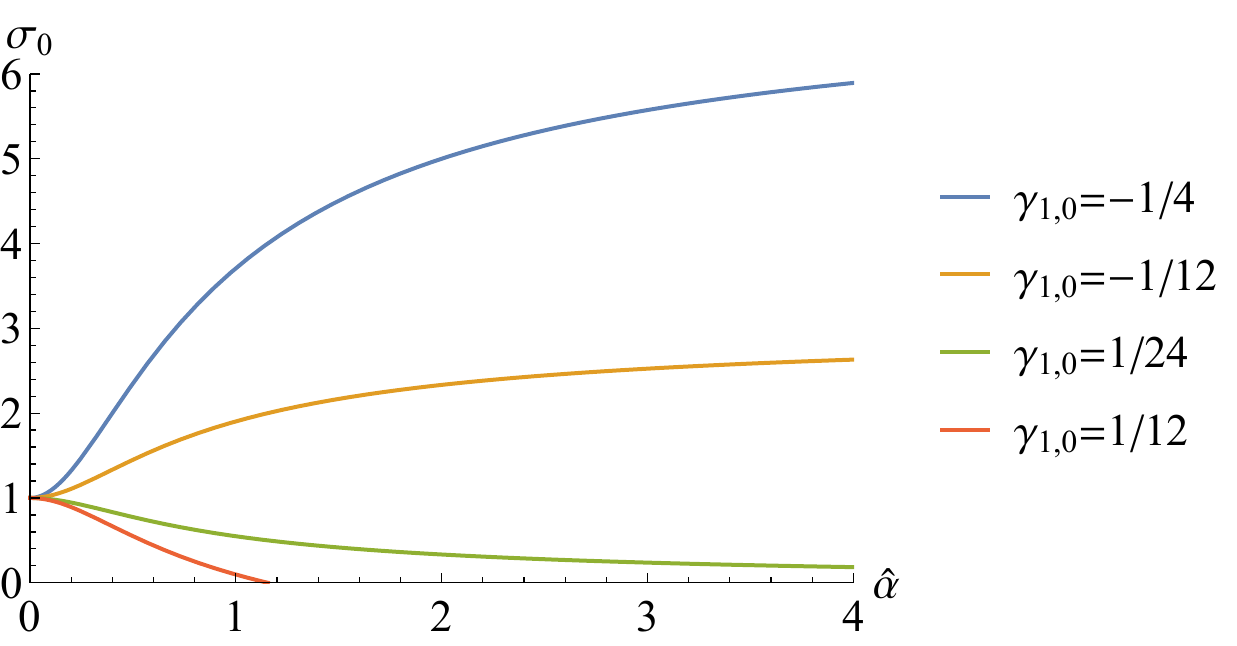}\ \\
\caption{\label{sigmavsalphabeta10} The DC conductivity as a function of $\hat{\alpha}$ for only $\gamma_{1,0}$ surviving.}}
\end{figure}

In terms of the expression of the DC conductivity \cite{Ritz:2008kh,Myers:2010pk}
\fa
\sigma_0=\sqrt{-g}g^{xx}\sqrt{-g^{tt}g^{uu}X_1X_5}\mid_{u=1}\,,
\label{DC}
\ffa
we can explicitly write it down when only $\gamma_{1,0}$ survives,
\begin{eqnarray}
\sigma_0=1-24\gamma_{1,0}-\frac{8\gamma_{1,0}}{\hat{\alpha}^2}+\frac{8\sqrt{1+6\hat{\alpha}^2}\gamma_{1,0}}{\hat{\alpha}^2}\,,
\end{eqnarray}
Figure \ref{sigmavsalphabeta10} shows $\sigma_0$ as a function of $\hat{\alpha}$ for sample values of $\gamma_{1,0}$.
We see that, for $\gamma_{1,0}\leq 1/24$, $\sigma_0$ is positive for all values of $\hat{\alpha}$.
Meanwhile, for $\gamma_{1,0}>1/24$, it vanishes for some finite $\hat{\alpha}$.
This can also be seen from the following: when $\hat{\alpha}\rightarrow+\infty$,
$\sigma_0=1-24\gamma_{1,0}$.
Therefore, a non-negative $\sigma_0$ gives a constraint on $\gamma_{1,0}$ as $\gamma_{1,0}\leq 1/24$.
\begin{figure}
\center{
\includegraphics[scale=0.5]{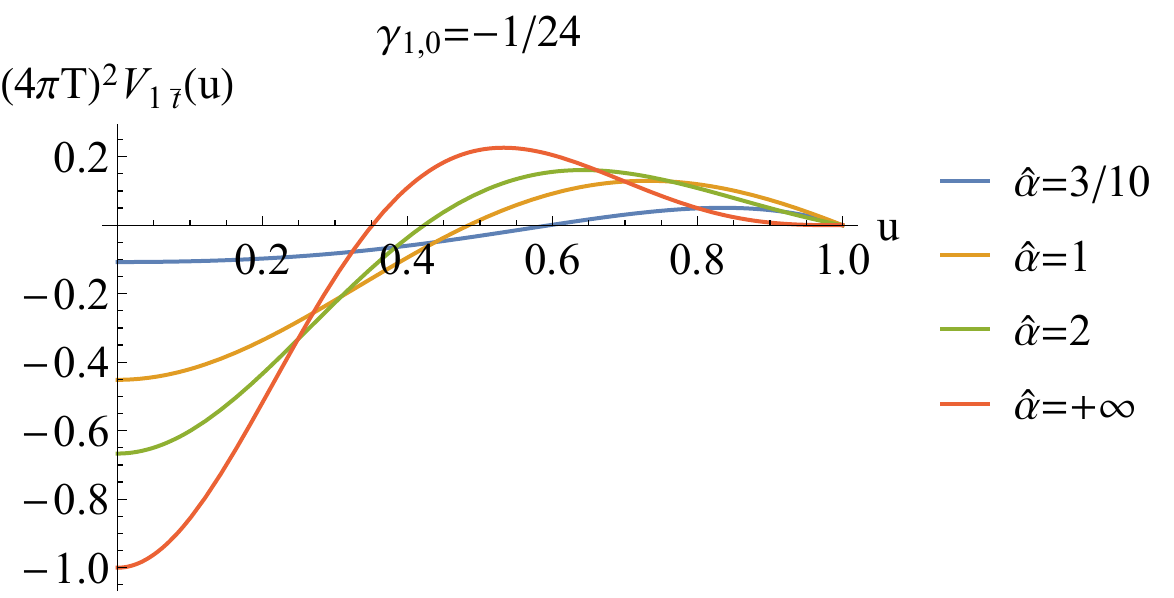}\ \hspace{0.8cm}
\includegraphics[scale=0.5]{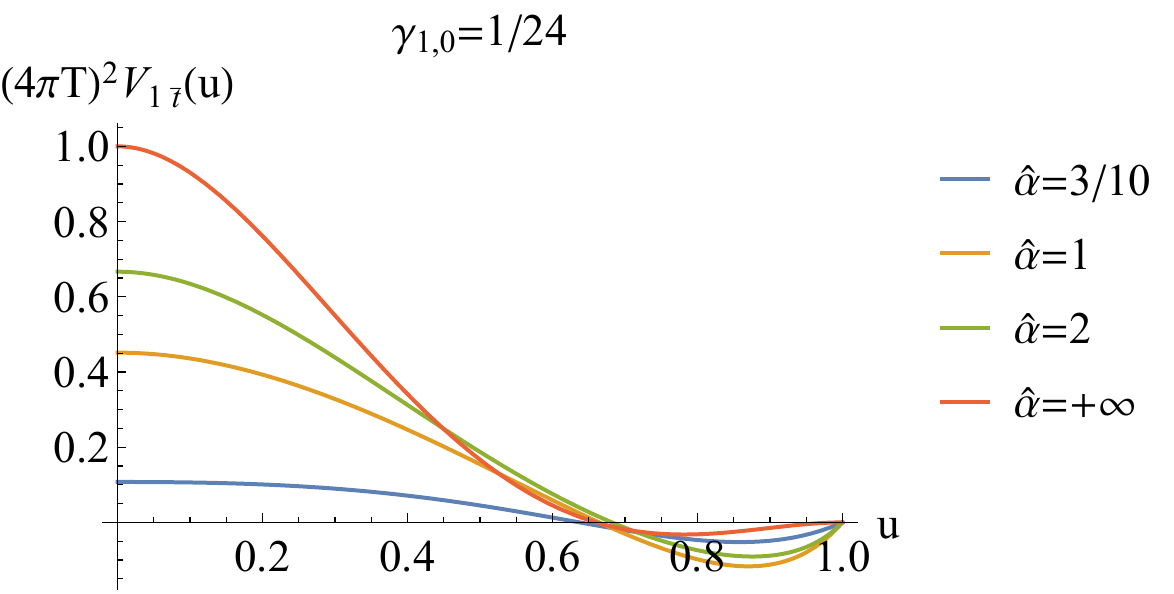}\ \hspace{0.8cm}\ \\
\includegraphics[scale=0.5]{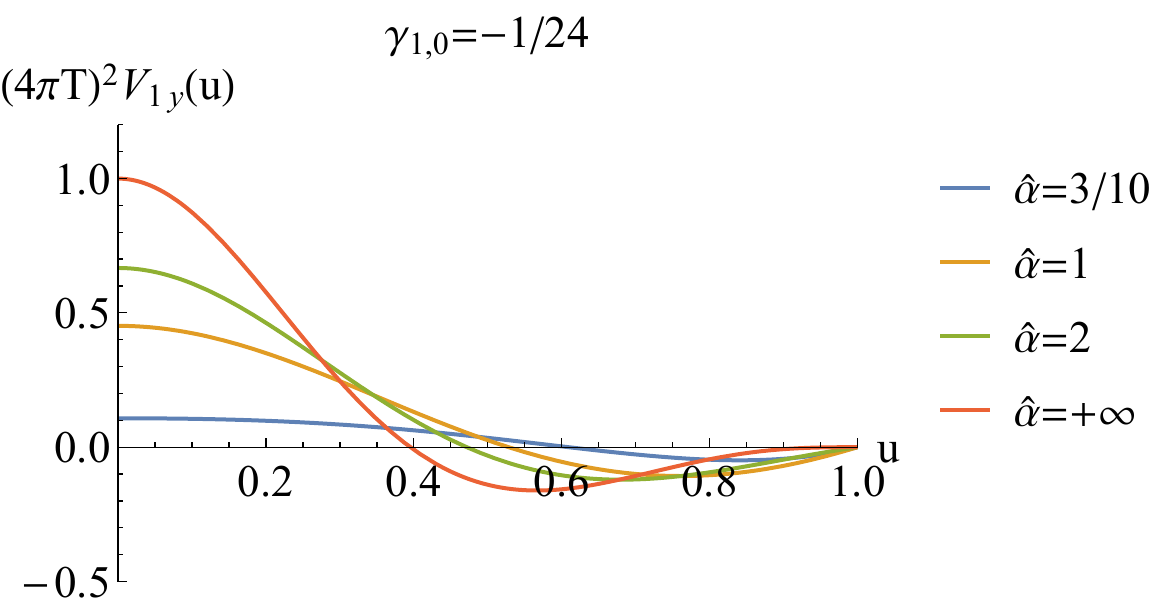}\ \hspace{0.8cm}
\includegraphics[scale=0.5]{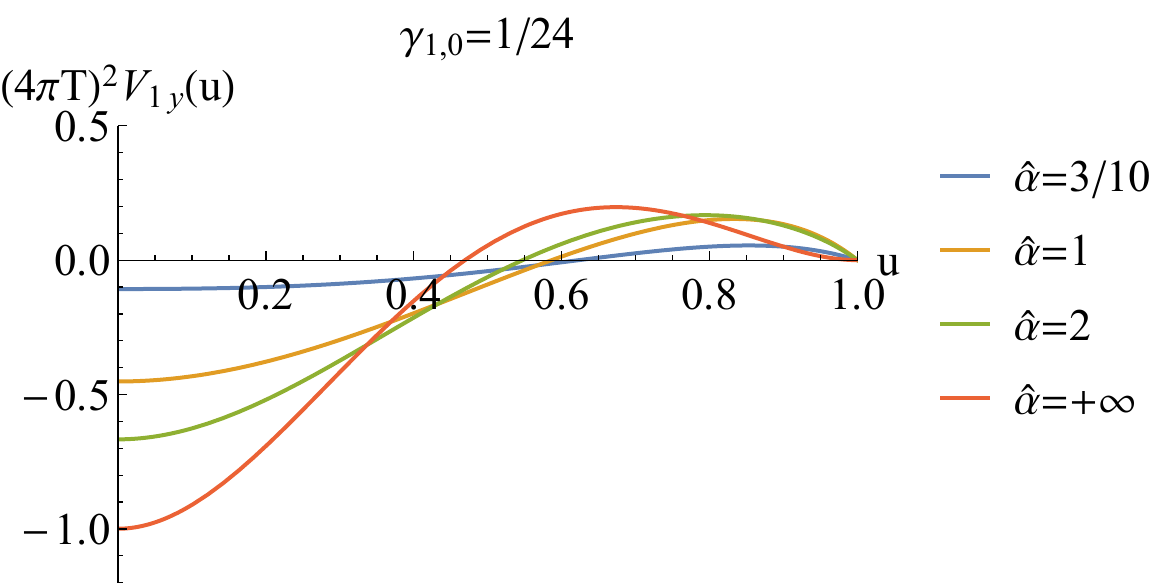}\ \hspace{0.8cm}\ \\
\caption{\label{V1tyvsugamma10} The potentials $V_{1\bar{t}}(u)$ (plots above) and $V_{1y}(u)$ (plots below)
with different $\gamma_{1,0}$ and $\hat{\alpha}$.}}
\end{figure}
\begin{figure}
\center{
\includegraphics[scale=0.5]{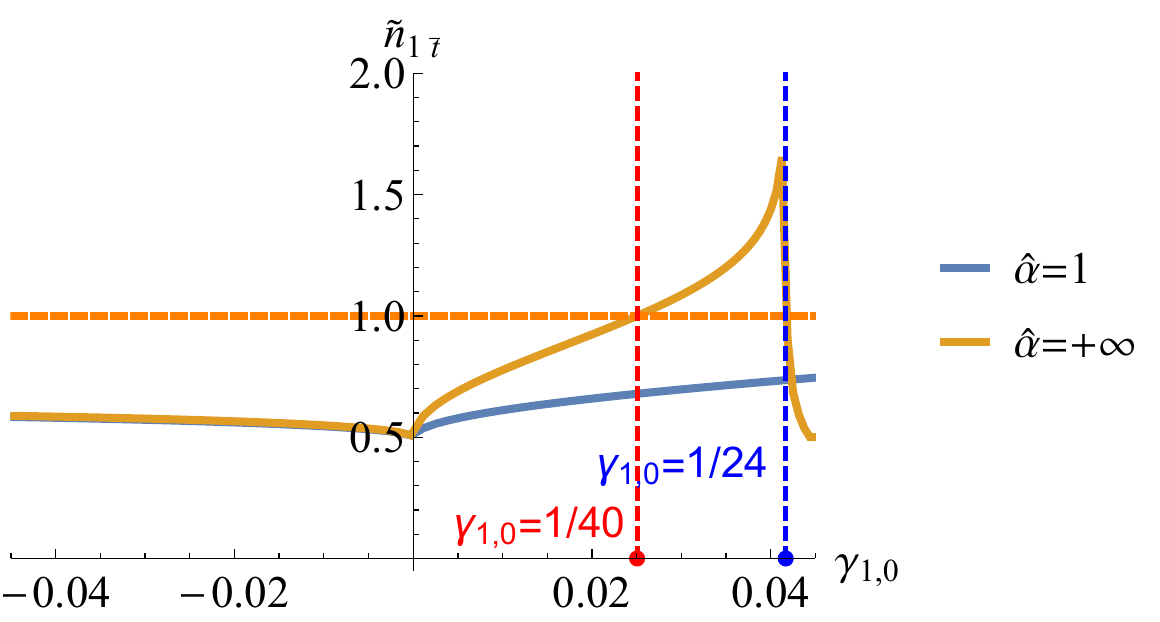}\ \hspace{0.8cm}
\includegraphics[scale=0.5]{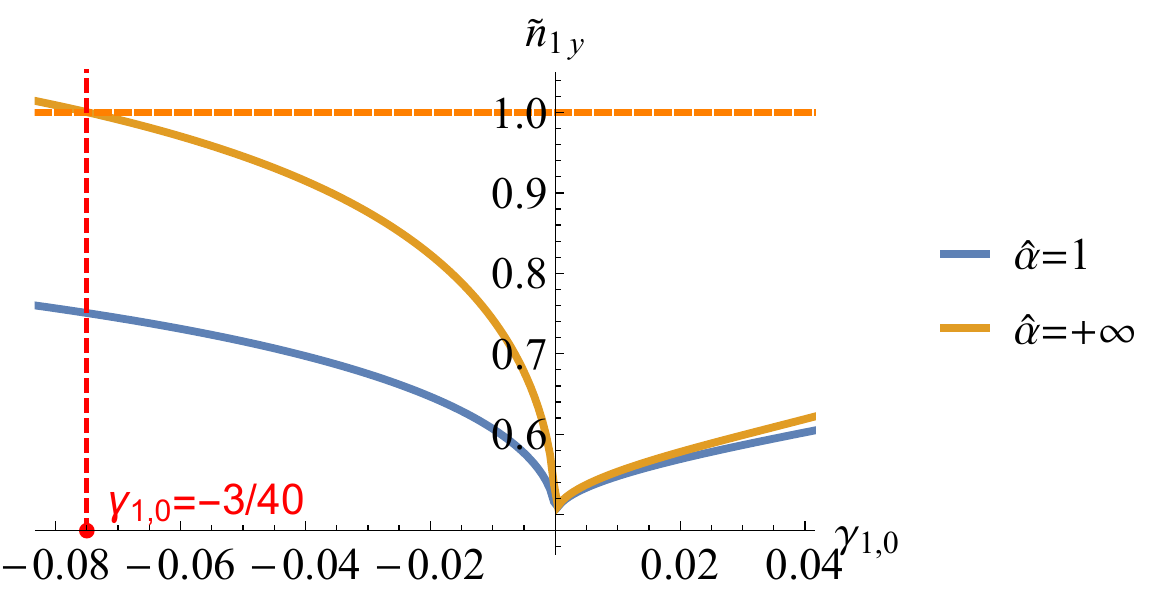}\ \hspace{0.8cm}\ \\
\caption{\label{n1tyvsgamma10} $\tilde{n}_{1\bar{t}}$ and $\tilde{n}_{1y}$ as a function of $\gamma_{1,0}$ for representative $\hat{\alpha}$.
}}
\end{figure}
\begin{figure}
\center{
\includegraphics[scale=0.5]{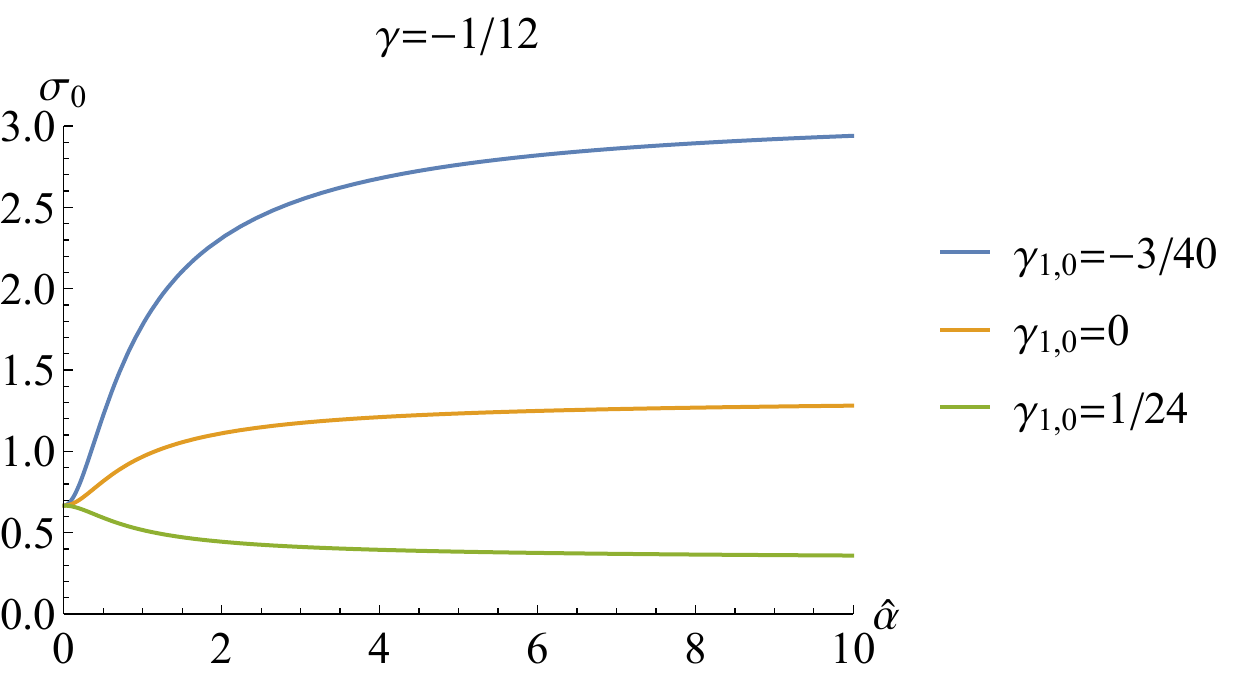}\ \hspace{0.8cm}
\includegraphics[scale=0.5]{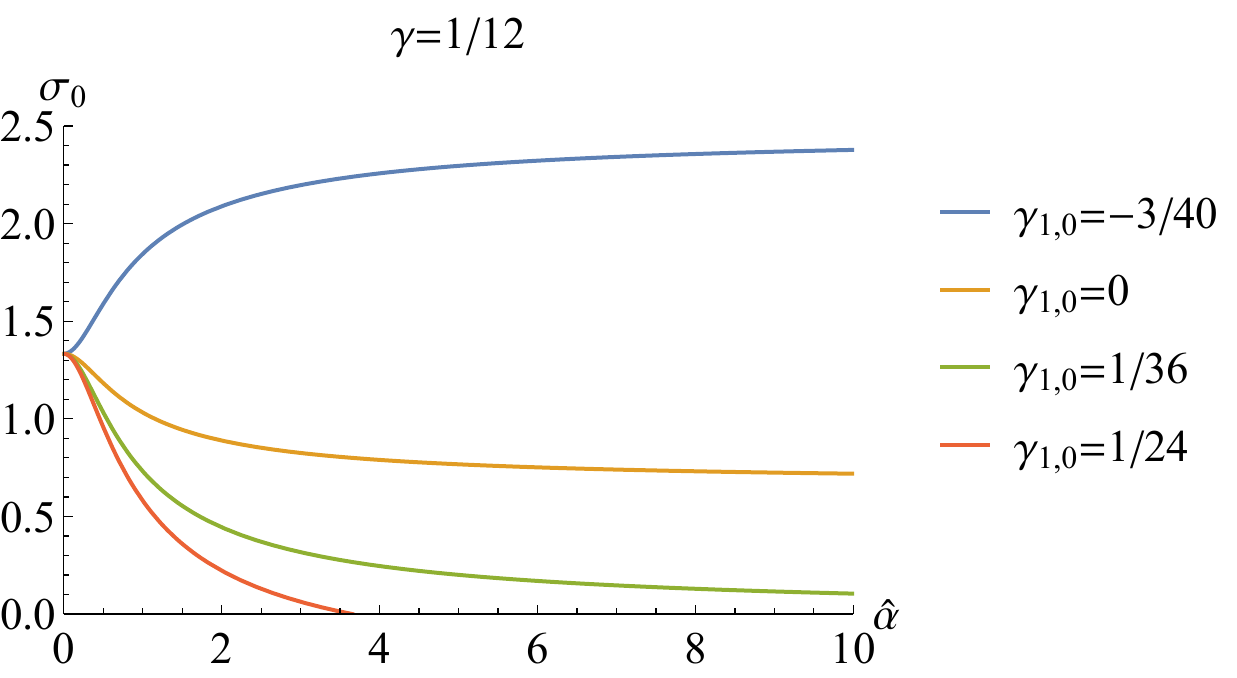}\ \hspace{0.8cm}\ \\
\caption{\label{sigmavsalphabeta10beta} The DC conductivity as a function of $\hat{\alpha}$ when both $\gamma_{1,0}$ and $\gamma$ are turned on.
}}
\end{figure}

Next, we turn to a discuss of the bounds of $\gamma_{1,0}$ imposed by the causality and the instabilities.
First, it is easy to find that in the limit of large momentum,
since $X_1=X_3=1-4\mathfrak{p}^2\hat{\alpha}^2\gamma_{1,0}u^2$ for only $\gamma_{1,0}$ we have surviving
$V_{0\bar{t}}=V_{0y}=f(u)$, which are the dominant terms.
Obviously, $V_{0\bar{t}}$ and $V_{0y}$ are independent of the parameter $\gamma_{1,0}$
and satisfy the constraint (\ref{V0i-constraint}).
Meanwhile for the case of the small momentum region, the dominant terms are $V_{1,i}$ ($i=\bar{t},y$),
which are shown for representative values of $\gamma_{1,0}$ and $\hat{\alpha}$ in Fig. \ref{V1tyvsugamma10}.
We can see that there is a negative minimum in $V_{1i}$.
So we need to analyze the zero energy bound state of the potentials,
which is \cite{Myers:2007we}
\fa
\label{zero-erergy-bound}
\tilde{n}_{1\bar{t}}=I/\pi+1/2\,,\,\,\,\,\,\,\,
I\equiv
\Big(n-\frac{1}{2}\Big)\pi
=\int_{u_0}^{u_1}\frac{\mathfrak{p}}{f(u)}\sqrt{-V_{1\bar{t}}(u)}\mathrm{d}u
\,,
\ffa
where $n$ is a positive integer and the potential well in the integral interval $[u_0,u_1]$ is negative.
Both $\tilde{n}_{1\bar{t}}$ and $\tilde{n}_{1y}$ as a function $\gamma_{1,0}$ for representative $\hat{\alpha}$
are exhibited in Fig. \ref{n1tyvsgamma10}.
The detailed analysis indicates that, when $\gamma_{1,0}$ belongs to the region $\gamma_{1,0}<-3/40$ and $1/40<\gamma_{1,0}<1/24$,
the $\tilde{n}_{1i}$ are greater than unit and unstable modes develop.
Combining the observation from DC conductivity,
we can infer that the allowed region for $\gamma_{1,0}$ is
\fa
\label{con_gamma10}
-3/40\leq\gamma_{1,0}\leq1/40\,.
\ffa
Also, we have checked that, for finite momentum, no unstable mode appears for the constraint \eqref{con_gamma10}.
Note that the lower bound of $\gamma_{1,0}$ is consistent with that found in \cite{Gouteraux:2016wxj}, but the upper bound
becomes tighter than that in \cite{Gouteraux:2016wxj}, which results from the instability of the mode $A_t$.
\begin{figure}
\center{
\includegraphics[scale=0.5]{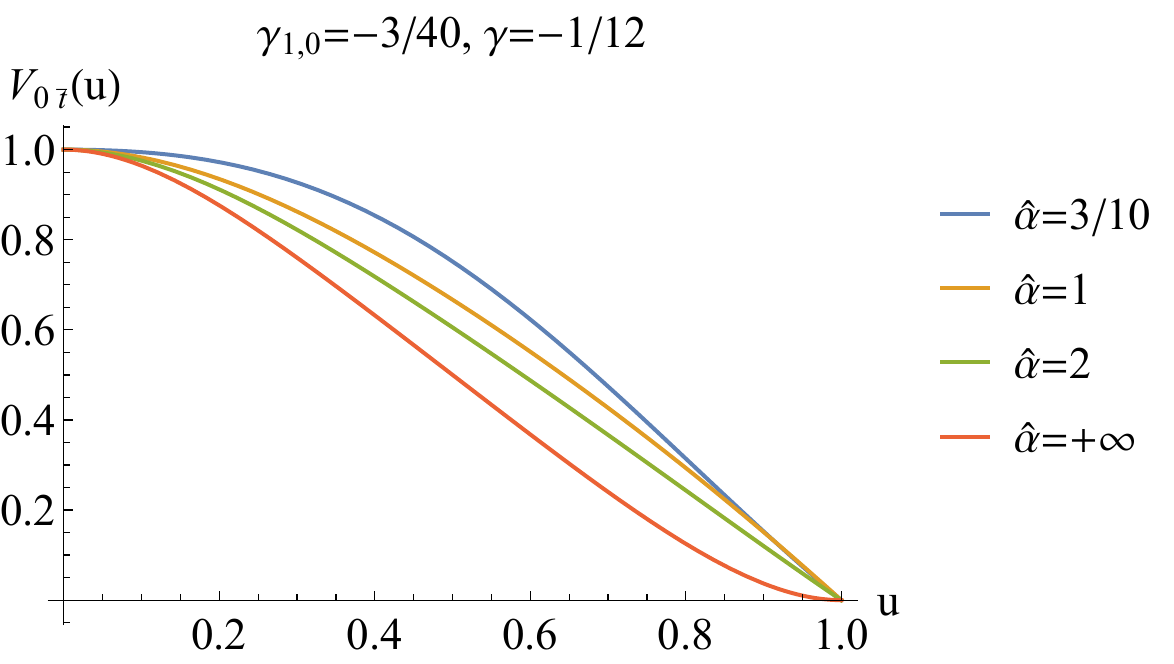}\ \hspace{0.8cm}
\includegraphics[scale=0.5]{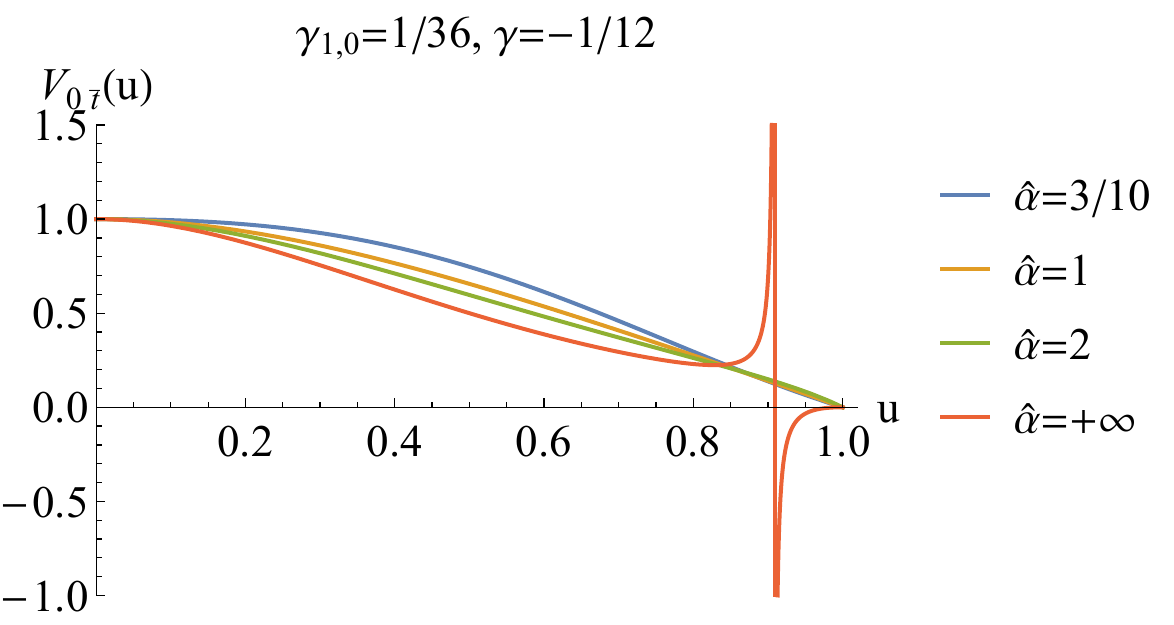}\ \hspace{0.8cm}\ \\
\includegraphics[scale=0.5]{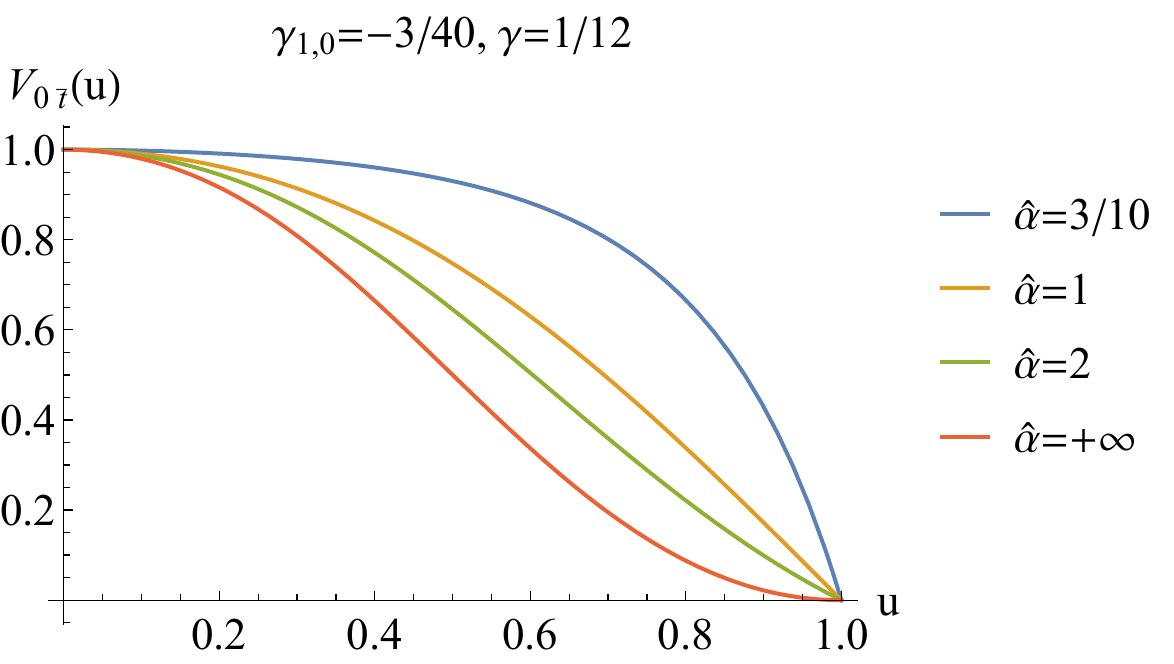}\ \hspace{0.8cm}
\includegraphics[scale=0.5]{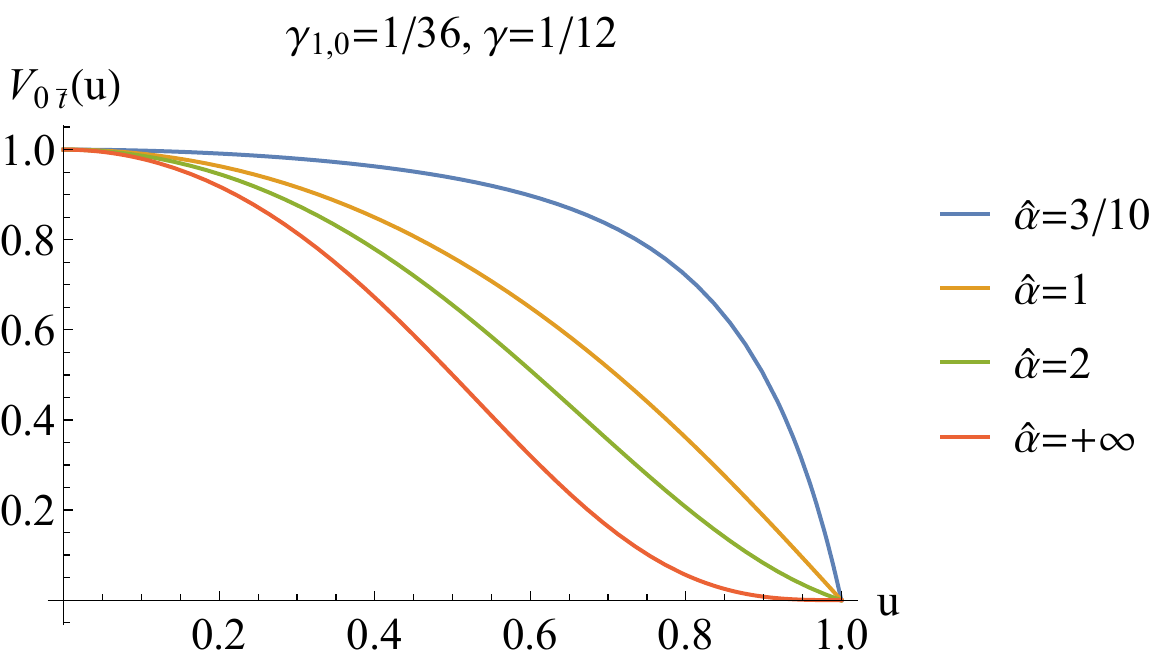}\ \hspace{0.8cm}\ \\
\caption{\label{V0tvsugamma10gamma} The potentials $V_{0\bar{t}}(u)$
with representative $\gamma_{1,0}$, $\gamma$ and $\hat{\alpha}$.
}}
\end{figure}
\begin{figure}
\center{
\includegraphics[scale=0.5]{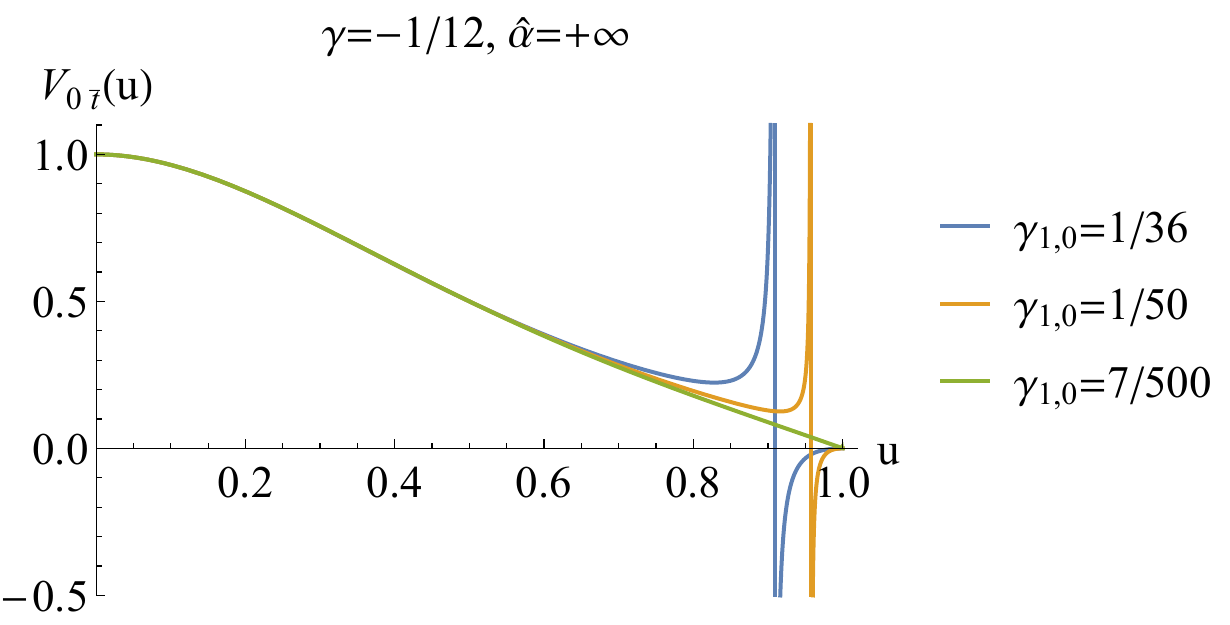}\ \hspace{0.8cm}\ \\
\caption{\label{V0tvsugamma10gammav5} The potentials $V_{0\bar{t}}(u)$
with $\gamma=-1/12$ and $\hat{\alpha}=+\infty$ and different $\gamma_{1,0}$.
}}
\end{figure}
\begin{figure}
\center{
\includegraphics[scale=0.5]{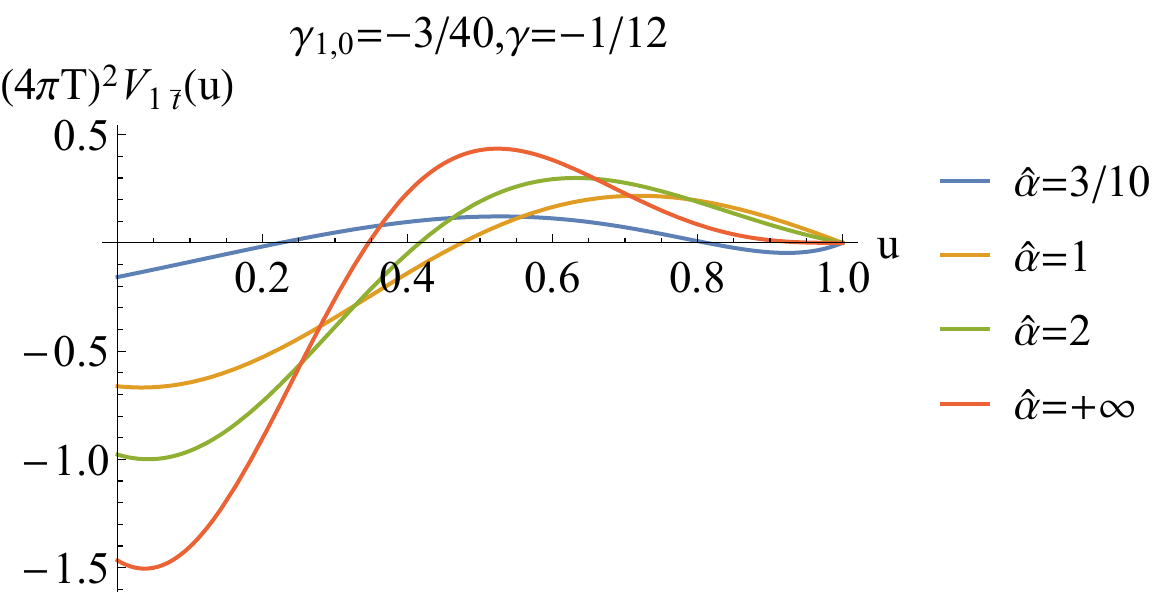}\ \hspace{0.8cm}
\includegraphics[scale=0.5]{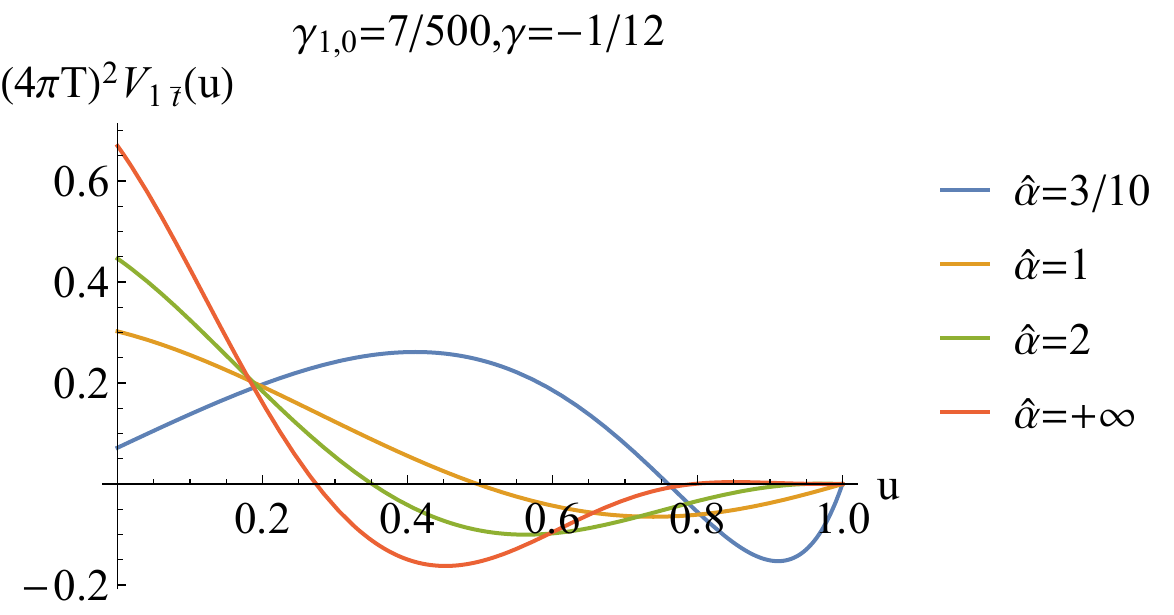}\ \hspace{0.8cm}\ \\
\includegraphics[scale=0.5]{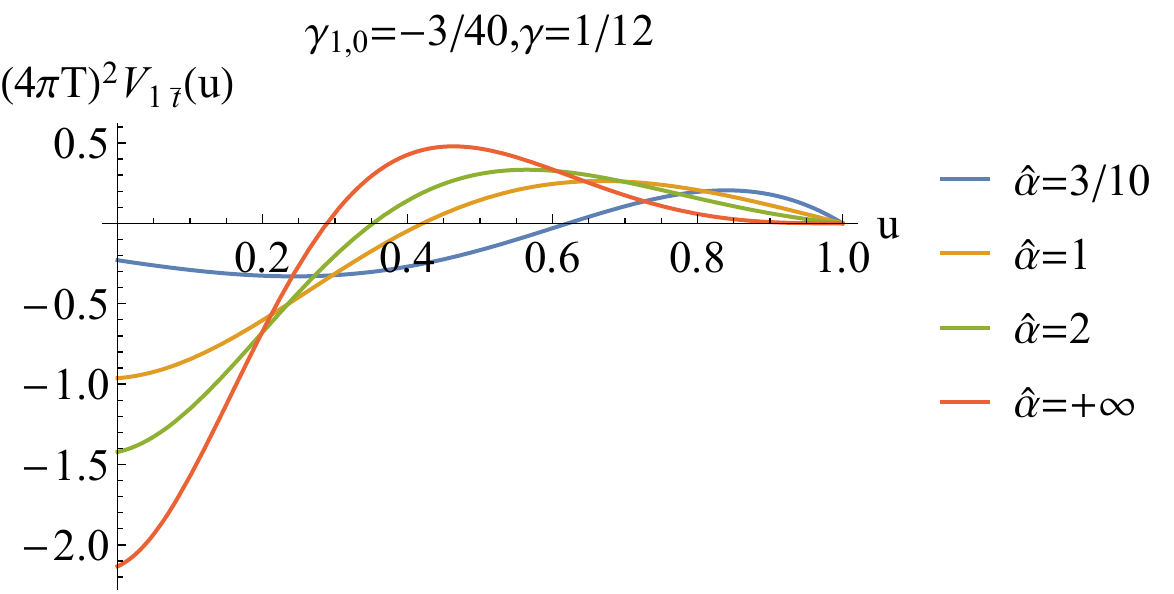}\ \hspace{0.8cm}
\includegraphics[scale=0.5]{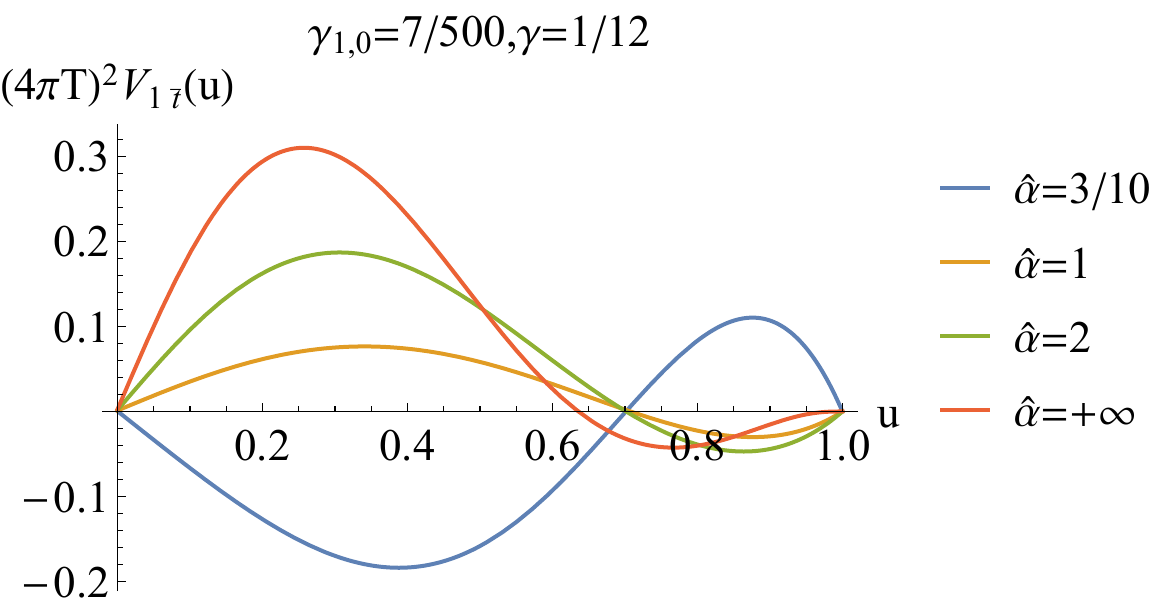}\ \hspace{0.8cm}\ \\
\caption{\label{V1tvsu_model_1} The potentials $(4\pi T)^2V_{1\bar{t}}(u)$
with representative $\gamma_{1,0}$, $\gamma$ and $\hat{\alpha}$.
}}
\end{figure}
\begin{figure}
\center{
\includegraphics[scale=0.5]{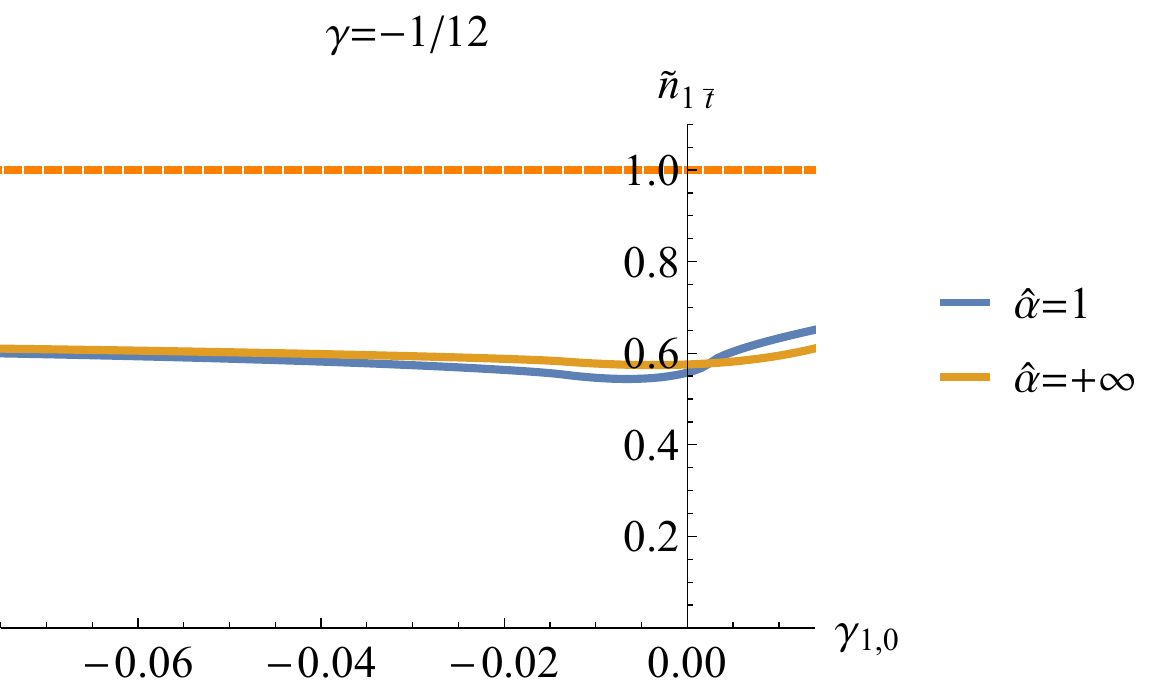}\ \hspace{0.8cm}
\includegraphics[scale=0.5]{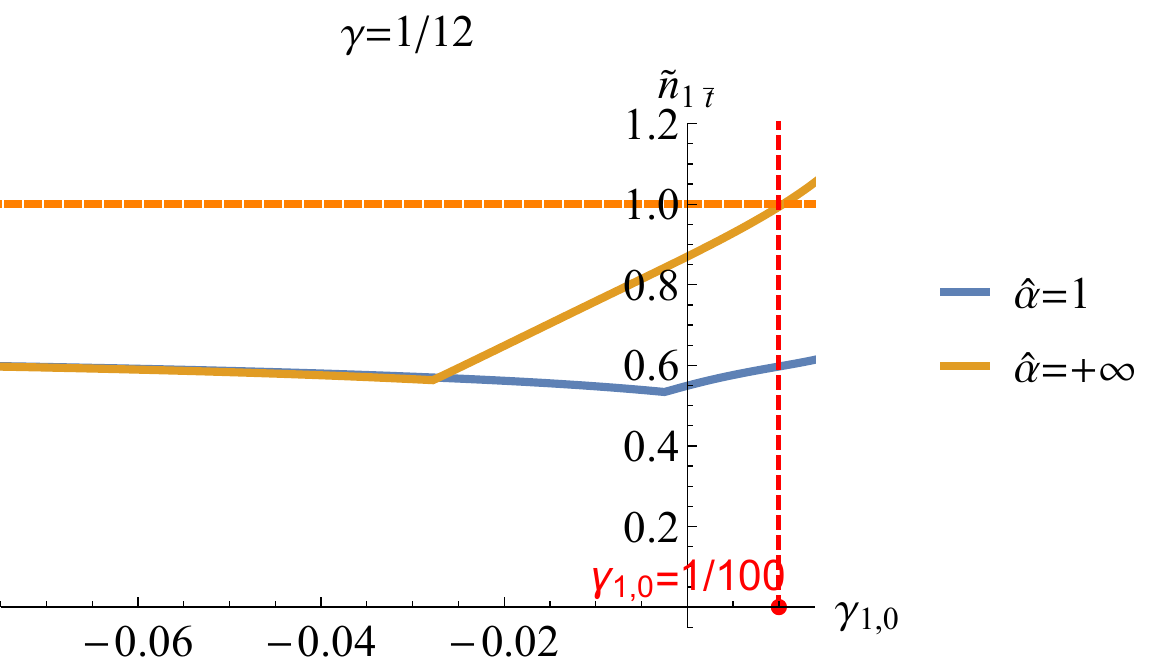}\ \hspace{0.8cm}\ \\
\caption{\label{n1tyvsgamma10_model_1} $\tilde{n}_{1\bar{t}}$ as the function $\gamma_{1,0}$ for representative $\gamma$ and $\hat{\alpha}$.
}}
\end{figure}

Now, we begin to discuss the bounds on the coupling when both $\gamma_{1,0}$ and $\gamma$ are turned on.
We mainly restrict $\gamma$ to the region $-1/12\leq\gamma\leq1/12$
and explore the constraint of $\gamma_{1,0}$.
First, in this case we derive the DC conductivity as
\begin{eqnarray}
&&
\sigma_0=1+\frac{2}{3}\Big(2+\frac{4(-1-2\hat{\alpha}^2+\sqrt{1+6\hat{\alpha}^2})}{\hat{\alpha}^2}\Big)\gamma
-4\hat{\alpha}^2\Big(2+\frac{-1-2\hat{\alpha}^2+\sqrt{1+6\hat{\alpha}^2}}{\hat{\alpha}^2}\Big)^2\gamma_{1,0}\,,
\nonumber
\\
\label{dc-gamma10-gamma}
\end{eqnarray}
which we plot as a function of $\hat{\alpha}$ for sample values of $\gamma_{1,0}$ and $\gamma$ in Fig. \ref{sigmavsalphabeta10beta}.
Since with the increases of $\hat{\alpha}$ the positive $\gamma$ lowers the DC conductivity,
it gives a tighter constraint on $\gamma_{1,0}$.
Specially, when $\gamma=1/12$, to have positive $\sigma_0$, $\gamma_{1,0}\leq 1/36$ should be imposed.
It can also be deduced thus: in the limit of $\hat{\alpha}\rightarrow+\infty$,
$\sigma_0=1-4\gamma-24\gamma_{1,0}$.
Second, we examine the potential $V_{0,\bar{t}}$,
which is shown in Fig. \ref{V0tvsugamma10gamma}.
We see that, for $\gamma_{1,0}=1/36$ and $\gamma=-1/12$,
an infinite positive and negative well appears in the limit of $\hat{\alpha}$,
which signals an instability.
This instability is due to turning on of $\gamma_{1,0}$ for $\gamma=-1/12$.
We find that, when we tune $\gamma_{1,0}$ to become smaller,
so that $\gamma_{1,0}\leq 7/500$,
the infinite well gradually disappears (see Fig. \ref{V0tvsugamma10gammav5}).
Therefore, if we set $-1/12\leq\gamma\leq 1/12$, then the constraint $-3/40\leq\gamma_{1,0}\leq 7/500$ should be imposed.
At the same time, it is easy to see that, for the above range of $\gamma$ and $\gamma_{1,0}$,
the potential $V_{0,y}$ satisfies the constraint (\ref{V0i-constraint}).
Third, we analyze the potential $V_{1\bar{t}}$,
which is shown in Fig. \ref{V1tvsu_model_1}.
We see that $V_{1\bar{t}}$ develops a negative minimum.
So to determine the range of parameter $\gamma_{1,0}$,
we study $\tilde{n}_{1\bar{t}}$ as a function of $\gamma_{1,0}$
for the representative values of $\gamma$ and $\hat{\alpha}$, which are plotted in Fig. \ref{n1tyvsgamma10_model_1}.
A detailed analysis indicates that, when $\gamma_{1,0}$ belongs to the region $-3/40\leq\gamma_{1,0}\leq1/100$,
no unstable mode appears.
Therefore, the constraint on $\gamma_{1,0}$ and $\gamma$ is
\fa
\label{con_gamma_gamma10}
-1/12\leq\gamma\leq1/12\,,~~~~~~-3/40\leq\gamma_{1,0}\leq 1/100\,.
\ffa

\subsubsection{Six derivative theory}

\begin{figure}
\center{
\includegraphics[scale=0.5]{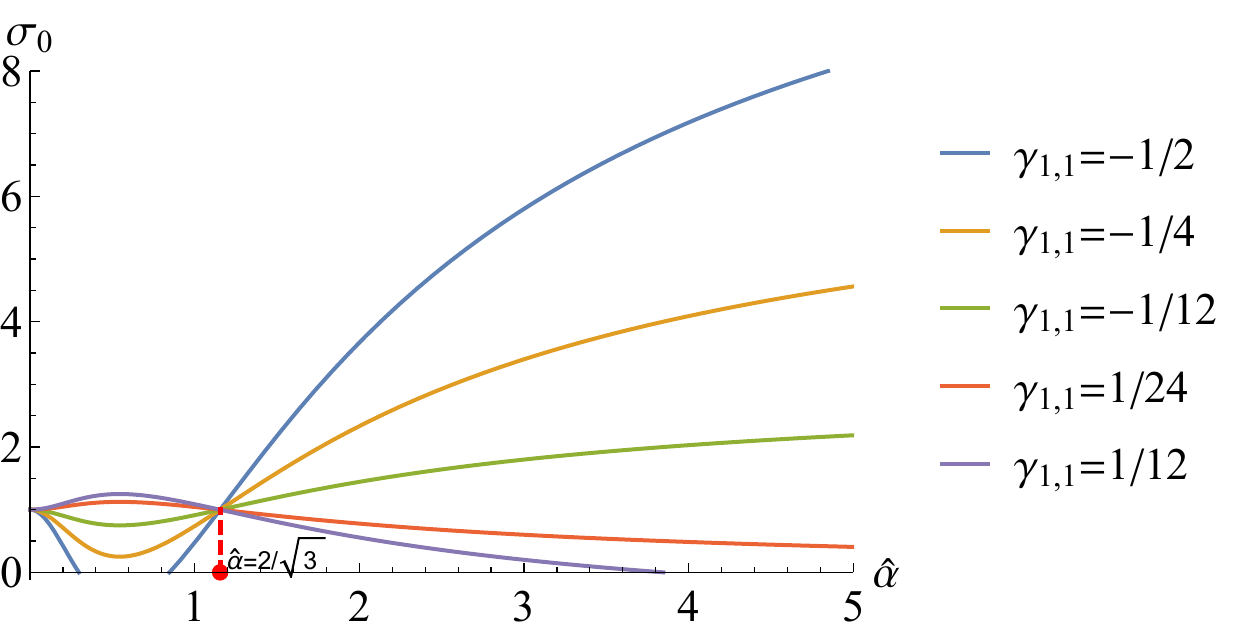}\ \hspace{0.8cm}\ \\
\caption{\label{sigmavsalphabeta11} The DC conductivity as a function of $\hat{\alpha}$ when only $\gamma_{1,1}$ are turned on.
}}
\end{figure}

In this subsection, we study the bounds on the coupling $\gamma_{1,1}$ with other coupling vanishing.
First, we derive the DC conductivity:
\begin{eqnarray}
\label{DC-gamma11}
\sigma_0=1-\frac{2}{3}\hat{\alpha}^2\Big(-2-\frac{4(-1-2\hat{\alpha}^2+\sqrt{1+6\hat{\alpha}^2})}{\hat{\alpha}^2}\Big)
\Big(2+\frac{-1-2\hat{\alpha}^2+\sqrt{1+6\hat{\alpha}^2}}{\hat{\alpha}^2}\Big)\gamma_{1,1}\,.
\end{eqnarray}
We plot it as a function of $\hat{\alpha}$ for sample values of $\gamma_{1,1}$ in FIG.\ref{sigmavsalphabeta11},
in which we see that there are lower and upper bounds set by DC conductivity.
By detailed analyzing, we find that $-1/3\leq\gamma_{1,1}\leq 1/24$.
Specially, the upper bound can be deduced from that in the limit of $\hat{\alpha}\rightarrow +\infty$,
$\sigma_0=1-24\gamma_{1,1}$.
\begin{figure}
\center{
\includegraphics[scale=0.5]{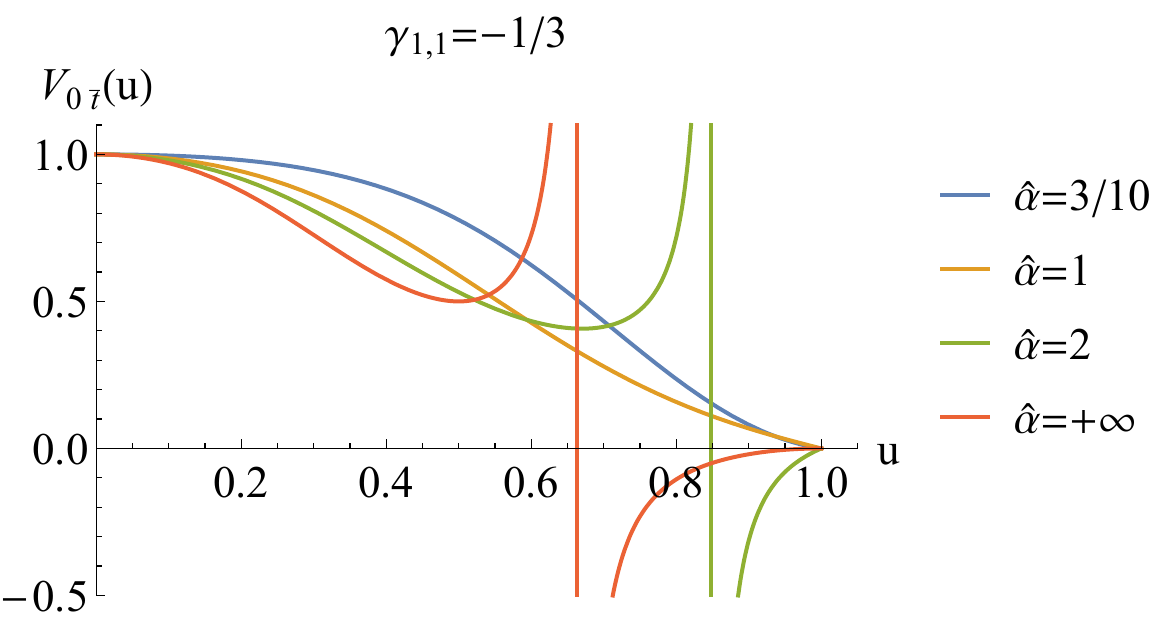}\ \hspace{0.8cm}
\includegraphics[scale=0.5]{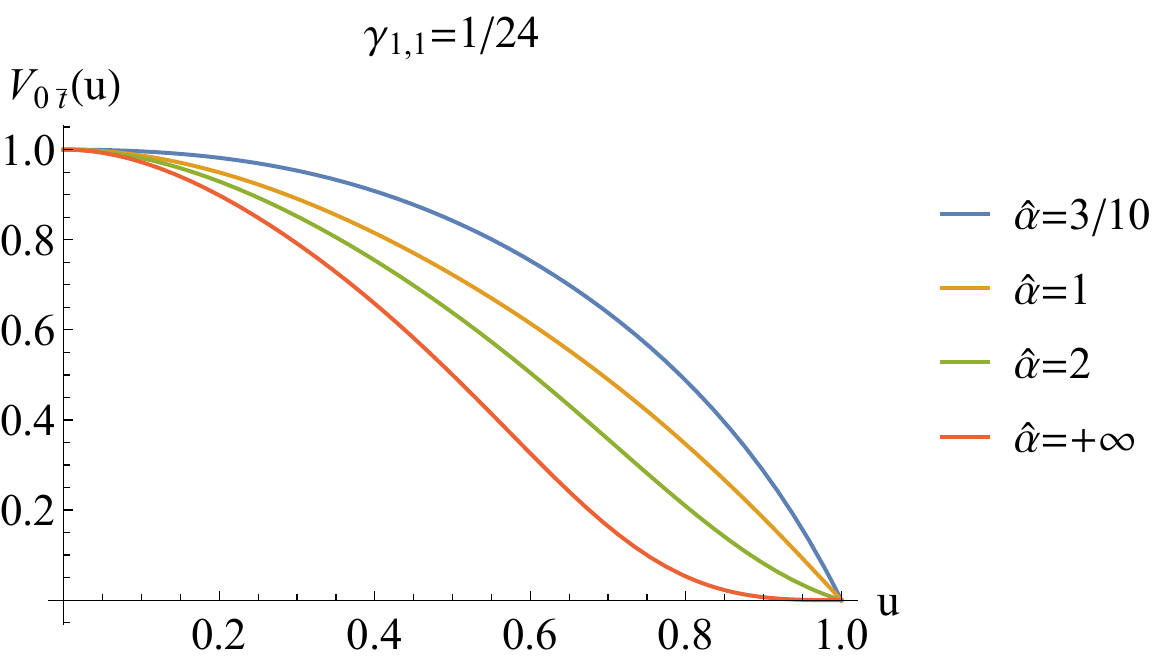}\ \hspace{0.8cm}\ \\
\caption{\label{V0tvsu_gamma11} The potentials $V_{0\bar{t}}(u)$ with representative $\gamma_{1,1}$ and $\hat{\alpha}$.
}}
\end{figure}

Then, we consider the constraint from $V_{0\bar{t}}$, which is shown in Fig.\ref{V0tvsu_gamma11}.
We see that, for $\gamma_{1,1}=-1/3$, with the increase of $\hat{\alpha}$, the condition (\ref{V0i-constraint}) is violated,
which indicates that a tighter lower bound should be imposed on $\gamma_{1,1}$.
A detailed examination indicates that $-1/50\leq\gamma_{1,1}\leq1/24$.
Also, we examine $V_{0y}$ for this range $\gamma_{1,1}\in[-1/50,1/24]$ and find that
it satisfies the condition (\ref{V0i-constraint}).
\begin{figure}
\center{
\includegraphics[scale=0.5]{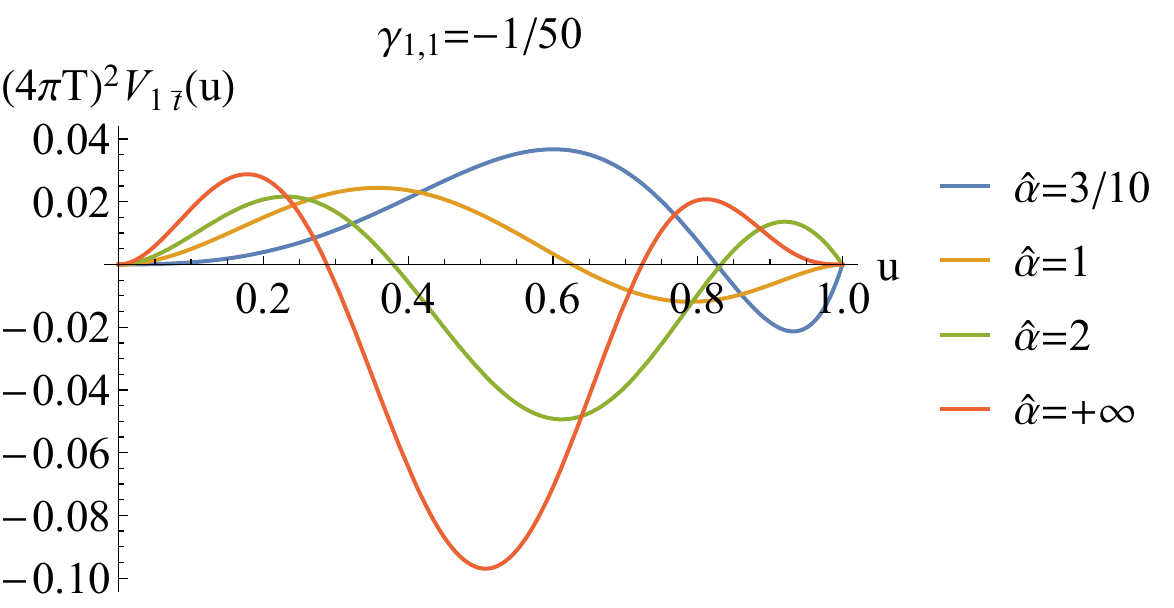}\ \hspace{0.8cm}
\includegraphics[scale=0.5]{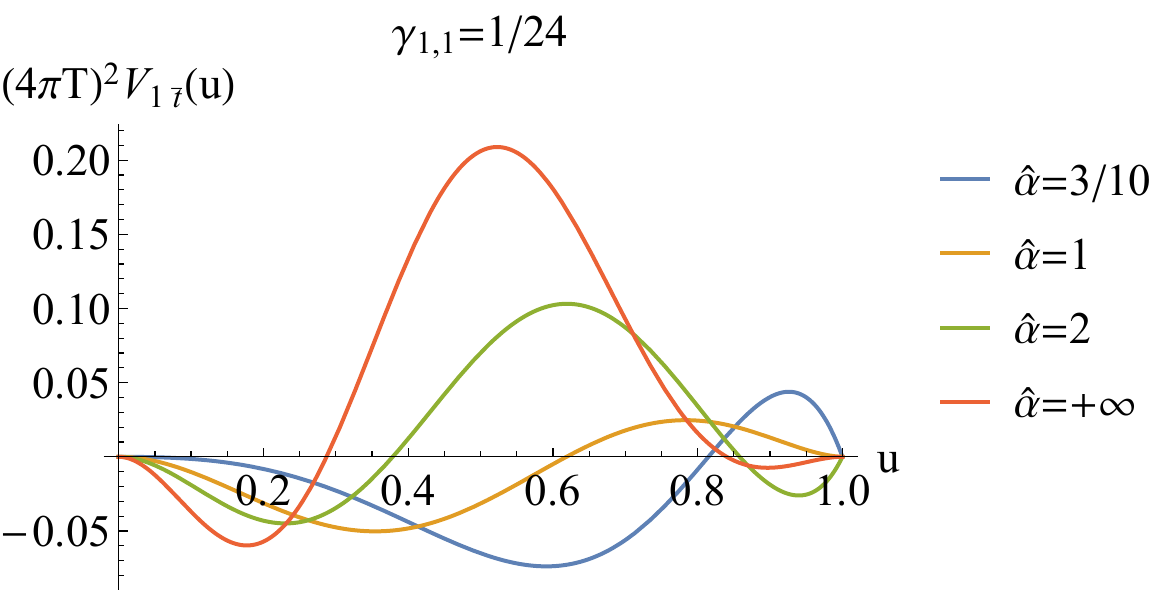}\ \hspace{0.8cm}\ \\
\caption{\label{V1tvsu_model_3} The potentials $(4\pi T)^2V_{1\bar{t}}(u)$
with representative $\gamma_{1,1}$ and $\hat{\alpha}$.
}}
\end{figure}
\begin{figure}
\center{
\includegraphics[scale=0.5]{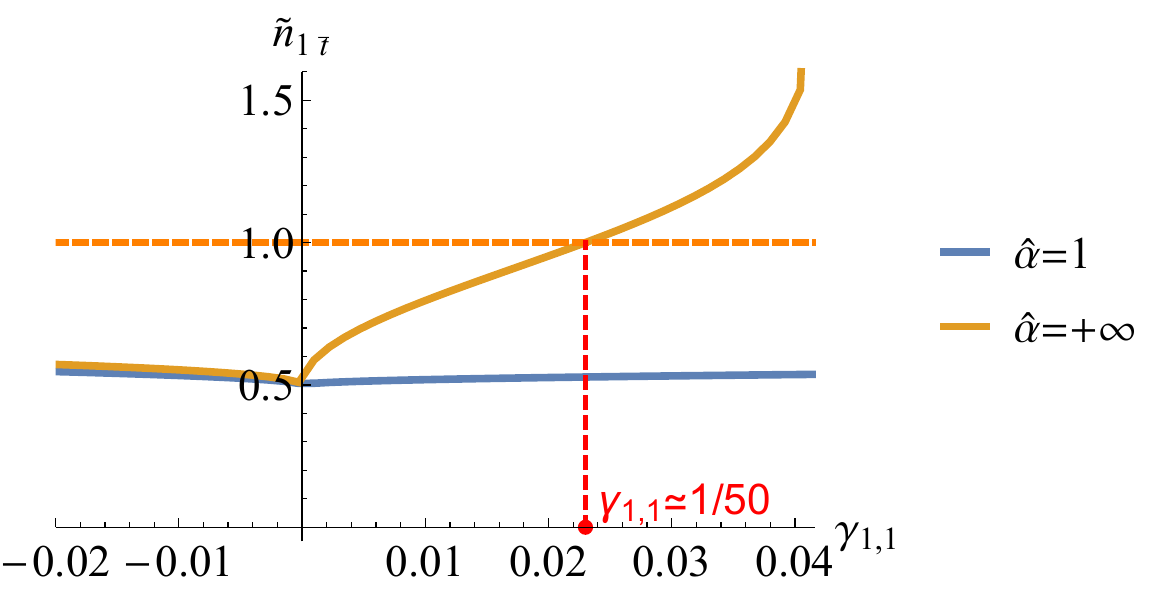}\ \hspace{0.8cm}\ \\
\caption{\label{n1tvsgamma11_model3} $\tilde{n}_{1\bar{t}}$ as the function $\gamma_{1,1}$ for representative $\hat{\alpha}$.
}}
\end{figure}

Now, we examine the potential $V_{1,\bar{t}}$, which we plot in Fig. \ref{V1tvsu_model_3}.
As in the previous case, a negative minimum appears in $V_{1,\bar{t}}$.
So we further plot $\tilde{n}_{1\bar{t}}$ as a function of $\gamma_{1,1}$
for the representative values of $\hat{\alpha}$,
which are shown in FIG.\ref{n1tvsgamma11_model3}.
We find that, for the range
\fa
\label{con_gamma11}
-1/50\leq\gamma_{1,1}\leq 1/50\,,
\ffa
no unstable mode appears.
A similar analysis also indicates that, for $\gamma_{1,1}$ satisfying the constraint \eqref{con_gamma11},
$\tilde{n}_{1y}\leq 1$.
In addition, this range of $\gamma_{1,1}$ is also a physically viable region for finite momentum.

\subsection{Bounds on the coupling at finite charge density}

In this section, we discuss the bounds on the coupling at finite charge density on top of the perturbative black brane geometry in Sect. \ref{sec-BBS}.
Since the perturbative equations of vectors involve a set of third order differential equations with high nonlinearity,
it is hard to decoupling them at finite charge density, like that at zero charge density.
Therefore, it is difficult to study the bounds on the coupling at finite charge density by the method of Schr\"odinger potentials at zero charge density as Appendix \ref{sec-Bounds-zero}
or the quasi-normal modes of vector modes. We hope that these problems can be worked out in the future.
Here, we only give the constraints on the coupling parameters at finite charge density from the requirement that the mass of the graviton is real.

\begin{figure}
\center{
\includegraphics[scale=0.5]{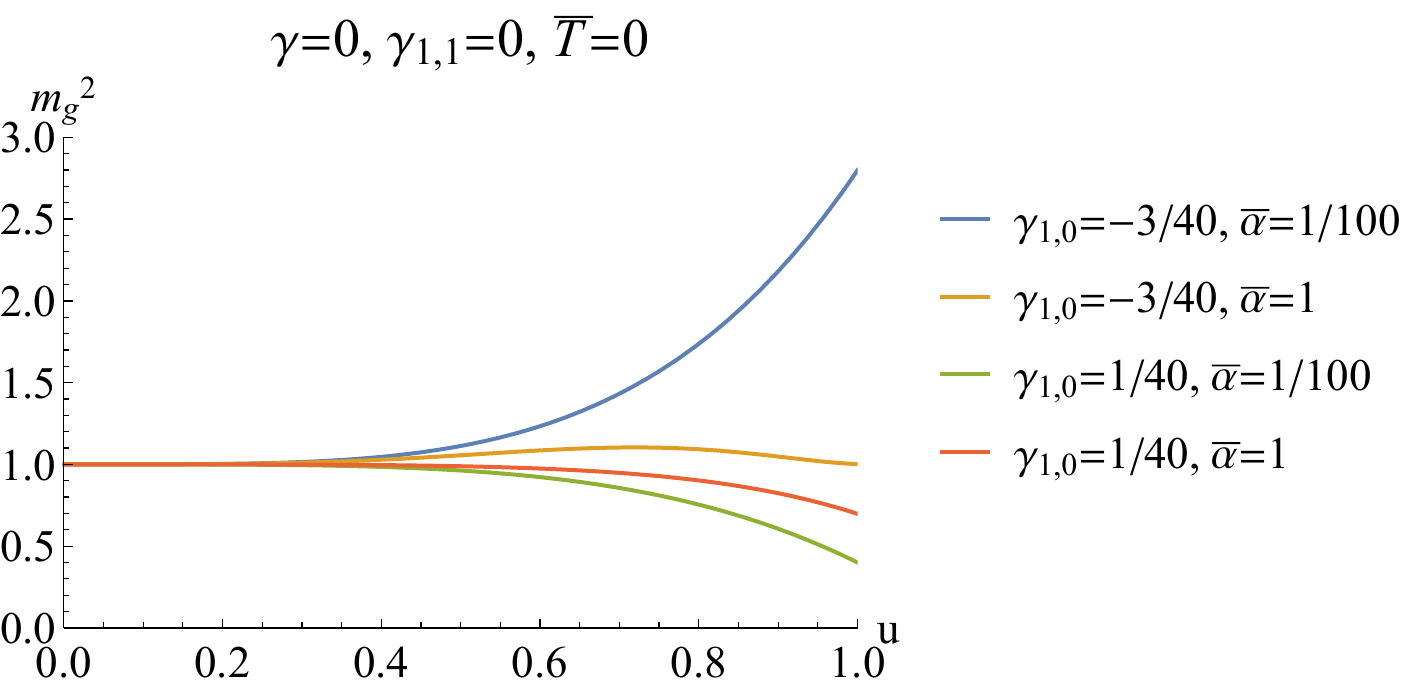}\ \hspace{0.8cm}\ \\
\caption{\label{mg_gamma10} $m_g^2$ as a function of $u$ at zero temperature for representative $\bar{\alpha}$ and $\gamma_{1,0}$ ($\gamma=0$ and $\gamma_{1,1}=0$).
}}
\end{figure}

It has been demonstrated in \cite{Blake:2013owa} (also refer to \cite{Baggioli:2014roa,Alberte:2015isw}) that the holographic lattices give the graviton an effective mass.
In our present model \eqref{action}, the effective graviton mass is
\fa
\label{gra-mass}
m_g^2=1-\frac{1}{2}\gamma_{1,0}I_{\mu\nu}^{\ \ \rho\sigma}F^{\mu\nu}F_{\rho\sigma}-\gamma_{1,1}C_{\mu\nu}^{\ \ \rho\sigma}F^{\mu\nu}F_{\rho\sigma}\,.
\ffa
Obviously, $m_g^2>0$ for the case of zero charge density.
In what follows, we shall discuss the bounds on the coupling parameters at finite charge density.

We first turn on $\gamma_{1,0}$.
Figure \ref{mg_gamma10} shows $m_g^2$ as a function of $u$ at zero temperature for representative $\bar{\alpha}$ and $\gamma_{1,0}$.
We can see that $m_g^2>0$ when $\gamma_{1,0}$ satisfies the constraint \eqref{con_gamma10}, which is the constraint at zero charge density.

\begin{figure}
\center{
\includegraphics[scale=0.5]{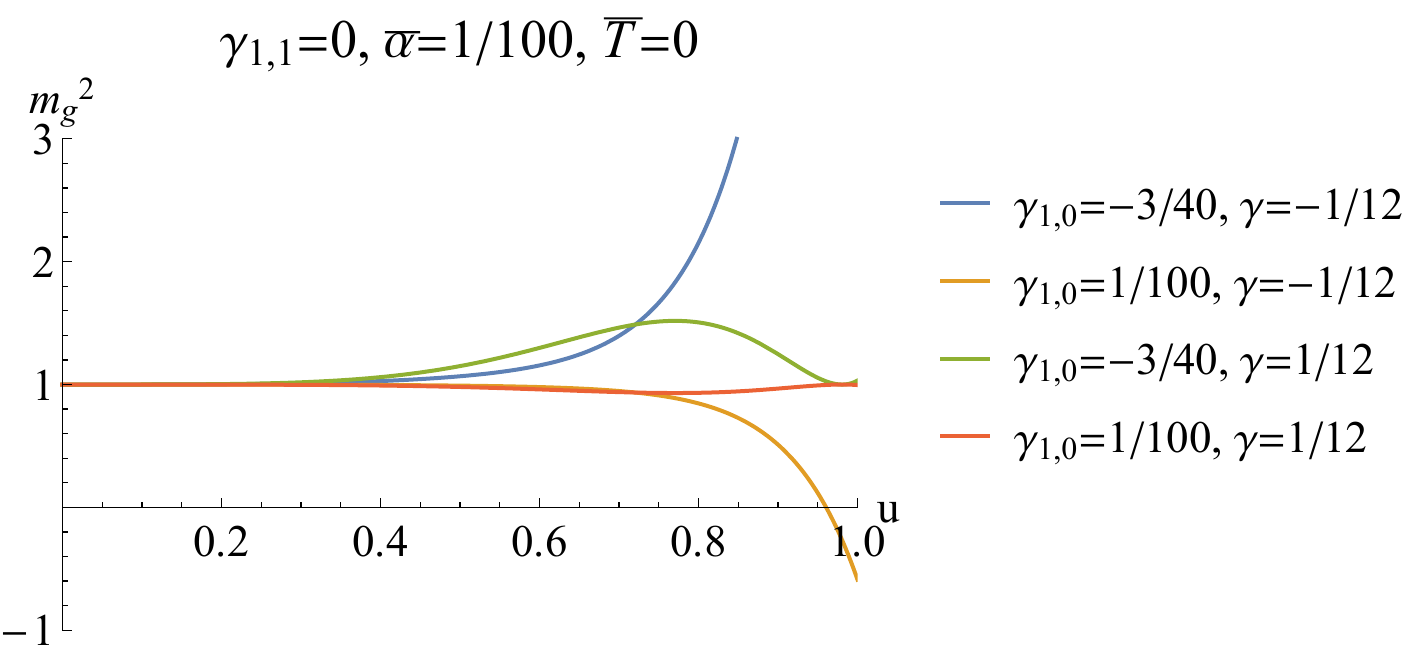}\ \hspace{0.8cm}
\includegraphics[scale=0.5]{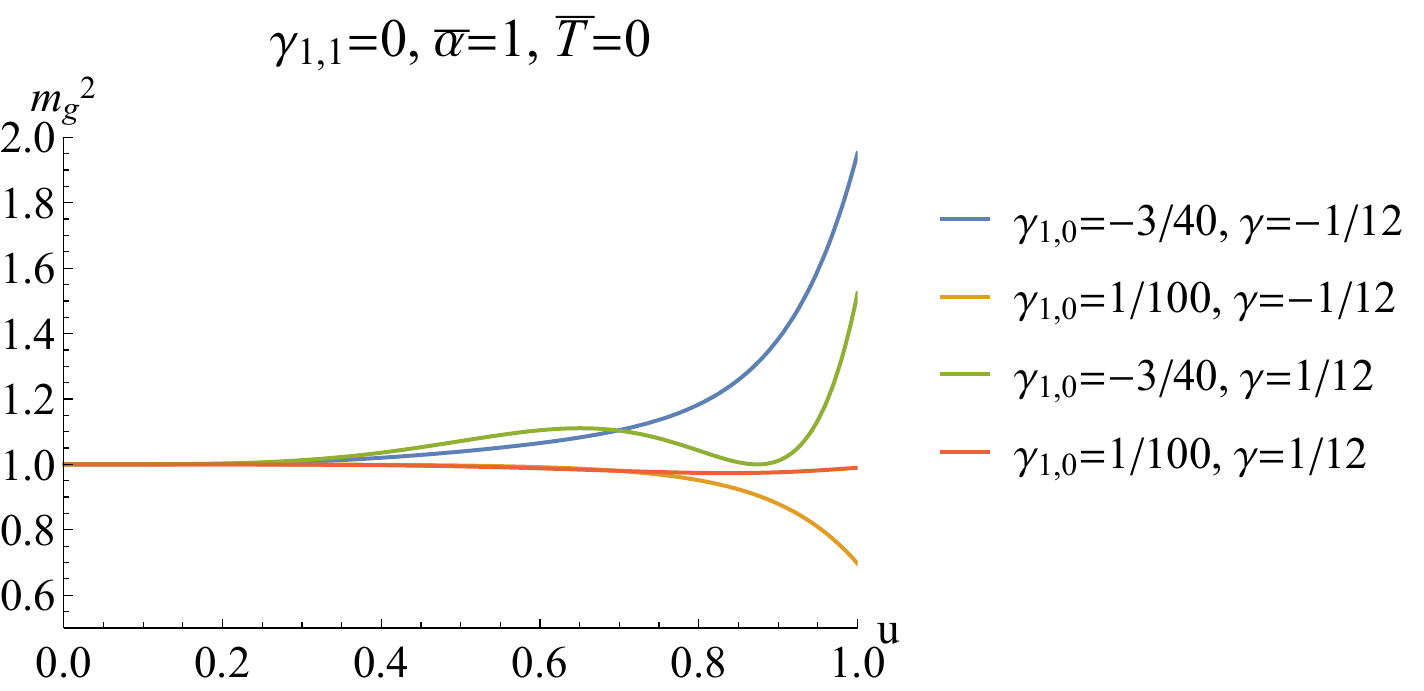}\ \hspace{0.8cm}\ \\
\caption{\label{mg_gamma10_gamma11} $m_g^2$ as a function of $u$ at zero temperature for representative $\bar{\alpha}$, $\gamma_{1,0}$ and $\gamma$ ($\gamma_{1,1}=0$).
}}
\end{figure}
\begin{figure}
\center{
\includegraphics[scale=0.5]{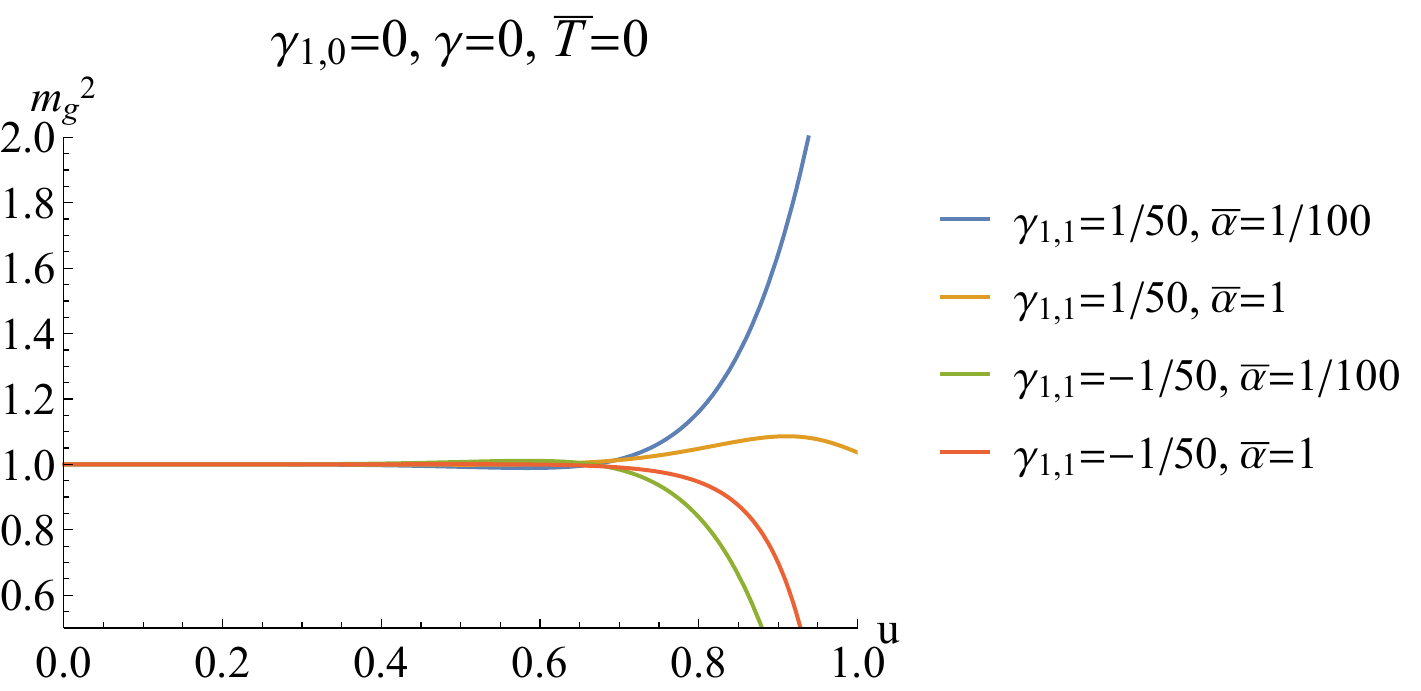}\ \hspace{0.8cm}\
\includegraphics[scale=0.5]{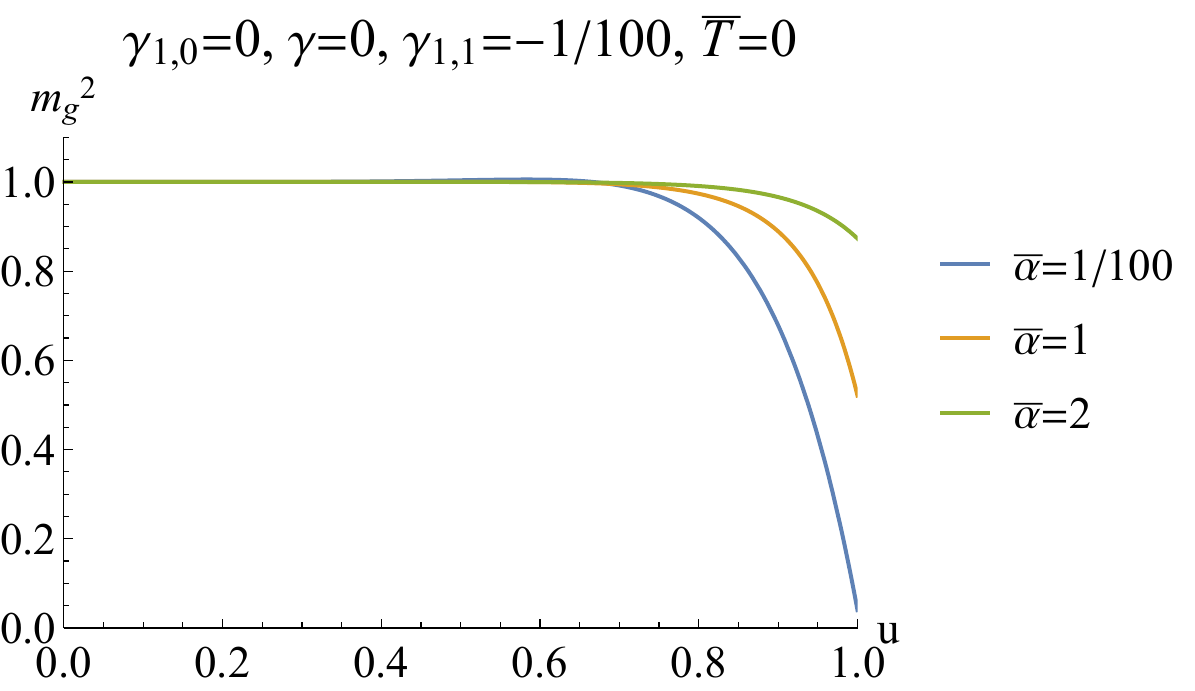}\ \hspace{0.8cm}\ \\
\caption{\label{mg_gamma11} Left plot: $m_g^2$ as a function of $u$ at zero temperature for representative $\bar{\alpha}$ and $\gamma_{1,1}$ ($\gamma_{1,0}=0$ and $\gamma=0$).
Right plot: $m_g^2$ as a function of $u$ at zero temperature for $\gamma_{1,1}=-1/100$ and different $\bar{\alpha}$ ($\gamma_{1,0}=0$ and $\gamma=0$).}}
\end{figure}

Second, we turn on both $\gamma_{1,0}$ and $\gamma$. Figure \ref{mg_gamma10_gamma11} shows that if $\gamma_{1,0}$ and $\gamma$ satisfy the constraint \eqref{con_gamma_gamma10}, set at zero charge density,
$m_g^2>0$ for large $\bar{\alpha}$ (right plot in Fig. \ref{mg_gamma10_gamma11}),
but $m_g^2>0$ is violated for small $\bar{\alpha}$ (left plot in Fig. \ref{mg_gamma10_gamma11}).
By detailed analysis, we constrain $\gamma_{1,0}$ and $\gamma$ in the region
\fa
\label{con_gamma_gamma10_finite}
-1/12\leq\gamma\leq1/12\,,~~~~~~-3/40\leq\gamma_{1,0}\leq 6/1000\,.
\ffa

Finally, we analyze the constraint on $\gamma_{1,1}$ ($\gamma_{1,0}$ and $\gamma$ are turned off).
The left plot in Fig. \ref{mg_gamma11} shows that, when $\gamma_{1,1}$ reaches its lower bound set at zero charge density,
$m_g^2$ becomes negative for $u$ approaching the horizon.
Further analysis indicates that if $\gamma_{1,1}$ satisfies
\fa
\label{con_gamma11_finite}
-1/100\leq\gamma_{1,1}\leq 1/50\,,
\ffa
$m_g^2>0$ (see the right plot in Fig. \ref{mg_gamma11}).

\end{appendix}

\end{document}